\documentclass[aps,preprint,onecolumn,secnumarabic,nobalancelastpage,amsmath,amssymb,
nofootinbib]{revtex4}

\usepackage{pst-plot}
\usepackage{pstricks,pstricks-add}
\usepackage[scriptsize,nooneline,hang]{caption}      
\usepackage[hang,nooneline,scriptsize]{subfigure}
     
\usepackage{graphics}
\usepackage{graphicx}     
\usepackage{url}
\usepackage{dcolumn}         
\usepackage{hyperref} 
\usepackage{bm}               
\usepackage{mathrsfs}
\usepackage{epstopdf}
\usepackage{hyperref}         
\usepackage{color}

\begin{document}

reprint{APS/123-QED}

\title{Shadow of a Charged Rotating Black Hole in $f(R)$ Gravity}

\author{Sara Dastan${}^1$}
\author{Reza Saffari${}^1$}%
\email{rsk@guilan.ac.ir}
\author{Saheb Soroushfar${}^2$}
 \affiliation{${}^1$ Department of Physics, University of Guilan, 41335-1914, Rasht, Iran.\\
${}^2$ Department of Mining and Technology, Yasouj University, 75761, Choram, Iran.
}%



\date{\today}

\begin{abstract}
We study the shadow of a charged rotating black hole in $f(R)$ gravity. This black hole is characterized by mass, $M$, spin, $a$, electric charge, $Q$ and $R_{0}$ which is proportional to cosmological constant. We analyze the image of the black hole shadow in four types 1) at $r\rightarrow \infty$, 2) at $r\rightarrow r_{O}$ in vacuum, 3) at $r\rightarrow \infty$ and 4) at $r\rightarrow r_{O}$ for an observer at the presence of plasma. Moreover, we investigate the effect of spin, charge and modification of gravity on the shape of shadow. In addition, we use two observables, the radius $R_{s}$ and the distortion parameter $\delta_{s}$, characterizing the apparent shape. We show that for all cases, the shadow becomes smaller with increasing electric charge. Also, by increasing the rotation parameters, circular symmetry of the image of black hole's shadow will change. Furthermore, in the presence of  plasma, plasma parameter also effects on size of the shadow.   
\end{abstract}

\maketitle

\section{INTRODUCTION}
Is it possible to analyze the shadow of an object which had not been observed yet? 
The study of black holes is as old as the theory of general relativity, with lots of research and literature in this field, but there are still many unknown issues. The main problem in the study of black holes, is lack of observational data. 
According to predictions of general relativity, there are two observational methods to collect information about the black holes, i.e. gravitational lensing and detection of gravitational waves. The possibility of the existence of a black hole in an area in space increases by detection of gravitational lensing effect and gravitational waves; such as, the observational \textit{LISA} project data \cite{Antonucci:2011zza}, which was proof of binary black holes.
Recently, observer are motivated to look for the possible black hole in the center of Milky Way. There are two projects which called \textit{Event Horizon Telescope} (EHT)~\cite{Doeleman:2008qh,EHT:2008qh} and European \textit{BlackHoleCam} (BHC)~\cite{BHC:2008qh}, have started to collect data in this field. The aim of these projects are obtaining another observational evidence for black holes, that it is called "shadow" of black holes. To understand the concept of shadow, we should provide a correct definition of it. Suppose a light source at $r_{L}$ and an observer at $r_{O}$, where $r_{L}>r_{O}$. Therefore, the light rays have two paths;
1) They are deflected by black hole and back to the light source, so we have brightness in the observer's sky, 2) They go to the event horizon and they do not come back to the light source, in this case, the observer's sky have darkness. The darkness in the observer's sky called shadow.
Already, several black holes were studied in pure gravity and at least properties of most of them have been analyzed theoretically, so theoretical study of shadow for these black holes are possible.
This approach effects on a better understanding of the theory of gravity and black holes, also, the shape of  different black holes’ shadow and the impression of various parameters can be investigated.
It is shown that the image of Schwarzschild black hole's shadow is circular and have a photon sphere~\cite{synge}, while the Kerr black hole, has a photon region, and it doesn't have a circular shadow image, which means that it can take into account for the deviation from circular symmetry~\cite{Chandrasekhar:1985kt}. Lensing in Schwarzschild black hole~\cite{Virbhadra:1999nm} and geometry of photon surface~\cite{Claudel:2000yi} had been studied too. So, shadow deviation from circular form, can determine the spin parameter of black holes.
In this story, there are many studies on different black holes, such as, Kerr~\cite{Bardeen}, Kerr-Newman~\cite{kerr-newman}, Kerr-NUT~\cite{Abdujabbarov:2012bn}, regular black hole~\cite{Li:2013jra}, multi-black hole~\cite{Yumoto:2012kz}, black holes in extended Chern-Simons modified gravity~\cite{Amarilla:2010zq}, Randall-Sundrum braneworld~\cite{Amarilla:2011fx} and Kerr black holes with scalar hair~\cite{Cunha:2015yba,Vincent:2016sjq}.
Deviation of circular symmetry and the changes in the size of shadow image is defined as $\delta_{s}$ and $R_{s}$, by Hioki and Maeda~\cite{Hioki:2009na}. 
However, in most of investigations on the study of black hole's shadow, the location of observer is at infinity. In this work, we study the case of observer at infinity and the case of observer at limited distance, following Ref.~\cite{Grenzebach:2014fha}.
The study of black holes, has been considered not only in general relativity, but also in extended and alternative theories such as, braneworld cosmology~\cite{Brax:2003fv}, Lovelock gravity~\cite{Lovelock:1971yv}, scalar tensor~\cite{Brans:1961sx} and $f(R)$ gravity.
In fact, these theories are replaced and refined to justify some subjects like, cosmic acceleration,
dark matter, cosmic inflation and the solar system abnormalities~\cite{Riess:1998cb,Buchdahl:1983zz,Starobinsky:1980te,Bamba:2008ja,Akbar:2006mq,Saffari:2007zt,Chakraborty:2014xla,Chakraborty:2015bja}. However, the properties of black holes always has been considered in these theories too, such as \cite{Soroushfar:2015wqa} and \cite{Soroushfar:2016nbu}.
In this paper, we investigate shadow of a charged rotating black hole in $f(R)$ gravity in the absence and the presence of plasma for an observer at infinity and especially in a limited distance. Here, we consider, there is no light source close to the black hole, which means, we use the light like geodesic as path of the incident light rays.
This paper is organized as follow, In Sect.\ref{gravity} and .\ref{geodesic}, we summarize the properties of $f(R)$ gravity and its geodesics. In Sect.\ref{infinite}, we calculate shadow for an observer at $r=\infty$. In Sect.\ref{limited}, we obtain an analytical formula for an observer in $r =r_{O}$ . In Sect.\ref{plasma} and .\ref{plasma limited}, we analyze these situations in the presence of plasma and our results conclude in Sect.\ref{conclusion}.

\section{FIELD EQUATIONS IN $f(R)$ MODIFIED GRAVITY}\label{gravity}

In this section, we study field equation and metric in f(R) gravity. The action with Maxwell term is
\begin{align}\label{1}
S=S_{g}+S_{M},
\end{align}
where, $S_{g}$ and $S_{M}$ are the gravitational action and the electromagnetic actions as
\begin{equation}\label{2}
S_{g}=\dfrac{1}{16 \pi} \int d^{D} x \sqrt{\vert g \vert }(R+f(R)),
\end{equation}
\begin{equation}\label{3}
S_{M}=\dfrac{-1}{16 \pi} \int d^{4} x \sqrt{-g }[F_{\mu\nu}F^{\mu\nu}],
\end{equation}
where, $R$ is the scalar curvature, $R + f(R)$, is the function defining the theory under consideration,  and $g$ is the determinant of the metric.
The Maxwell and field equation are
\begin{equation}\label{4}
\nabla_{\mu}F^{\mu\nu}=0,
\end{equation}

\begin{align}\label{trace}
R_{\mu\nu}\big(1+ f'(R)\big)-\frac{1}{2}\big(R+f(R)\big)g_{\mu\nu}
+\big(g_{\mu\nu}\nabla^{2}-\nabla_{\mu}\nabla_{\nu}\big)f'(R)=2T_{\mu\nu},
\end{align}
where, $\nabla$ is the usual covariant derivative, $R_{\mu\nu}$, is the Ricci tensor and the stress-energy tensor of the electromagnetic field is given by
\begin{equation}\label{6}
T_{\mu\nu}=F_{\mu\rho}F_{\nu}^{\rho}-\dfrac{g_{\mu\nu}}{4}F_{\rho\sigma}F^{\rho\sigma},
\end{equation}
with
\begin{equation}\label{7}
T^{\mu} _{\mu}=0 .
\end{equation}
the constant curvature scalar $R=R_{0}$ and The trace of Eq.~(\ref{trace}), leads to
\begin{equation}\label{8}
R_{0}\big(1+f'(R_{0})\big)-2\big(R_{0}+f(R_{0})\big)=0,
\end{equation}
which introduces the negative constant curvature scalar as
\begin{equation}\label{9}
R_{0}=\dfrac{2f(R_{0})}{f'(R_{0})-1} .
\end{equation}
Using this relation in equation~(\ref{trace}) gives the Ricci tensor
\begin{align}\label{10}
R_{\mu\nu}=\dfrac{1}{2}\big(\dfrac{f(R_{0})}{f'(R_{0})-1}\big)g_{\mu\nu}
+\dfrac{2}{\big(1+f'(R_{0})\big)}T_{\mu\nu} .
\end{align}
Finally, Alexis Larranaga~\cite{Larranaga:2011fv}, introduced the axisymmetric ansatz in Boyer--Lindquist--type coordinates $(t,r,\theta,\varphi)$ inspired by the Kerr-Newman-AdS black hole solution as
\begin{align}\label{11}
ds^{2}=-\dfrac{\Delta_{r}}{\rho^{2}} \big[dt-\dfrac{a sin^{2}\theta
d\varphi}{\Xi}\big]^{2}
+\dfrac{\rho^{2}}{\Delta_{r}}dr^{2}+\dfrac{\rho^{2}}{\Delta_{\theta}}d\theta^{2}
+\dfrac{\Delta_{\theta}sin^{2}\theta}{\rho^{2}}\big[adt-\dfrac{r^{2}+a^{2}}{\Xi}d\varphi
\big]^{2}.
\end{align}
where
\begin{align}\label{12}
\Delta_{r}=(r^{2}+a^{2}) \big(1+\dfrac{R_{0}}{12}r^{2} \big)-2Mr+\dfrac{Q^{2}}{(1+f'(R_{0}))},
\end{align}
\begin{align}\label{13}
\Xi=1-\dfrac{R_{0}}{12}a^{2}, \qquad
\rho^{2}=r^{2}+a^{2}cos^{2}\theta, \qquad
\Delta_{\theta}=1-\dfrac{R_{0}}{12}a^{2}cos^{2}\theta ,
\end{align}
which, $ a $ is the angular momentum per mass of the black hole, $R_{0}$ is a constant proper to cosmological constant ($R_{0}=-4\Lambda$) and $ Q $ is the electric charge.
\section{The geodesic equations}\label{geodesic}

In this section, we study the geodesic equation and introduce effective potential.
The Hamilton--Jacobi equation is
\begin{equation}\label{14}
\dfrac{\partial S}{\partial\tau} +\frac{1}{2} \ g^{ij}\dfrac{\partial S}{\partial x^{i}}\dfrac{\partial S}{\partial x^{j}}=0, 
\end{equation}
Eq.~(\ref{14}), can be solved with an ansatz for the action
\begin{equation}\label{15}
S=\frac{1}{2}\varepsilon \tau - Et+L_{z}\phi +S_{\theta}(\theta) + S_{r} (r).
\end{equation}
 The angular momentum $L$ and the energy $E$, are the constants of motion as
\begin{equation}\label{16}
g_{tt}\dot{t}+g_{t \varphi}\dot{\varphi}=-E,  \qquad g_{\varphi \varphi}\dot\varphi +g_{t \varphi}\dot{t} =L.
\end{equation}
By substituting Eq.~(\ref{15}) and ~(\ref{16}) in Eq.~(\ref{14}), we get
\begin{align}\label{17}
\Delta_{\theta}(\dfrac{ds}{d\theta})^{2}+\varepsilon a^{2}cos^{2}\theta -\dfrac{2aEL\Xi - E^{2}a^{2}sin^{2}\theta}{\Delta_{\theta}}+\dfrac{L^{2}\Xi^{2}}{
\Delta_{\theta}sin^{2}\theta}=-\Delta_{r}(\dfrac{ds}{dr})^{2}-\nonumber\\ \varepsilon r^{2}+\dfrac{(a^{2}+r^{2})^{2}E^{2}+a^{2}L^{2} \Xi^{2}-2aEL\Xi (r^{2}+a^{2})}{\Delta_{r}} ,
\end{align}
where each side depends on $r$ or $\theta$ only. We derive the equations of motion using the separation ansatz Eq.~(\ref{15}), and with the help of the Carter constant~\cite{B.Carter}
\begin{align}\label{18}
\rho^{4}(\dfrac{dr}{d\tau})^{2}=-\Delta_{r}(K+\varepsilon r^{2})+\big[(a^{2}+r^{2})E-aL\Xi \big]^{2}=R(r),
\end{align}
\begin{align}\label{19}
\rho^{4}(\dfrac{d\theta}{d\tau})^{2}=\Delta_{\theta}(K-\varepsilon a^{2}cos^{2}\theta)-\dfrac{1}{sin^{2}\theta}\big(aE sin^{2}\theta -L\Xi \big)^{2} =\Theta(\theta),
\end{align}
\begin{align}\label{20}
\rho^{2}(\dfrac{d\varphi}{d\tau})=\dfrac{aE\Xi (a^{2}+r^{2})-a^{2}\Xi^{2}L}{\Delta_{r}}-\dfrac{1}{\Delta_{\theta}sin^{2}\theta}(a\Xi E sin^{2}\theta -\Xi^{2}L),
\end{align}
\begin{align}\label{21}
\rho^{2}(\dfrac{dt}{d\tau})=\dfrac{E(r^{2}+a^{2})^{2}-aL\Xi(r^{2}+a^{2})}{\Delta_{r}}-\dfrac{sin^{2}\theta}{\Delta_{\theta}}(E a^{2}-\dfrac{L\Xi a}{sin^{2}\theta}).
\end{align}

In following, we investigate null geodesic, so $\varepsilon$=0, and we have
\begin{align}\label{18.1}
\rho^{4}(\dfrac{dr}{d\tau})^{2}=-\Delta_{r} K+\big[(a^{2}+r^{2})E-aL\Xi \big]^{2}=R(r),
\end{align}
\begin{align}\label{19.1}
\rho^{4}(\dfrac{d\theta}{d\tau})^{2}=\Delta_{\theta}(K)-\dfrac{1}{sin^{2}\theta}\big(aE sin^{2}\theta -L\Xi \big)^{2} =\Theta(\theta),
\end{align}
\begin{align}\label{20.1}
\rho^{2}(\dfrac{d\varphi}{d\tau})=\dfrac{aE\Xi (a^{2}+r^{2})-a^{2}\Xi^{2}L}{\Delta_{r}}-\dfrac{1}{\Delta_{\theta}sin^{2}\theta}(a\Xi E sin^{2}\theta -\Xi^{2}L),
\end{align}
\begin{align}\label{21.1}
\rho^{2}(\dfrac{dt}{d\tau})=\dfrac{E(r^{2}+a^{2})^{2}-aL\Xi(r^{2}+a^{2})}{\Delta_{r}}-\dfrac{sin^{2}\theta}{\Delta_{\theta}}(E a^{2}-\dfrac{L\Xi a}{sin^{2}\theta}).
\end{align}
Now, we introduce dimensionless quantities such that $\xi=\frac{L}{E}$ and $\eta=\frac{K}{E^{2}}$, which are constant along the geodesics, so Eq.~(\ref{18.1}) becomes
\begin{align}\label{18.2}
\rho^{4}(\dfrac{dr}{d\tau})^{2}=-\Delta_{r} (E^{2}\eta)+\big[(a^{2}+r^{2})E-a(E\xi)\Xi \big]^{2}=R(r).
\end{align}
For the radial motion of particles, the effective potential is substantial tool which can be obtained by using the equation~\cite{Wei:2013kza}
\begin{align}\label{18.3}
\rho^{4}(\dfrac{dr}{d\tau})^{2}+V_{eff}=0.
\end{align}
So,
\begin{align}\label{18.4}
V_{eff}=\Delta_{r} (E^{2}\eta)-\big[(a^{2}+r^{2})E-a(E\xi)\Xi \big]^{2}.
\end{align}
Circular orbits of the photons are important to find out $\xi$ and $\eta$~\cite{Bardeen:1972fi}
\begin{equation}\label{18.5}
V_{eff}=0,  \qquad \frac{dV_{eff}}{dr}=0.
\end{equation}
The condition in Eq.~(\ref{18.5}) is equal to $R(r)=0$ and $\dfrac{dR(r)}{dr}=0$
Using Eqs.~(\ref{18.4}) and (\ref{18.5}), we can obtain the parameters $\eta$ and $\xi$.

\section{The Shadow for an observer in $r=\infty$}\label{infinite}
In this section, we want to analyze the shadow of black holes for an observer in $r=\infty$, so we introduce the celestial coordinate $\alpha$ and $\beta$, which are~\cite{Vazquez:2003zm}
\begin{equation}\label{alpha}
\alpha =\lim_{r_{O}\longrightarrow\infty} (r^{2}_{O} \sin (\theta_{O})\dfrac{d\varphi}{dr}), 
\end{equation}
and
\begin{equation}\label{beta}
\beta =\lim_{r_{O}\longrightarrow\infty} r^{2}_{O}\dfrac{d\theta}{dr},
\end{equation}
where, $\theta_{O}$ is the inclination angle between the rotation axis of the black hole and  the line of sight of the observer, also considering an observer far away from the black hole, we have  $r_{O}\rightarrow \infty$.  
Using Eq.~(\ref{18.1})--~(\ref{20.1}), and take the limit of a faraway observer, the celestial coordinates take the forms
\begin{equation}\label{22}
\alpha = -\xi \csc\theta_{O}, 
\end{equation}
and
\begin{equation}\label{23}
\beta = \pm \sqrt{\eta+a^{2}\cos ^{2}\theta_{O}-\xi^{2}\cot ^{2}\theta_{O}}.
\end{equation}
For an observer located in the equatorial plane of the black hole, i.e. $\theta_{O}=\dfrac{\pi}{2}$, the gravitational effects are maximum, and $\alpha$ and $\beta$ become
\begin{equation}\label{22.1}
\alpha = -\xi ,
\end{equation}
and
\begin{equation}\label{23.1}
\beta = \pm \sqrt{\eta}.
\end{equation}
For an observer at infinity, we put $\Lambda=0$, because when $\Lambda=0$, $\Delta_{r}$ convert to second order equation in terms of $r$
\begin{equation}\label{second order deltar}
\Delta_{r}=(r^{2}+a^{2})-2Mr+\frac{Q^{2}}{(1+f^{'}(R_{0})},
\end{equation}
and the horizons can be obtained as
\begin{equation}\label{horizon}
r_{\pm}=M\pm\sqrt{M^{2}-\frac{Q^{2}}{(1+f^{'}(R_{0}))}-a^{2}},
\end{equation}
where, $a^{2}\leq a_{max}^{2}$, in which $a_{max}^{2}=M^{2}-\frac{Q^{2}}{(1+f^{'}(R_{0}))}$. In the case of, $a^{2}>a_{max}^{2}$, instead of black hole, we have a naked singularity, and the case $a=a_{max}$, is called extremal black hole. Since, outside of the event horizon, ($\bigtriangleup_{r}$) is greater than zero, $\partial_{r}$ is spacelike, so, communication is possible here. This region is called domain of outer communication and our observer is located in this region. In fact, when $\Lambda\neq0$, the space time is not flat asymptotically. Therefore, the observer at the domain of outer communication is apart from an observer who placed at $\infty$, by cosmological horizon when, $\Lambda>0$~\cite{Grenzebach:2014fha}.

In following, considering the above conditions, we plot $\beta$ in terms of $\alpha$ to obtain the counter of the black hole's shadow. These plots are represented for different values of rotation parameter ($a=0$, $a=0.5$, $a=0.7$ and $a=1$) and electric charge $Q$, in Fig.\ref{infinity}.
It can be seen that by increasing the electric charge, the shadows become smaller and, also increasing of spin parameter $a$, leads to change in symmetry of the image of shadows.
\begin{figure}[h]
	\centering
	\subfigure[]{
		\includegraphics[width=0.3\textwidth]{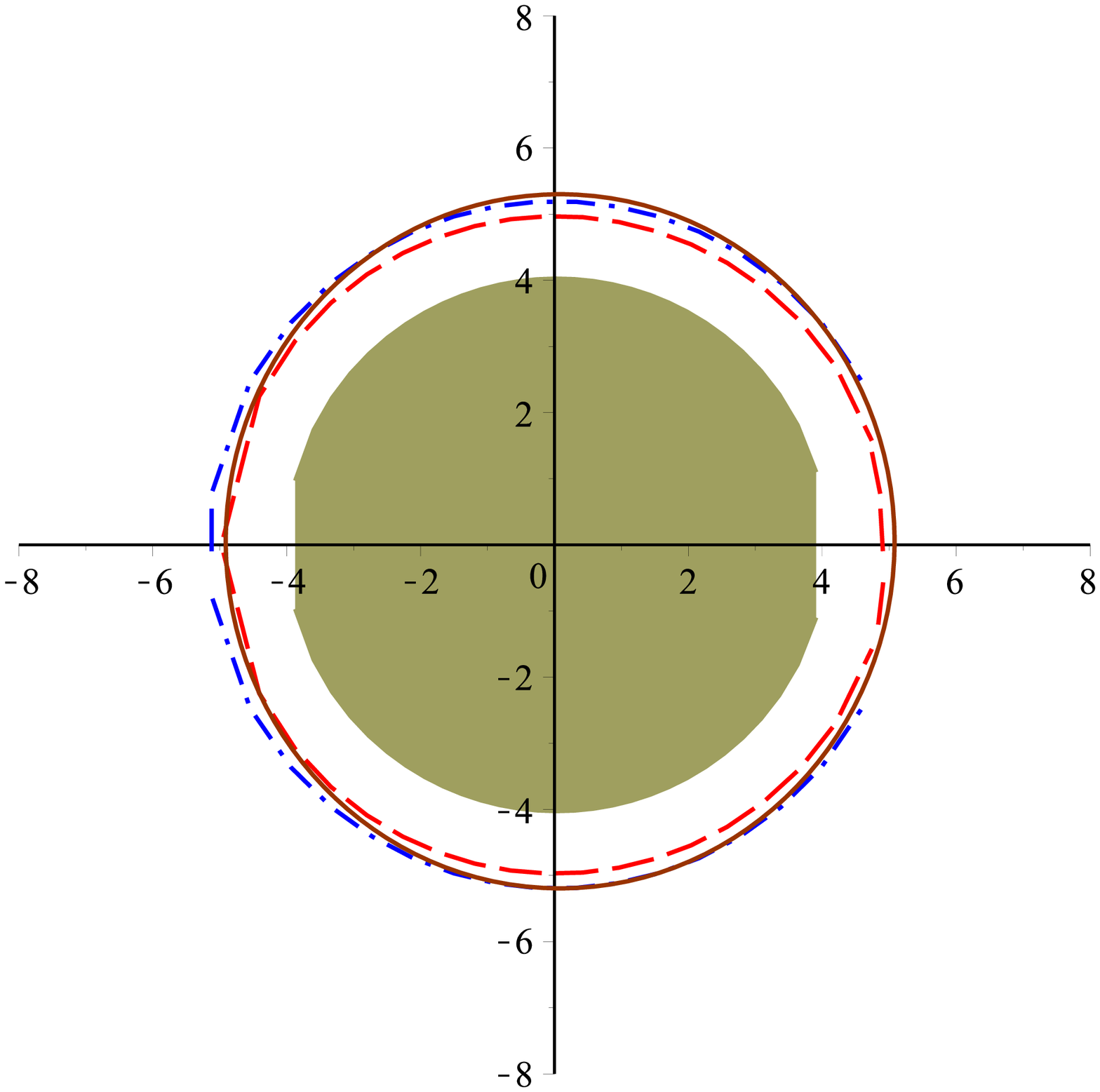}
	}
	\subfigure[]{
		\includegraphics[width=0.3\textwidth]{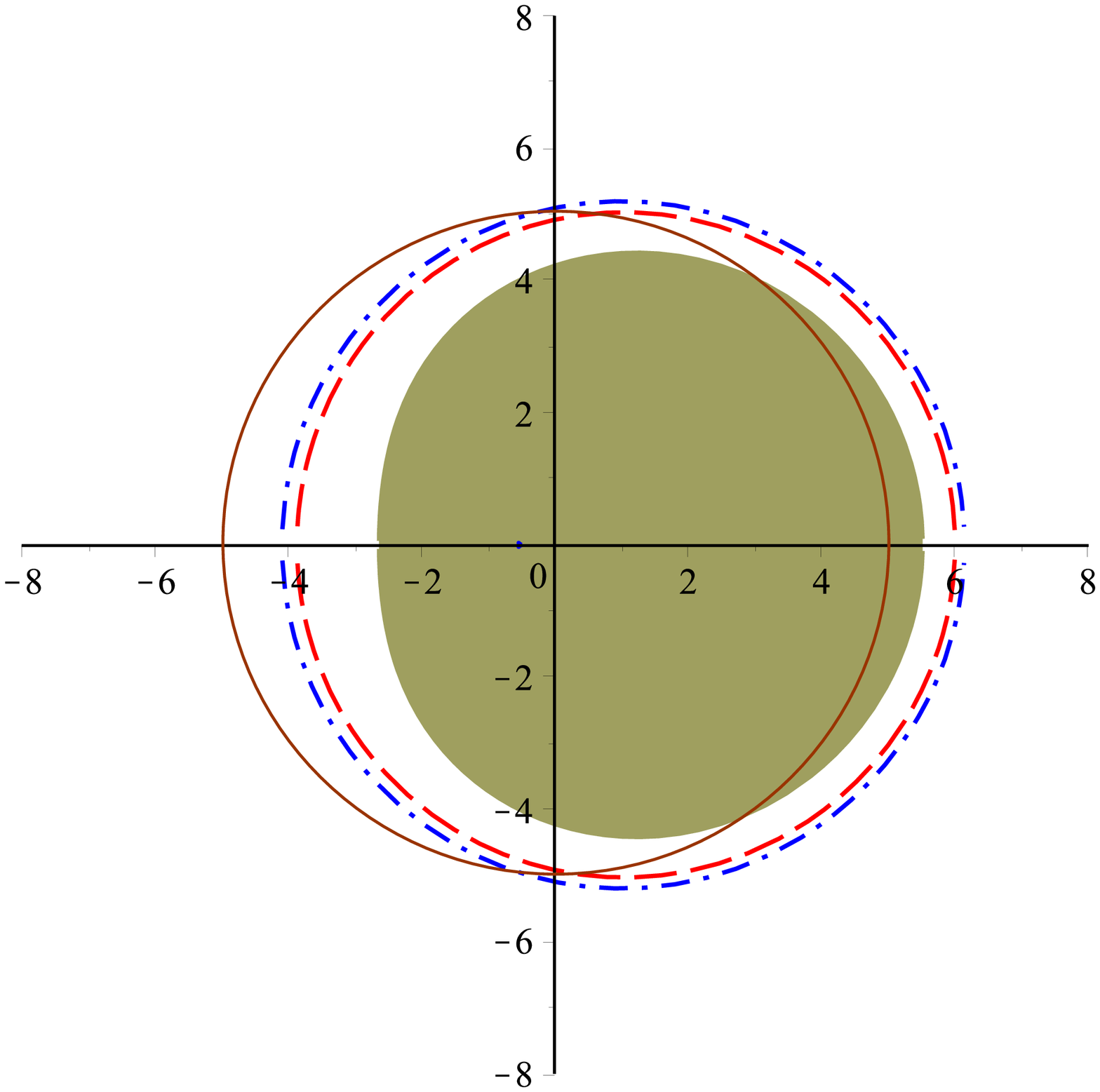}
	}
	\subfigure[]{
		\includegraphics[width=0.3\textwidth]{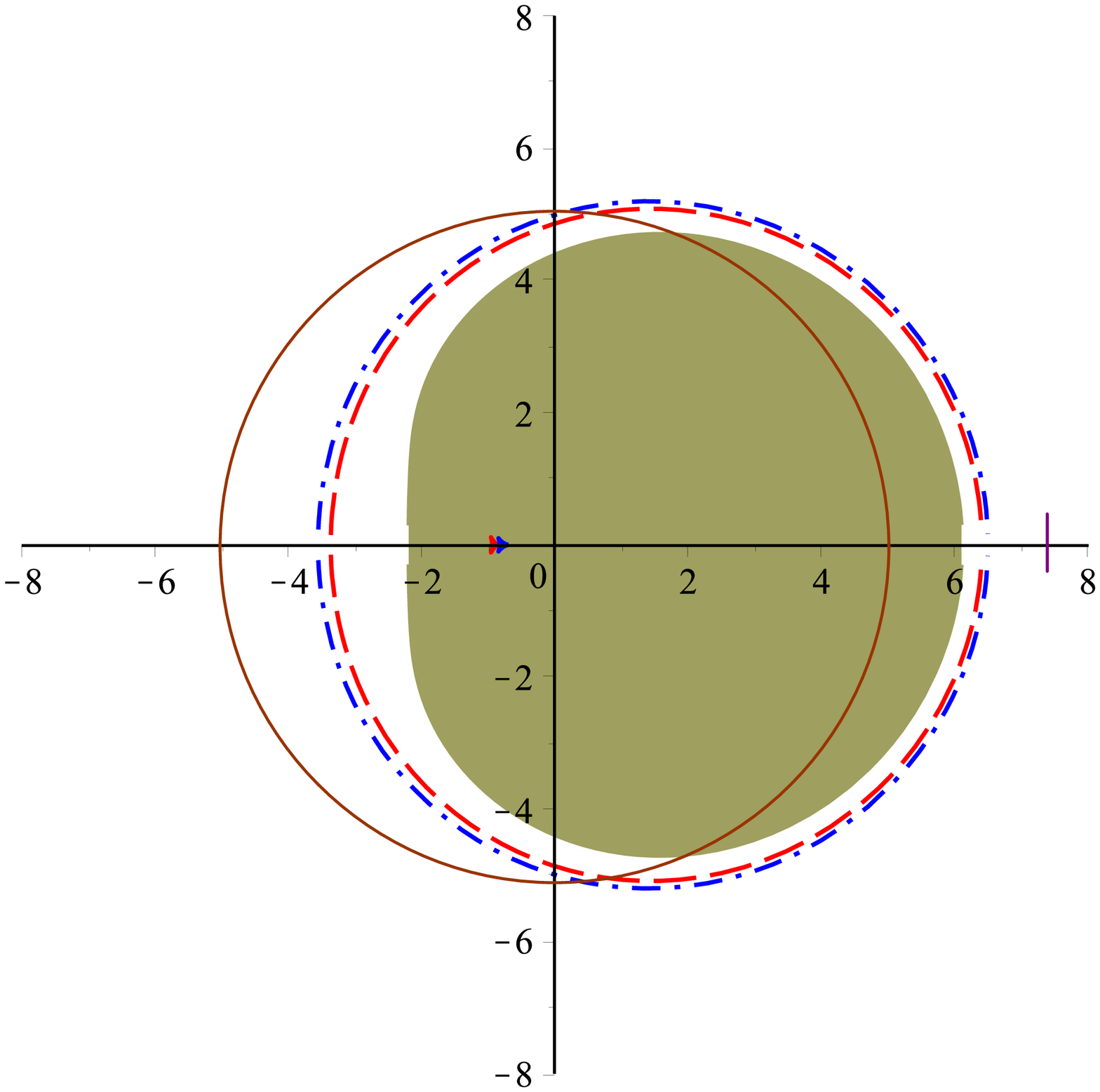}
	}
	\subfigure[]{
		\includegraphics[width=0.3\textwidth]{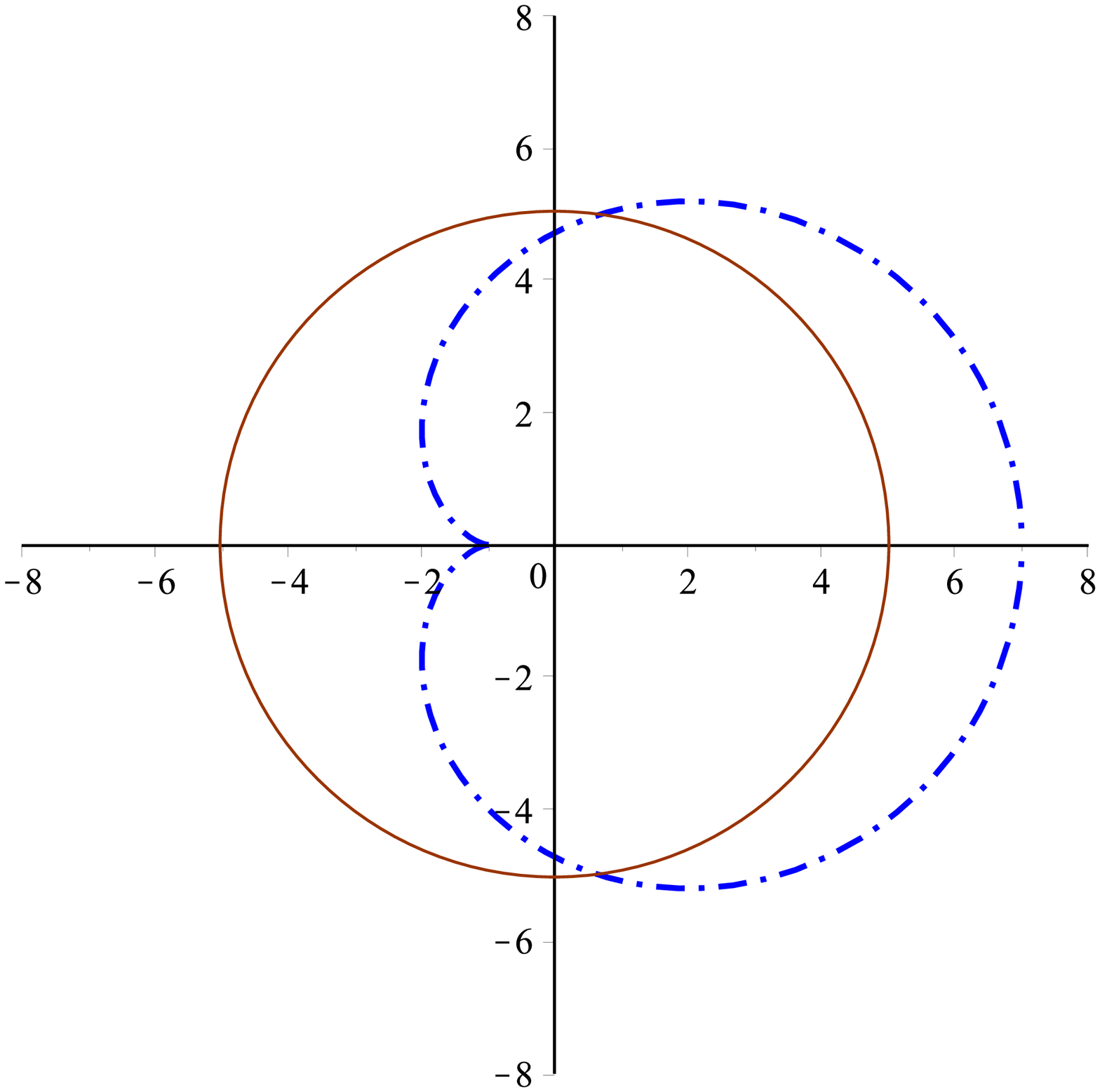}
	}
	\caption{\label{infinity}The image of black hole's shadow in $f(R)$ gravity, $a=0$, $a=0.5$, $a=0.7$ and $a=1$ for (a), (b), (c) and (d) respectively. For each value of $a$, $Q=0$(blue dash-dot line), $Q=\frac{Q_{c}}{2}$(red dash line) and $Q=Q_{c}$(green filled solid line) for (a), (b), (c)but for (d) only, $Q=0$ is considered. The brown solid line is reference circle. The detail of parameters is shown in table~\ref{tab1} (see Appendix).}
	\label{infinity}
 \label{pic:shadow}
\end{figure}

In addition, for studying the size and deviation of black hole's shadow, we introduce two observable $R_{s}$ and $\delta_{s}$ parameters. $R_{s}$ indicates the size of shadow, and $\delta_{s}$ explains the deviation of the shadow from circular.
We consider three points top, bottom and rightmost of the shadow (see Fig.\ref{deviation}~\cite{Abdujabbarov:2015xqa}), which expressed respectively by $( \alpha_{t},\beta_{t})$,$(\alpha_{b},\beta_{b})$ and $(\alpha_{r},0)$, so we have 
  \begin{equation}\label{23.1}
 R_{s}=\dfrac{(\alpha_{t}-\alpha_{r})^{2}+\beta^{2}_{t}}{2(\alpha_{t}-\alpha_{r})}.
\end{equation}
Furthermore, $\delta_{s}$ is indicated by $(\bar{\alpha}_{p}, 0)$ and $(\alpha_{p}, 0)$ as
\begin{equation}\label{23.2}
\delta_{s}=\dfrac{(\bar{\alpha}_{p}-\alpha_{p})}{R_{s}},
\end{equation}
In which, ($\alpha_{p}, 0$) and ( $\bar{\alpha}_{p}, 0$), are the points, where the contour of the shadow and reference circle cut the horizontal axis at opposite side of $(\alpha_{r}, 0)$. In Fig. \ref{RS}, the observable $R_{s}$ and $\delta_{s}$ for different values of $f^{'}(R_{0})$ is plotted. We can see that, by increasing $f^{'}(R_{0})$, $R_{s}$  increases and $\delta_{s}$ decreases. 
Thus, the larger value of parameter $f^{'}(R_{0})$ leads to increasing in the size and decreasing in the distortion of the shadow.
\begin{figure}[h]\label{deviation}
	\centering
	{
		\includegraphics[width=0.4\textwidth]{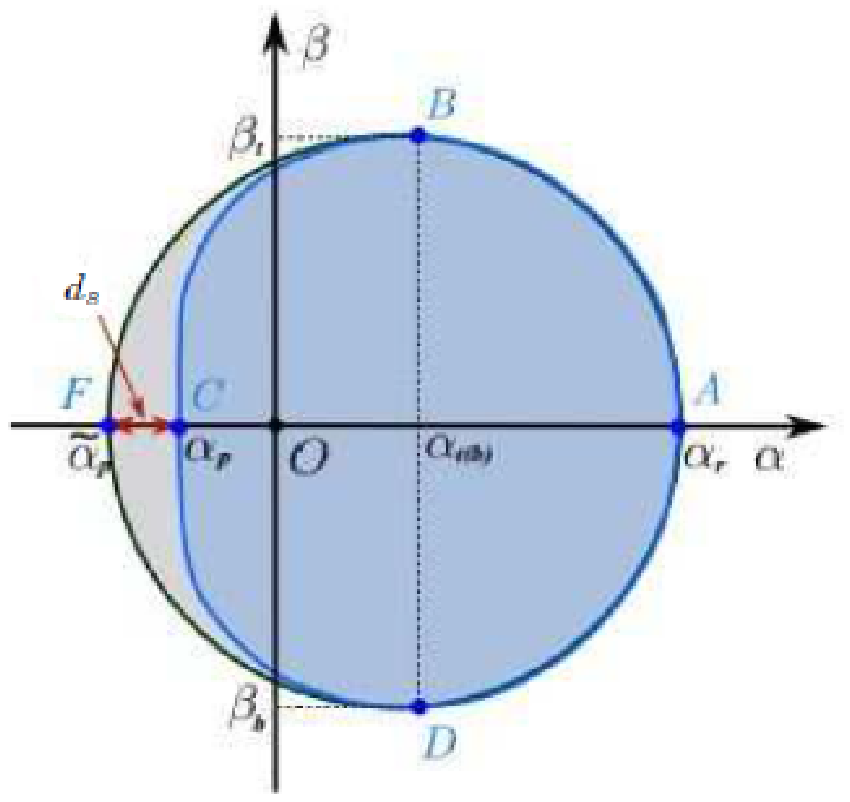}
	}
	\caption{\label{deviation}The black hole shadow and reference circle. $d_{s}$ is the distance between the left point of the shadow and the reference circle.}
 \label{pic:shadow}
\end{figure}

\begin{figure}[h]\label{RS}
	\centering
	\subfigure[$R_{s}$ and $f^{'}(R_{0})$ for $\theta_{o}=\frac{\pi}{2}$ ]{
		\includegraphics[width=0.4\textwidth]{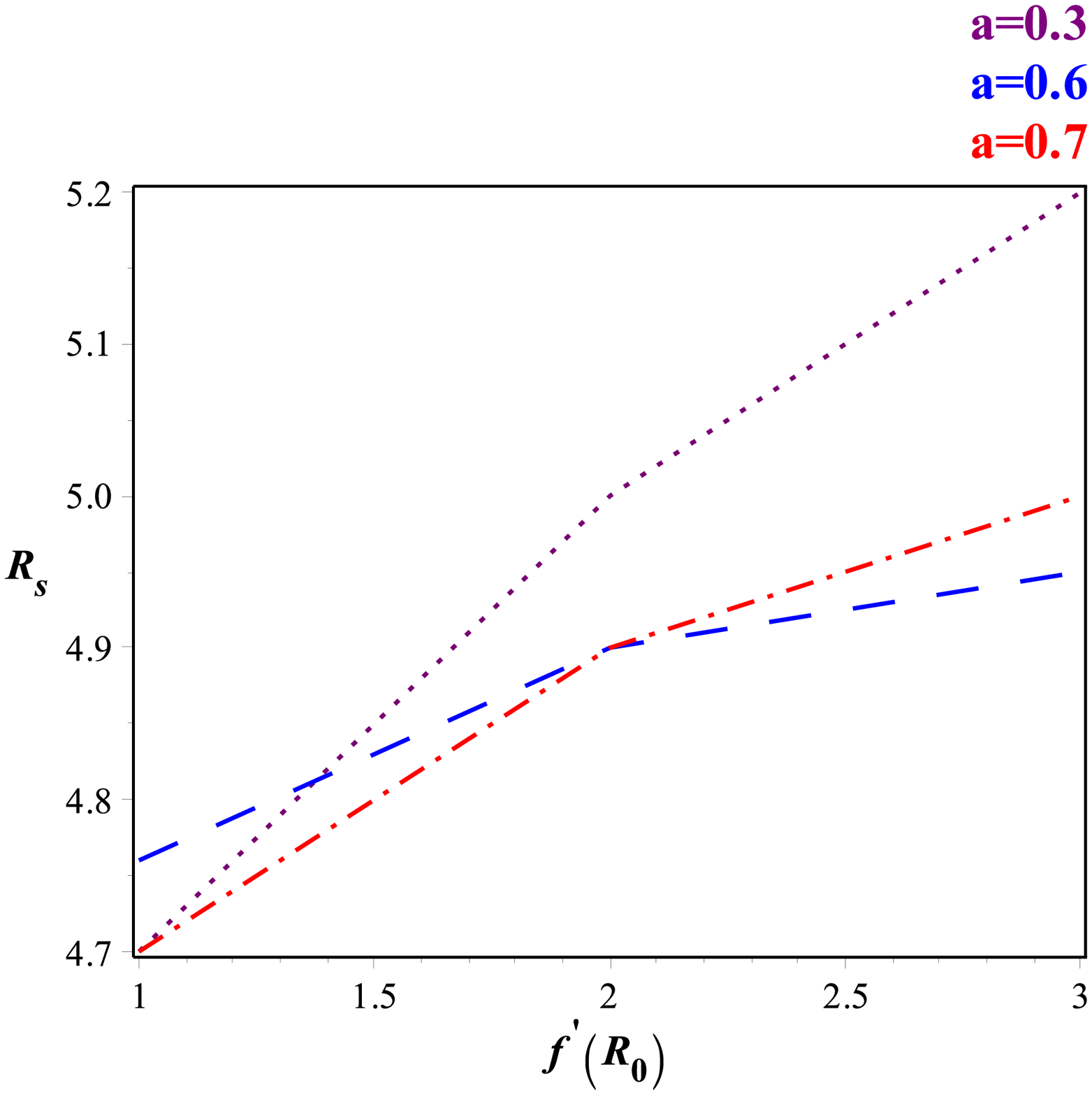}
	}
	\subfigure[$\delta_{s}$ and $f^{'}(R_{0})$ for $\theta_{o}=\frac{\pi}{2}$]{
		\includegraphics[width=0.4\textwidth]{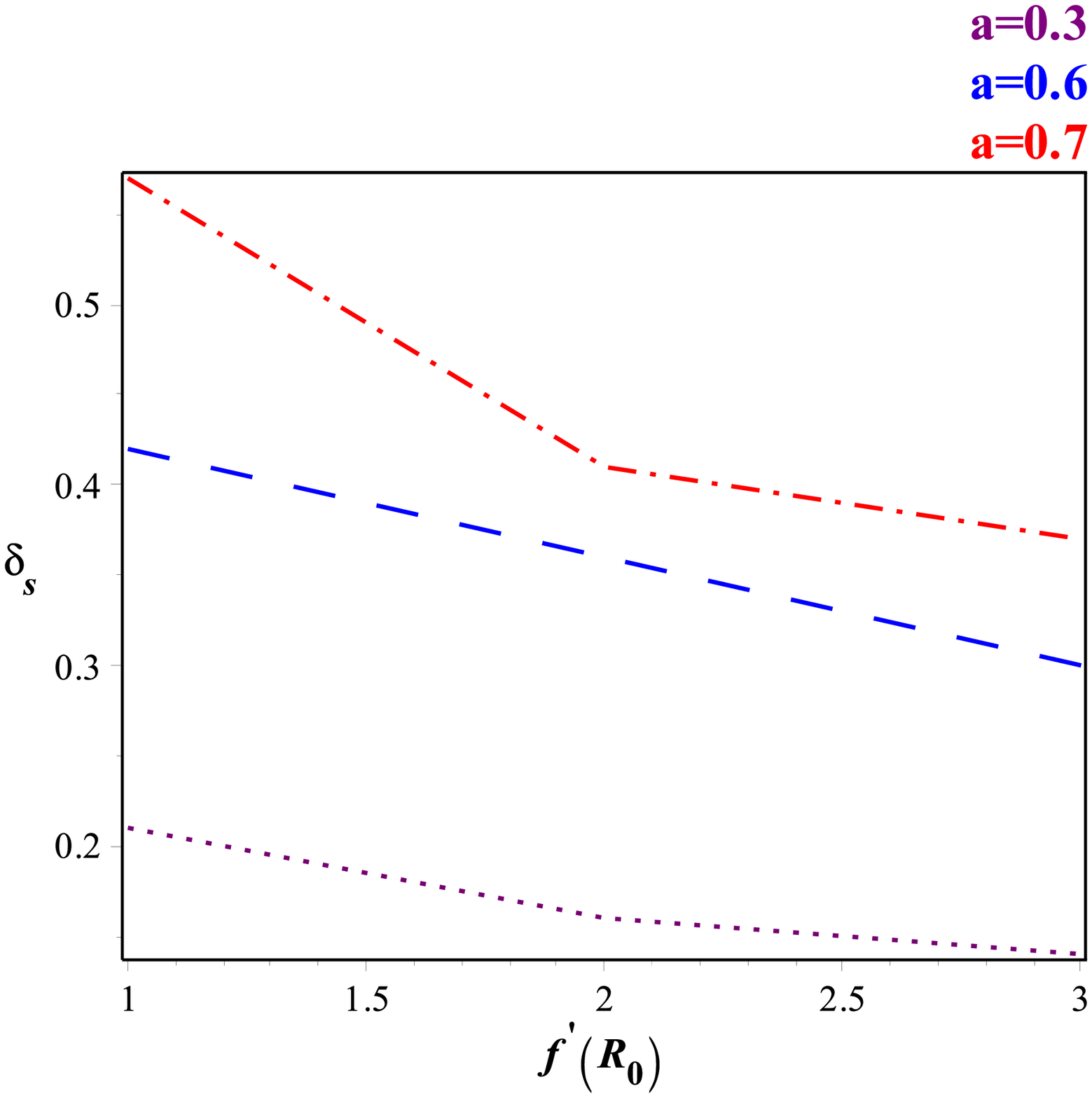}
	}
	\caption{\label{RS}Changes of $R_{s}$ and $\delta_{s}$ for an observer at $r \rightarrow \infty$. $a=0.3$, $a=0.6$ and $a=0.7$ for the purple (dot line), the blue (dash line) and the red (dash-dot line) respectively.}
 \label{pic:shadow}
\end{figure}

\section{Shadow of black hole for an observer at $r=r_{O}$}\label{limited}
In this section, we study shadow of black hole for an observer at $r=r_{O}$. We are interested in spherical lightlike geodesics, i.e. lightlike geodesics that stay on a sphere $r=constant$. The region which is filled by these geodesics are called \textit{photon region}. Photon region around the black hole is essential for building shadow, in fact, we can say that the shadow is an image of photon region. We put our observer at $(r_{O}$, $\theta_{O})$, in the domain of outer communication, in this case $\Lambda>0$ and we consider a light source at $r=r_{L}$ where  $r_{L}\geq r_{O}$. In introduction, we explain how the shadow forms.
Now, we have an observer at $(r_{O}$, $\theta_{O})$ and our purpose is to calculate the shadow of black hole in this situation. At first, we choose an orthonormal tetrad~\cite{Griffiths} at the observer's sky (see Fig.\ref{observer}) as
\begin{figure}[h]\label{observer}
	\centering
	{
		\includegraphics[width=0.5\textwidth]{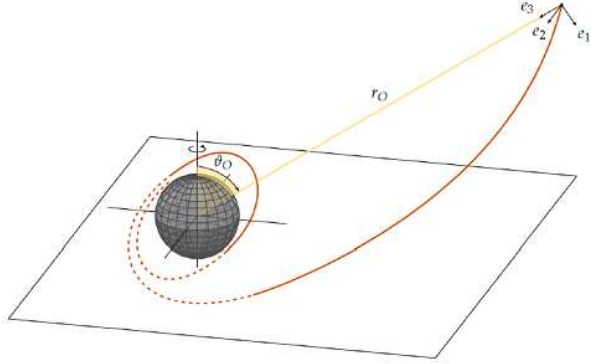}
	}
	\caption{\label{observer}An observer at $(r_{o}$,$\theta_{o})$ with an orthonormal tetrad $(e_{0}, e_{1}, e_{2}, e_{3})$ according to Eqs.~(\ref{24}).}
 \label{pic:shadow}
\end{figure}
\begin{equation}\label{24}
\left\{
\begin{array}{l}
e_{0}=\frac{\frac{(a^{2}+r^{2})}{\Xi}\partial_{t}+a\partial_{\varphi}}{\sqrt{\Delta}_{r} \rho}\mid_{(r_{o},\theta_{o})},\\
e_{1}=\frac{\sqrt{\Delta}_{\theta}}{\rho}\partial_{\theta}\mid_{(r_{o},\theta_{o})}, \\
e_{2}=-(\frac{\frac{a \sin ^{2}\theta }{\Xi}\partial_{t}+\partial_{\varphi}}{\sqrt{\Delta}_{\theta} \rho \sin \theta})\mid_{(r_{o},\theta_{o})},\\
e_{3}=-\frac{\sqrt{\Delta}_{r}}{\rho}\partial_{r}\mid_{(r_{o},\theta_{o})} .
\end{array} \right.
\end{equation}
Note that, these orthonormal tetrad can be changed by different position of an observer.
Since, put our observer in the domain outer communication, so $\triangle_{r}$ is positive, and the coefficients in Eq.~(\ref{24}) are real. The vector $e_{3}$ gives the spatial direction to the center of black hole (see Fig. \ref{observer}).
In addition, each light ray $\lambda(s)$ has the coordinate $r(s),\theta(s),\varphi(s),t(s)$, so the tangent vector at the position of the observer can be written as
\begin{equation}\label{25}
\dot{\lambda} =\dot{r}\partial_{r} +\dot{\theta}\partial_{\theta}+\dot{\varphi}\partial_{\varphi}+\dot{t}\partial_{t},
 \end{equation}
 or 
 \begin{equation}\label{26}
\dot{\lambda} = \Omega (-e_{0} + \sin \omega \cos \psi e_{1}  +\sin \omega \sin \psi e_{2}+ \cos \theta e_{3}) ,
 \end{equation}
 where, $\Omega$ is a scalar factor and it can be obtained from Eqs.~(\ref{11}),~(\ref{24}) and ~(\ref{25}) as
\begin{equation}\label{27}
\Omega = g(\dot{\lambda},e_{0}) = \frac{a L-\frac{(a^{2}+r^{2})}{\Xi}E}{\sqrt{\Delta}_{r} \rho}\mid_{(r_{O},\theta_{O})}.
\end{equation}
With comparison coefficients of $\partial_{\varphi}$ and $\partial_{r}$ in Eqs.~(\ref{25}) and ~(\ref{26}), we have
\begin{equation}\label{28}
\sin \psi=-\frac{\rho\sqrt{\Delta}_{\theta} \sin \theta}{\sin \omega}(\frac{\dot{\varphi}}{\Omega}+\frac{a}{\rho \sqrt{\Delta_{r}}})\mid_{(r_{O},\theta_{O})},
\end{equation}
\begin{equation}\label{29}
\cos \omega =-\frac{\rho\dot{r}}{\sqrt{\Delta_{r}}\Omega}\mid_{(r_{O},\theta_{O})},
\end{equation}
where $\psi$ and $\omega$ are the celestial coordinate. Also, as previous section, the constants of motion for this light ray are $\xi$ and $\eta$, as
\begin{align}\label{18.5.1}
\eta=\frac{16r^{2}\Delta_{r}}{\Delta_{r}^{'2}},
\end{align}
and
\begin{align}\label{18.5.2}
\xi=\frac{(a^{2}+r^{2})}{\Xi a}-\frac{4r\Delta_{r}}{\Xi a\Delta_{r}^{'}},
\end{align}
where $\Delta_{r}^{'}$ represents the derivative of $\Delta_{r}$ respect to $r$. Using Eqs.~(\ref{18.1}),~(\ref{20.1}),~(\ref{27}),~(\ref{18.5.1}) and ~(\ref{18.5.2}), we have
\begin{eqnarray}\label{28.1}
\sin \psi=\frac{\Xi}{\sin\theta\sqrt{\Delta_{\theta}\eta} }(\Xi\xi -a\sin^{2}\theta )+\frac{\sqrt{\Delta}_{r}}{\sin \theta\sqrt{\Delta_{\theta}}}\frac{(a\Xi\sin^{2}\theta-\Xi^{2}\xi)}{(a\xi -\frac{r^{2}+a^{2}}{\Xi})}\frac{\sqrt{\frac{1}{\Xi^{2}}-1}}{\frac{1}{\Xi}}\nonumber\\
+\frac{\sqrt{\Delta_{\theta}}}{\sqrt{\Delta_{r}}}\frac{\sin \theta}{\sin \omega}a(\Xi^{2}-1)\mid_{\theta_{O}},
\end{eqnarray}
and
\begin{equation}\label{29.1}
\sin \omega =\frac{\sqrt{\Delta_{r}\eta}}{\frac{1}{\Xi}((r^{2}+a^{2})-a\xi\Xi)}-\frac{\sqrt{\frac{1}{\Xi^{2}}-1}}{\frac{1}{\Xi}}\mid_{r_{O}}.
\end{equation}
Next, we use the Eqs.~(\ref{18.5.1})--~(\ref{29.1}) and stereographic projection from the celestial sphere onto a plane (see Fig.\ref{streo}) to plot the images of black hole's shadow. In fact, Eq.~(\ref{28.1}) and ~(\ref{29.1}) gives the counter of the shadow of black hole. In fact, the counter of the shadow demonstrated the light rays which approach to the spherical light like geodesic with radius $r_{p}$.
\begin{figure}[h]\label{streo}
	\centering
	{
		\includegraphics[width=0.7\textwidth]{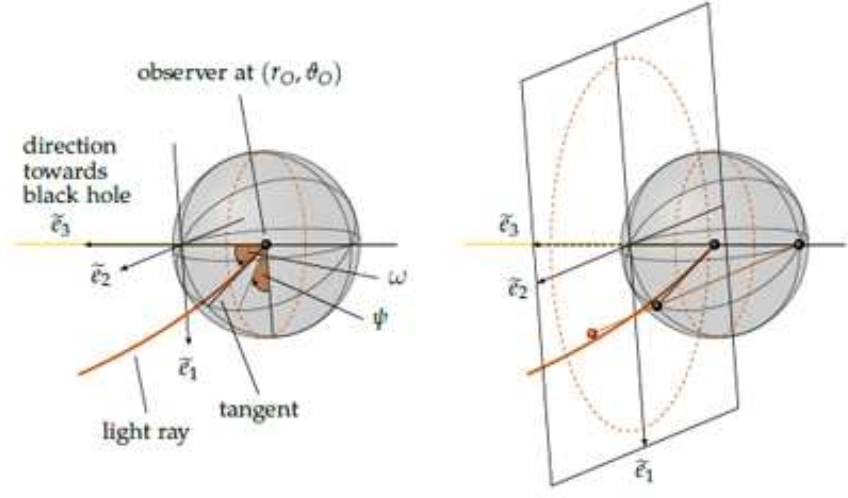}
	}
	\caption{\label{streo}For each light ray, we choose $\omega$ and $\psi$ from Eq.~(\ref{26}) (left figure). The red ball in the right figure presents stereographic projection of the point $\omega$ and $\psi$ on the celestial plane. The red dotted circles show the celestial equator $\theta=\frac{\pi}{2}$ and its projection.}
 \label{pic:shadow}
\end{figure}

Moreover, the cartesian coordinate are obtained by
\begin{eqnarray}\label{31}
x(r_{p})&=&−2\tan (\frac{\omega(r_{p})}{2}) \sin (\psi (r_{p})), \nonumber \\ 
y(r_{p})&=&−2\tan (\frac{\omega(r_{p})}{2}) \cos  (\psi (r_{p})),
\end{eqnarray}
So, these equations are used for plotting the image of black hole's shadow. Some examples of these plots are shown in (Figs. \ref{fig2} - \ref{fig4}). In this figures, $\theta_{O}=\frac{\pi}{2}$ and the observer is located in the domain of outer communication.
\begin{figure}[h]\label{fig2}
	\centering
	\subfigure[]
	{
		\includegraphics[width=0.3\textwidth]{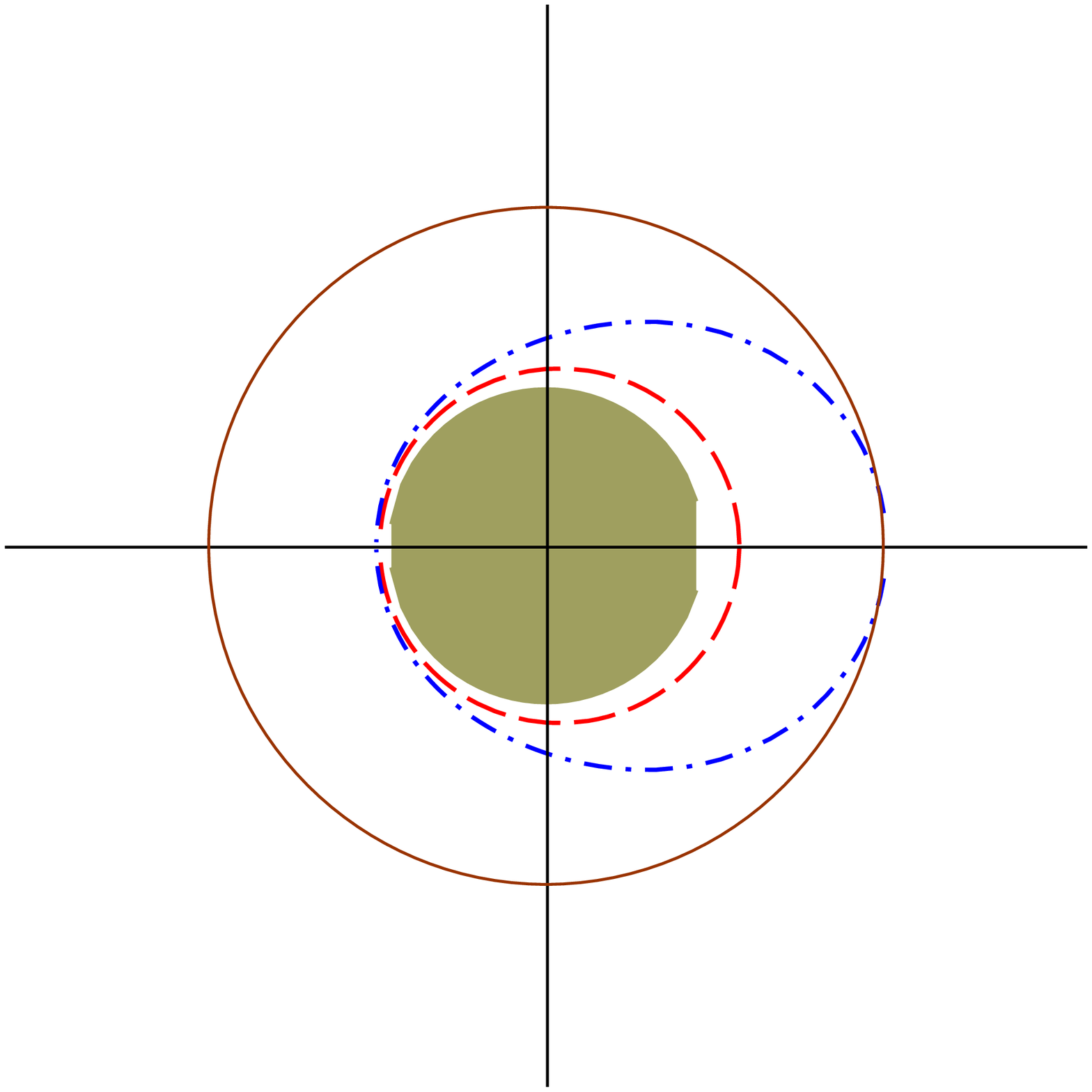}
	}
	\subfigure[]
	{
		\includegraphics[width=0.3\textwidth]{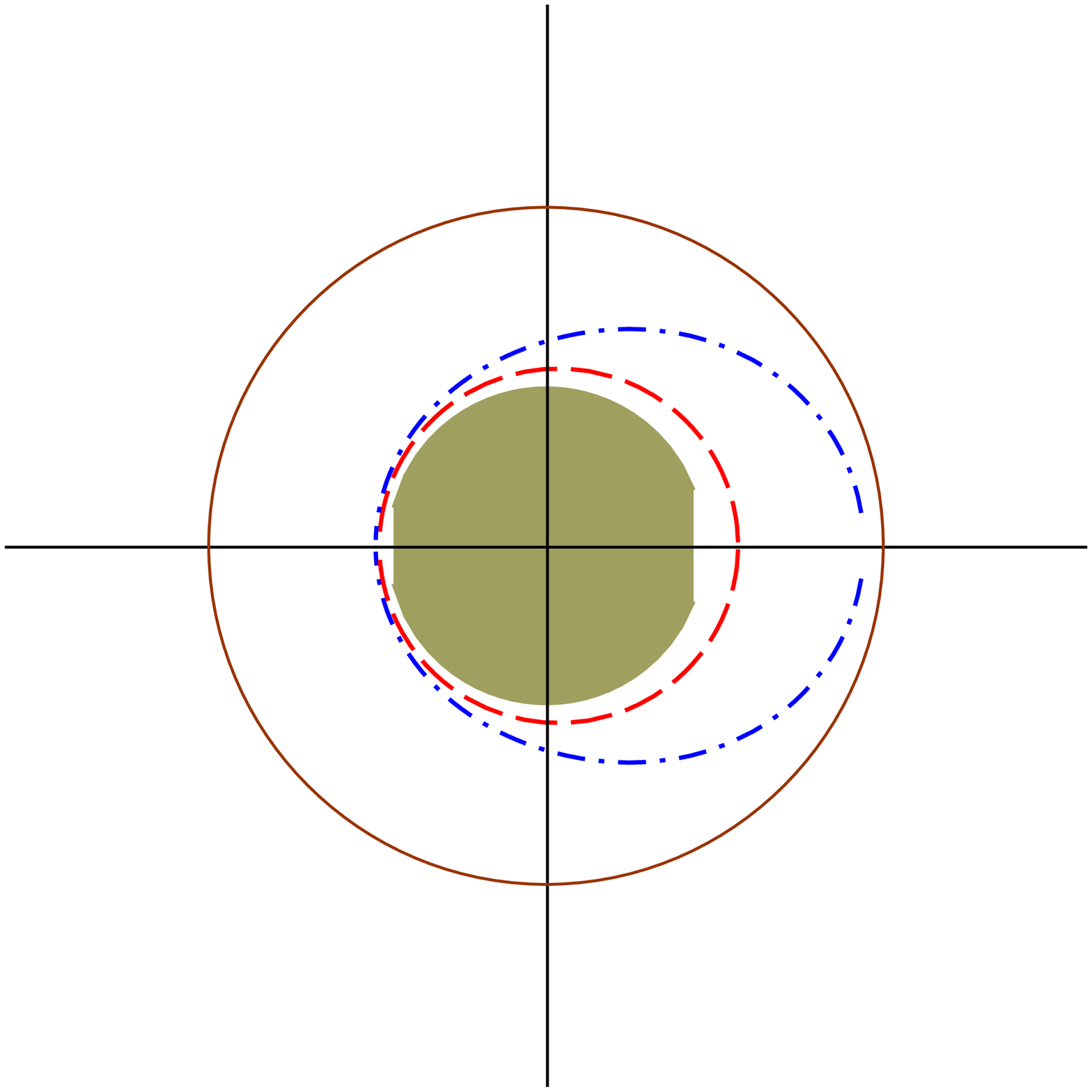}
	}
	\subfigure[]
	{
		\includegraphics[width=0.3\textwidth]{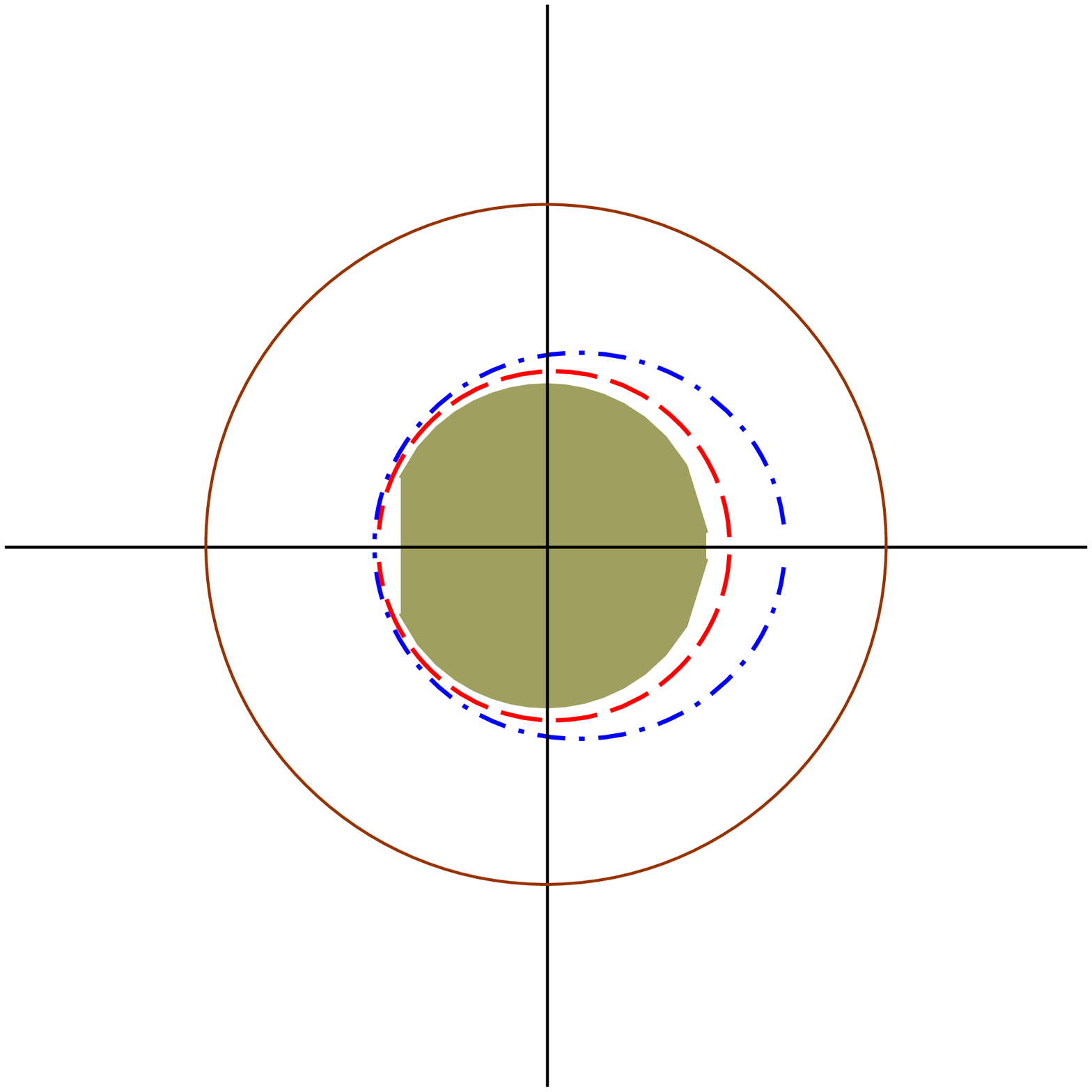}
	}
	\caption{\label{fig2}Shadow of black holes for $\Lambda=0$. The electric charge, ($Q=0, a_{max}=1$), ($Q=0.75, a_{max}=0.86$) and ($Q=1.35, a_{max}=0.43$) in (a), (b) and (c) respectively. Each figure is shown for different value of $a_{max}$. In the green solid filled pictures $a=\dfrac{2}{100} a_{max}$, in the dash red picture $a=\dfrac{2}{5} a_{max}$ and in the blue dash-dot plot $a=\dfrac{4}{5} a_{max}$. The brown solid line is reference circle. The detail of parameters is shown in table~\ref{tab2} (see Appendix).}
 \label{pic:shadow}
\end{figure}

\begin{figure}[h]\label{fig3}
	\centering
	\subfigure[]
	{
		\includegraphics[width=0.3\textwidth]{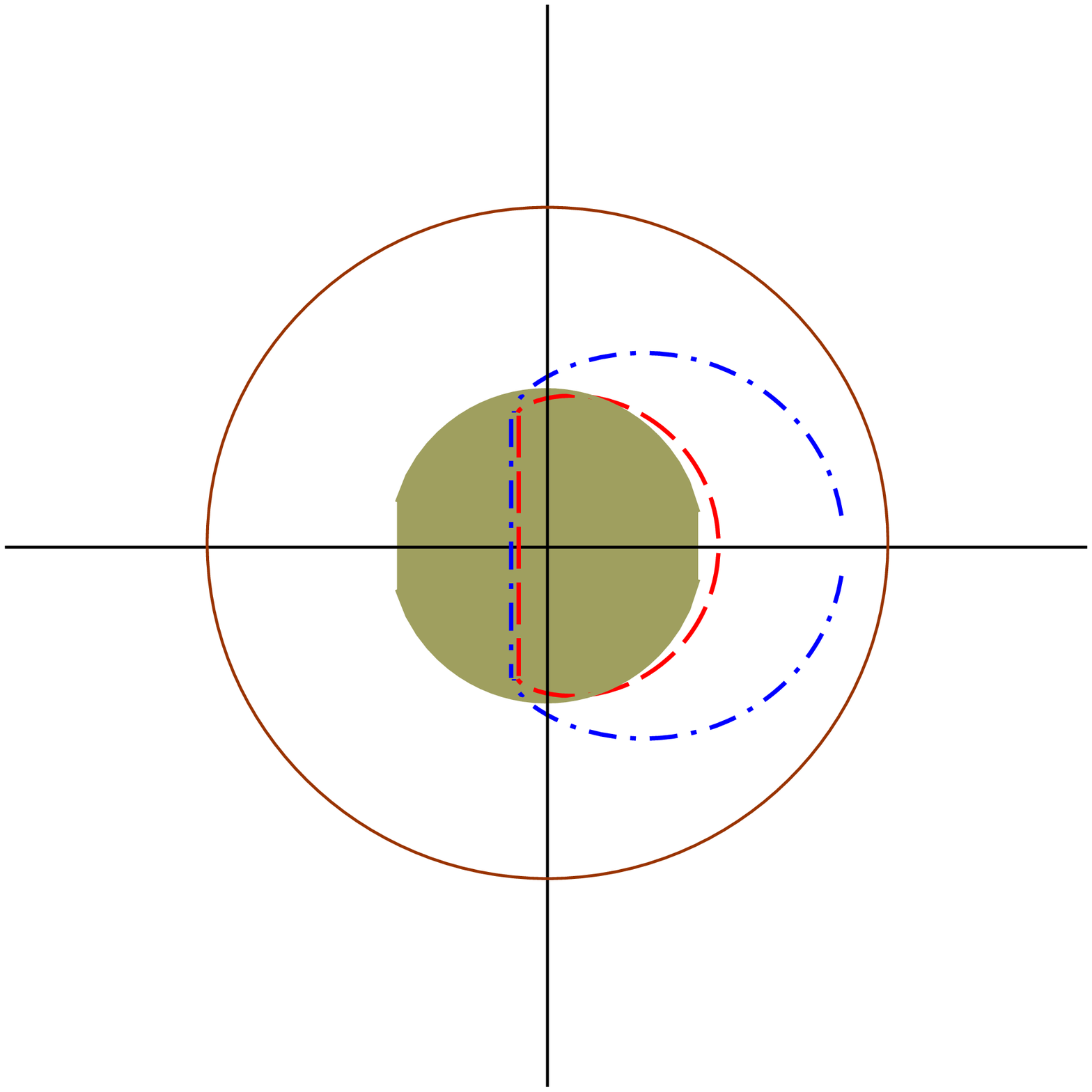}
	}
	\subfigure[]
	{
		\includegraphics[width=0.3\textwidth]{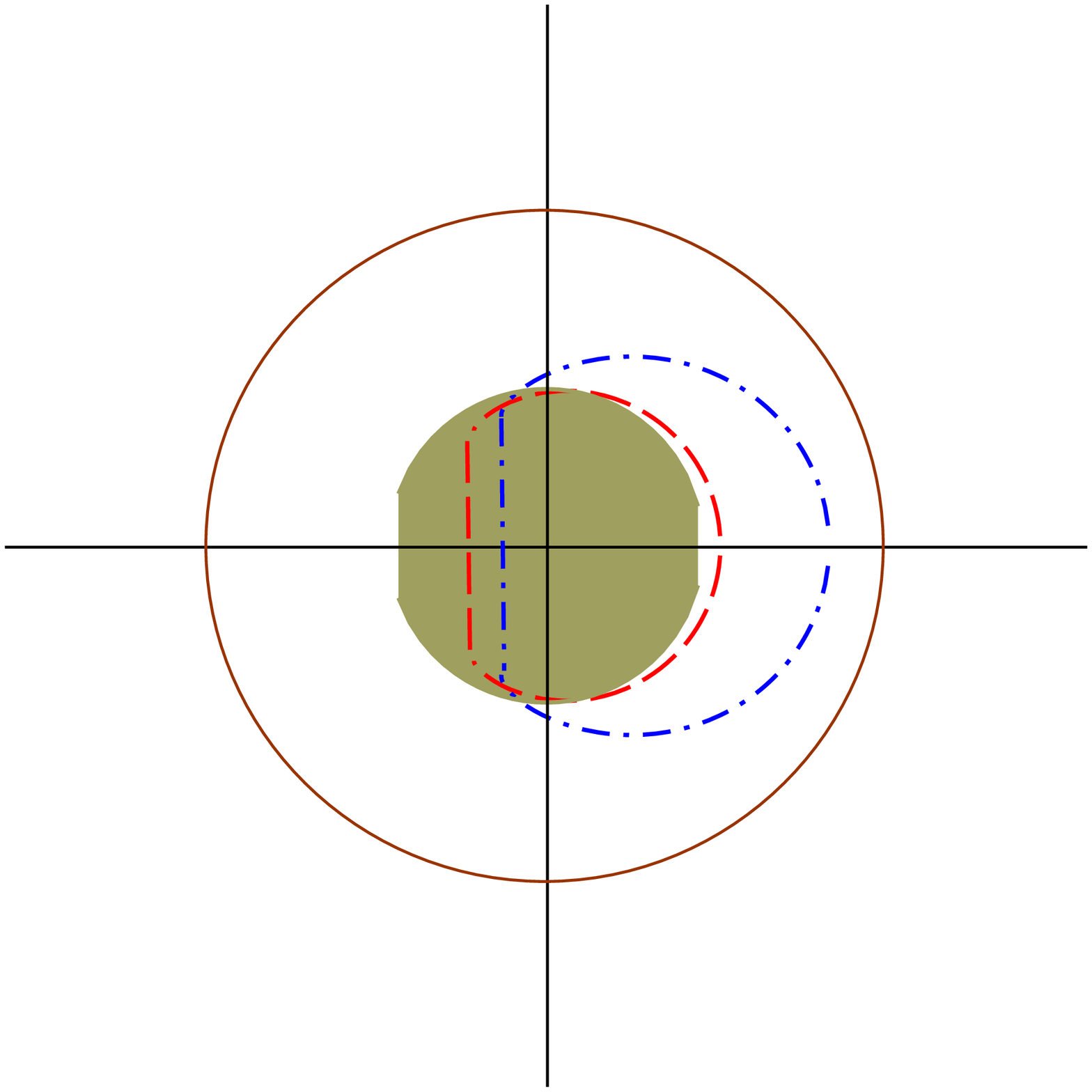}
	}
	\subfigure[]
	{
		\includegraphics[width=0.3\textwidth]{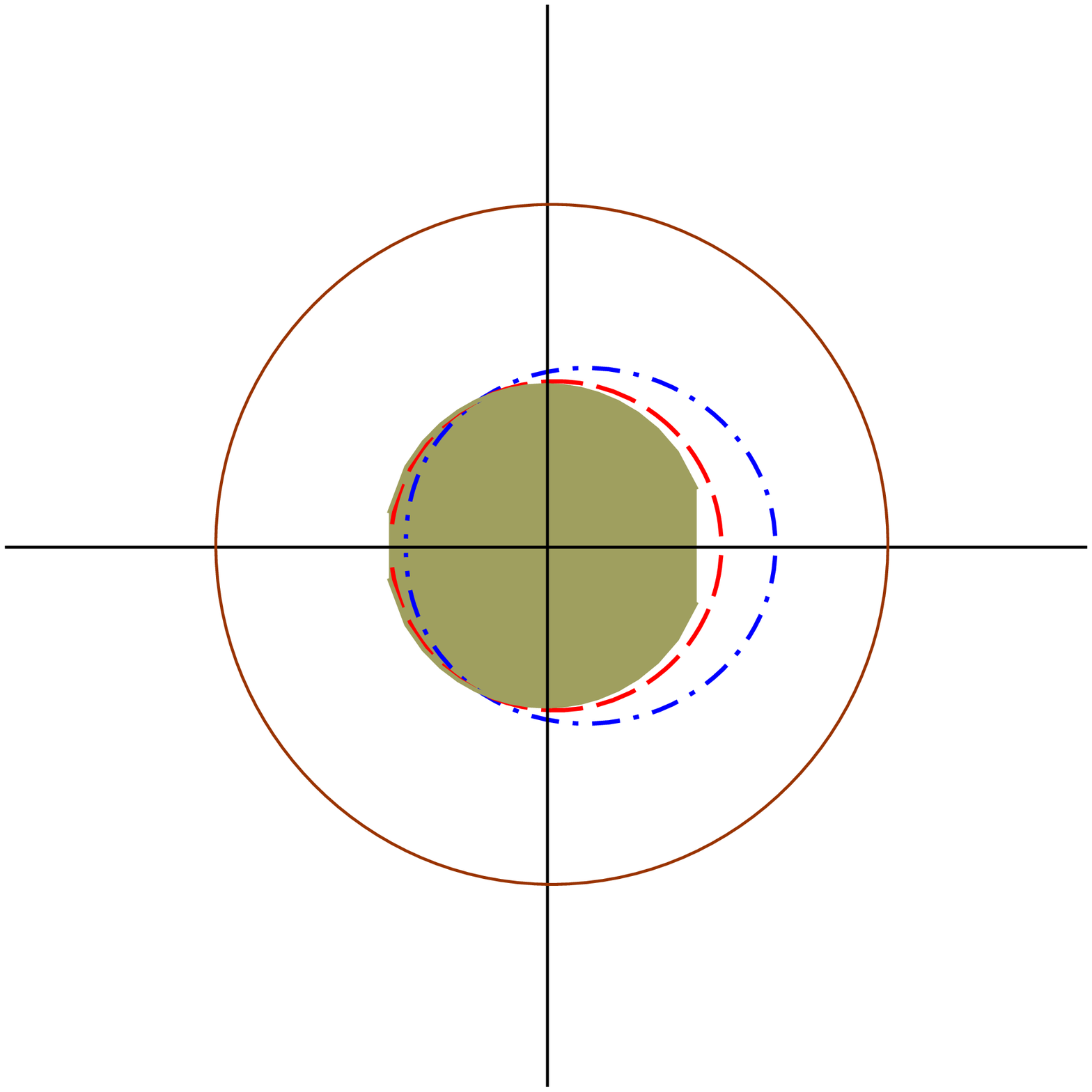}
	}
	\caption{\label{fig3}Shadow of black holes for $\Lambda=10^{-2}$. The electric charge, ($Q=0, a_{max}=1$), ($Q=0.75, a_{max}=0.86$) and ($Q=1.35, a_{max}=0.44$) for (a), (b) and (c) respectively. Each figure is shown for different value of $a_{max}$. In the green solid filled pictures $a=\dfrac{2}{100} a_{max}$, in the dash red picture $a=\dfrac{2}{5} a_{max}$ and in the blue dash-dot plot $a=\dfrac{4}{5} a_{max}$. The brown solid line is reference circle. The detail of parameters is shown in table~\ref{tab2} (see Appendix).}
 \label{pic:shadow}
\end{figure}

\begin{figure}[h]\label{fig4}
	\centering
	\subfigure[]
	{
		\includegraphics[width=0.3\textwidth]{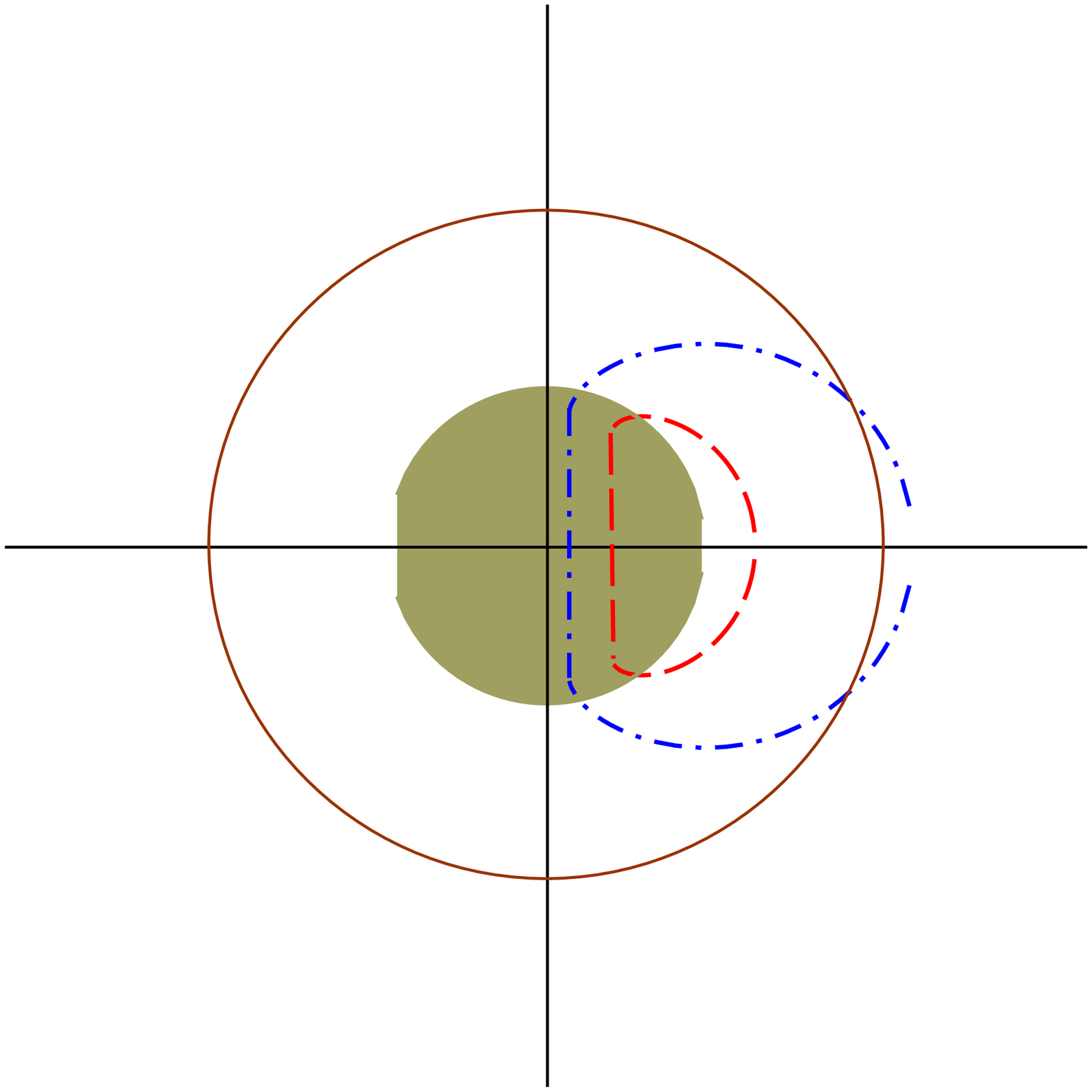}
	}
	\subfigure[]
	{
		\includegraphics[width=0.3\textwidth]{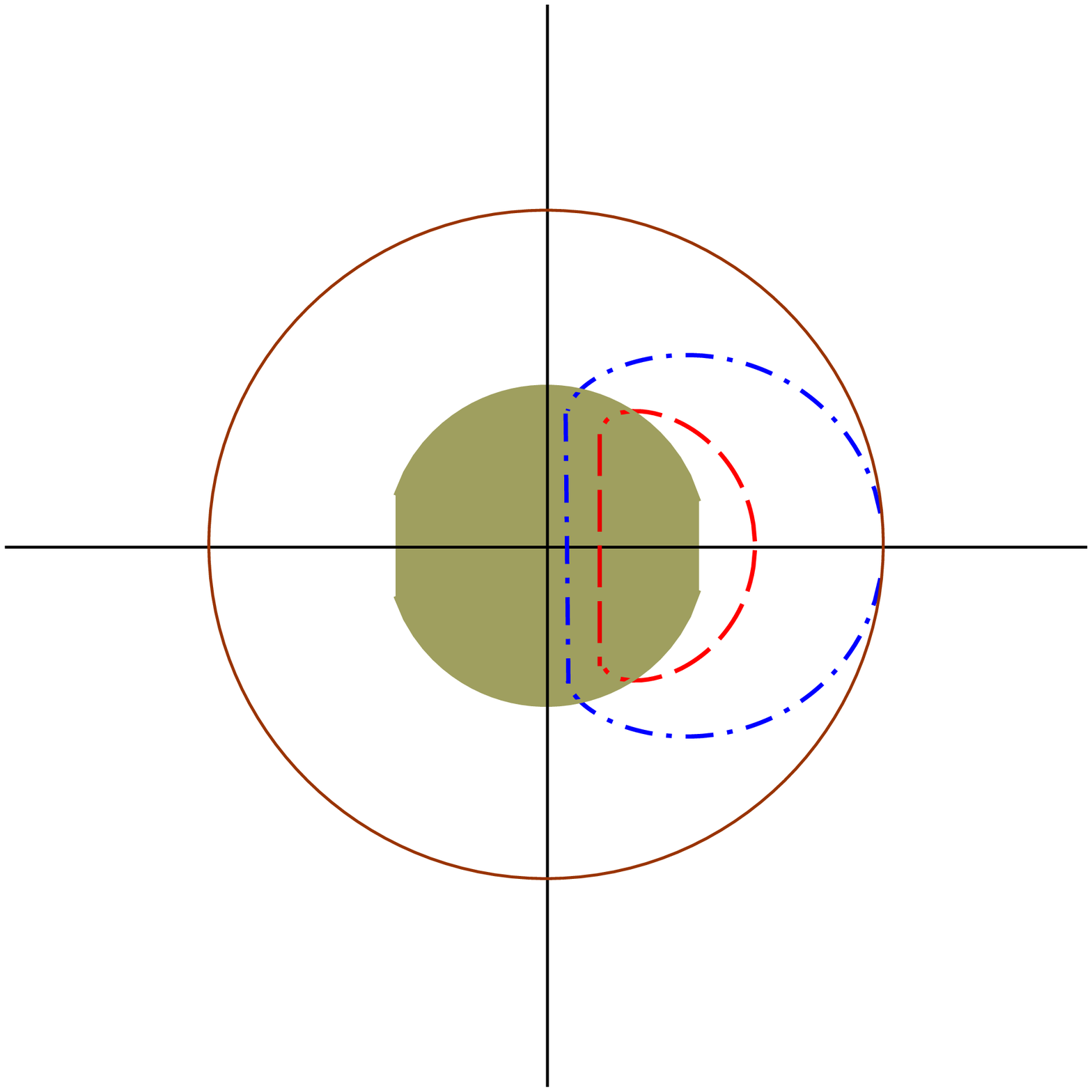}
	}
	\subfigure[]
	{
		\includegraphics[width=0.3\textwidth]{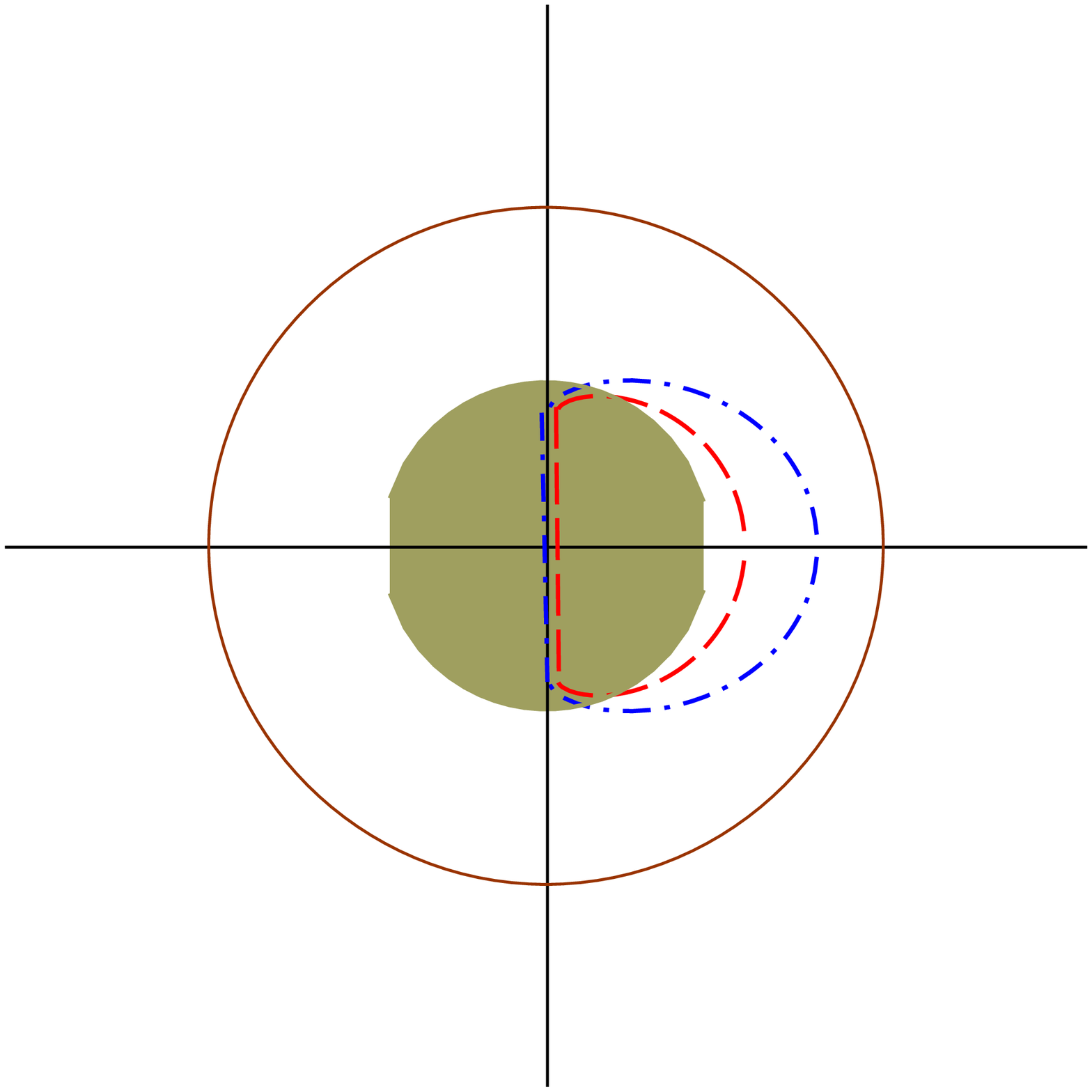}
	}
	\caption{\label{fig4}Shadow of black holes for $\Lambda=6\times10^{-2}$. The electric charge, ($Q=0, a_{max}=1.02$), ($Q=0.75, a_{max}=0.88$) and ($Q=1.35, a_{max}=0.46$) in (a), (b) and (c) respectively. Each figure is shown for different value of $a_{max}$. In the green solid filled pictures $a=\dfrac{2}{100} a_{max}$, in the dash red picture $a=\dfrac{2}{5} a_{max}$ and in the blue dash-dot plot $a=\dfrac{4}{5} a_{max}$. The brown solid line is reference circle. The detail of parameters is shown in table~\ref{tab2} (see Appendix).}
 \label{pic:shadow}
\end{figure}
It can be seen from Figs. \ref{fig2} - \ref{fig4}, by increasing the spin parameter, the shadow of black holes become more asymmetric into vertical axis. 

Also, in Fig. \ref{theta}, the effect of $\theta_{O}$ on the image of black hole's shadow is demonstrated. It can observe from this figure that by decreasing $\theta_{O}$ from $\frac{\pi}{2}$ to limit of 0, the asymmetric decreases.
\begin{figure}[h]\label{theta}
	\centering
	\subfigure[$\theta_{O}\simeq0$]{
		\includegraphics[width=0.3\textwidth]{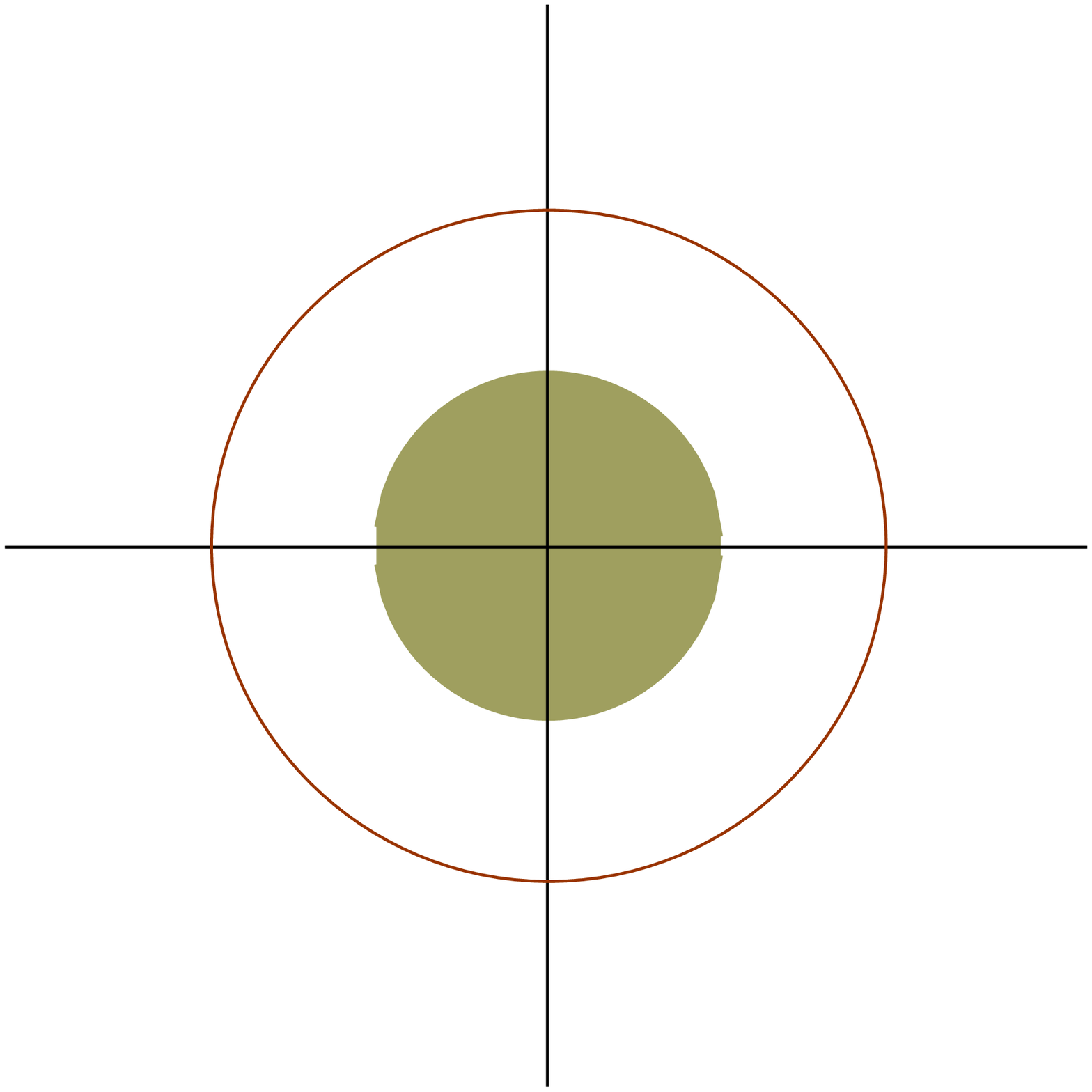}
	}
	\subfigure[$\theta_{O}$=$\frac{\pi}{6}$ ]{
		\includegraphics[width=0.3\textwidth]{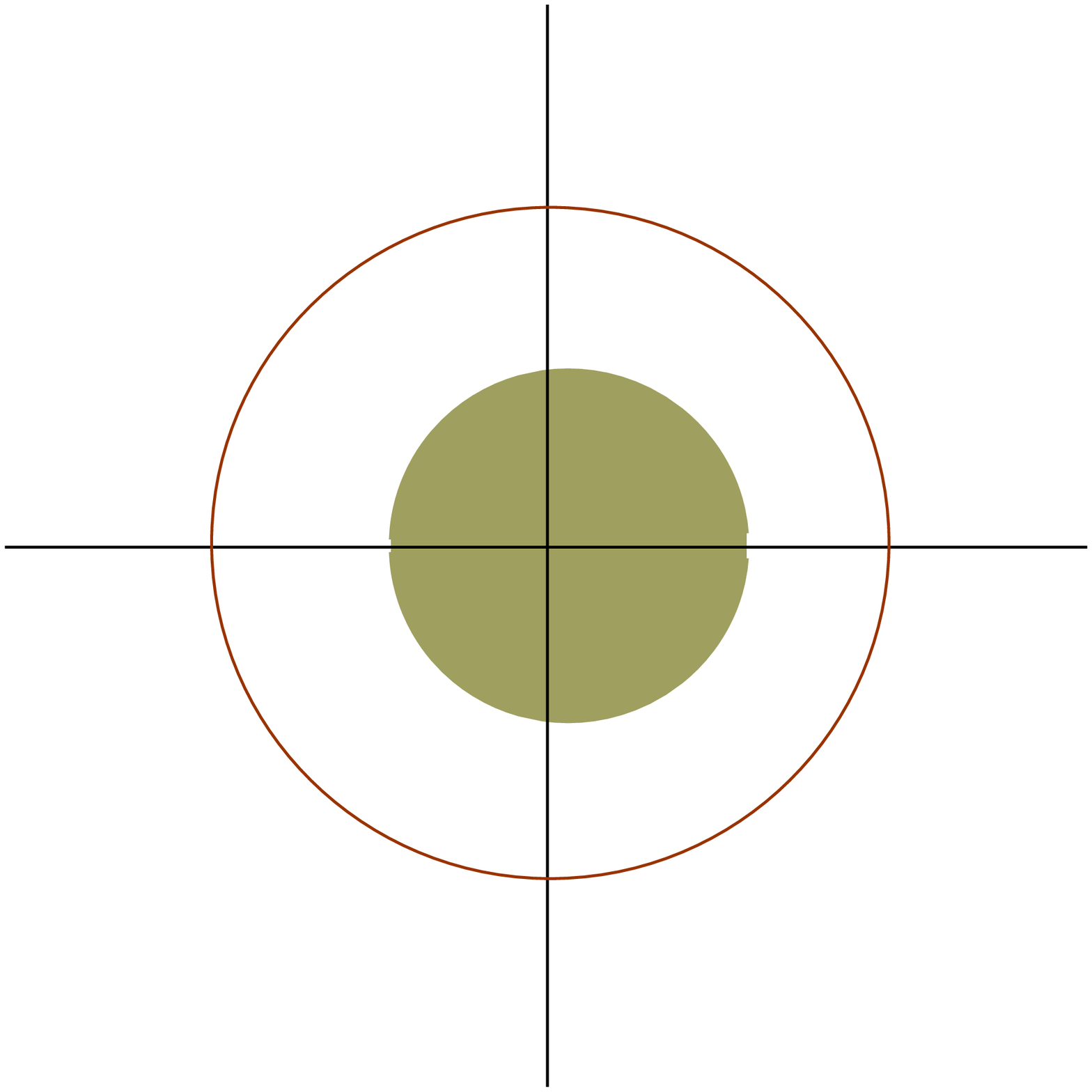}
	}
	\subfigure[$\theta_{O}$=$\frac{\pi}{4}$]{
		\includegraphics[width=0.3\textwidth]{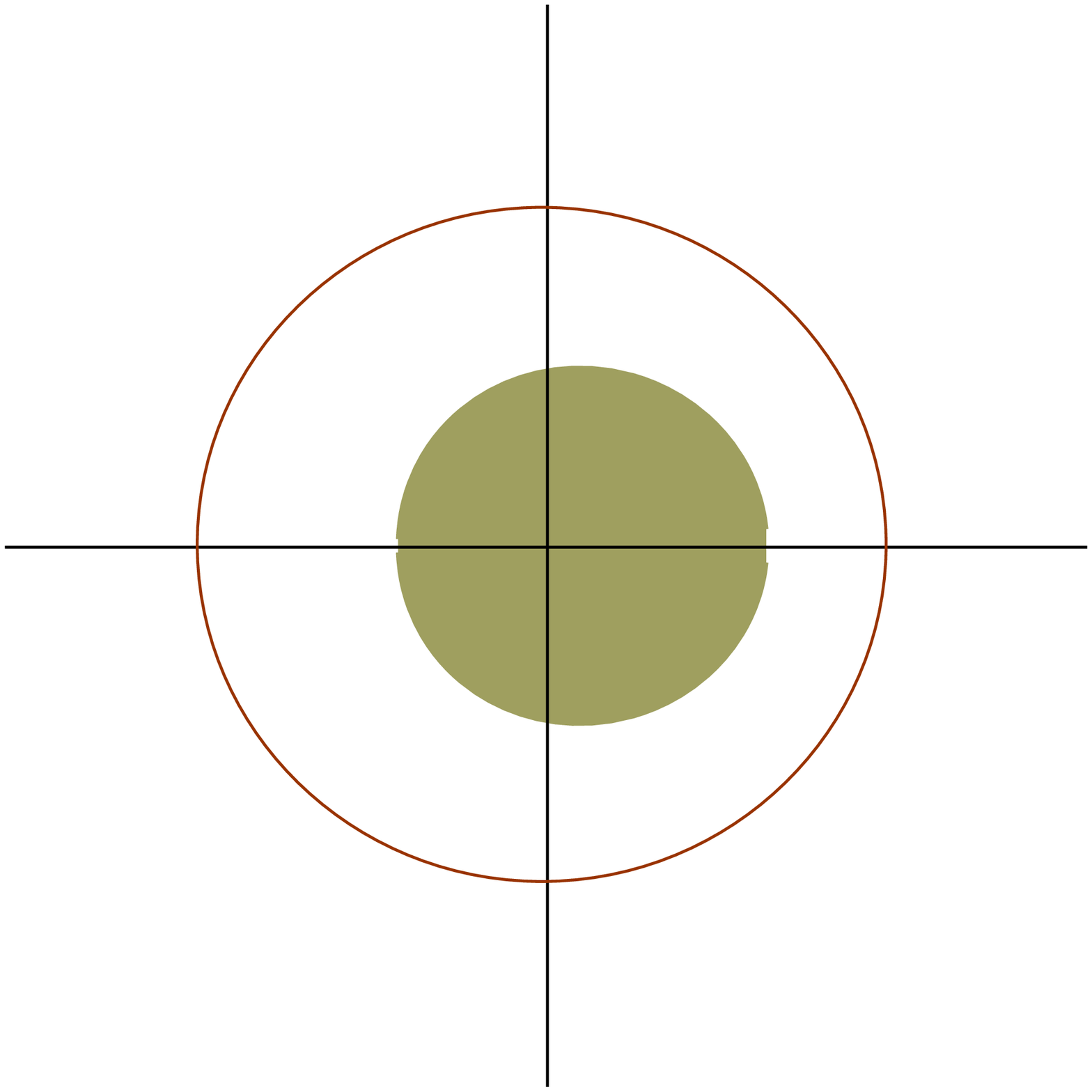}
	}
	\subfigure[$\theta_{O}$=$\frac{\pi}{3}$]{
		\includegraphics[width=0.3\textwidth]{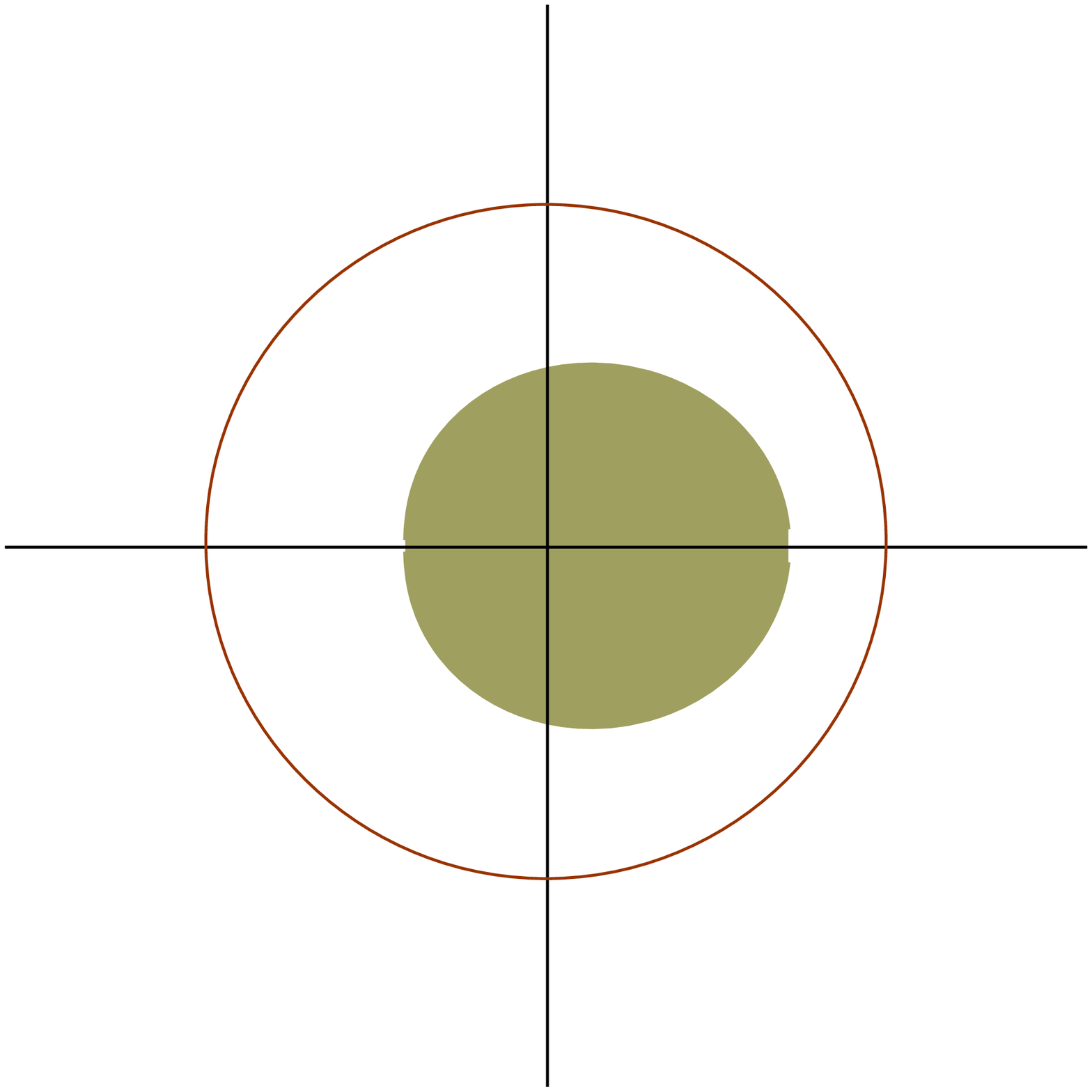}
	}
	\subfigure[$\theta_{O}$=$\frac{\pi}{2}$]{
		\includegraphics[width=0.3\textwidth]{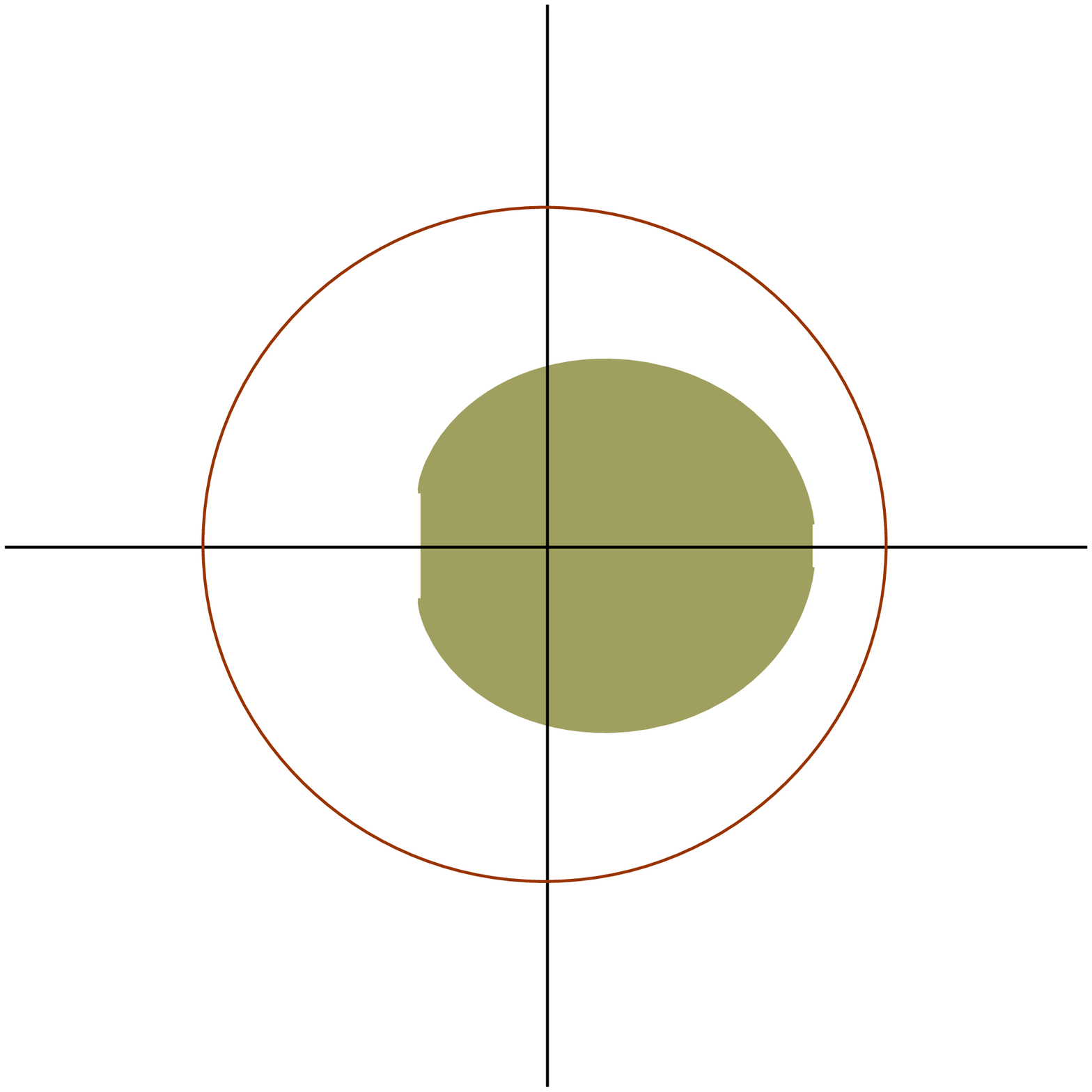}
	}
	\caption{\label{theta}Shadow of a black hole for an observer at $r_{O}=5M$ and different inclination angles $\theta_{O}$, with fixed $Q=1.35$,$\Lambda$ = $10^{-2}$ and $a=\frac{85}{100}a_{max}$. The brown solid line is reference circle.}
 \label{pic:shadow}
\end{figure}
Moreover, the effect of $\Lambda$ on the size and shape of shadow is shown in Fig. \ref{fig5}. It is obvious from Fig. \ref{fig5}, that for $\Lambda$ from 0 to $6\times10^{-2}$ and for the observer placed at $\theta_{O}=\frac{\pi}{2}$ and $r_{O}=5M$, the size of shadow become smaller. But in Fig. \ref{fig5}(d), since $a=0$, these changes don't appear.

\begin{figure}[h]\label{fig5}
	\centering
	
	\subfigure[]{
		\includegraphics[width=0.4\textwidth]{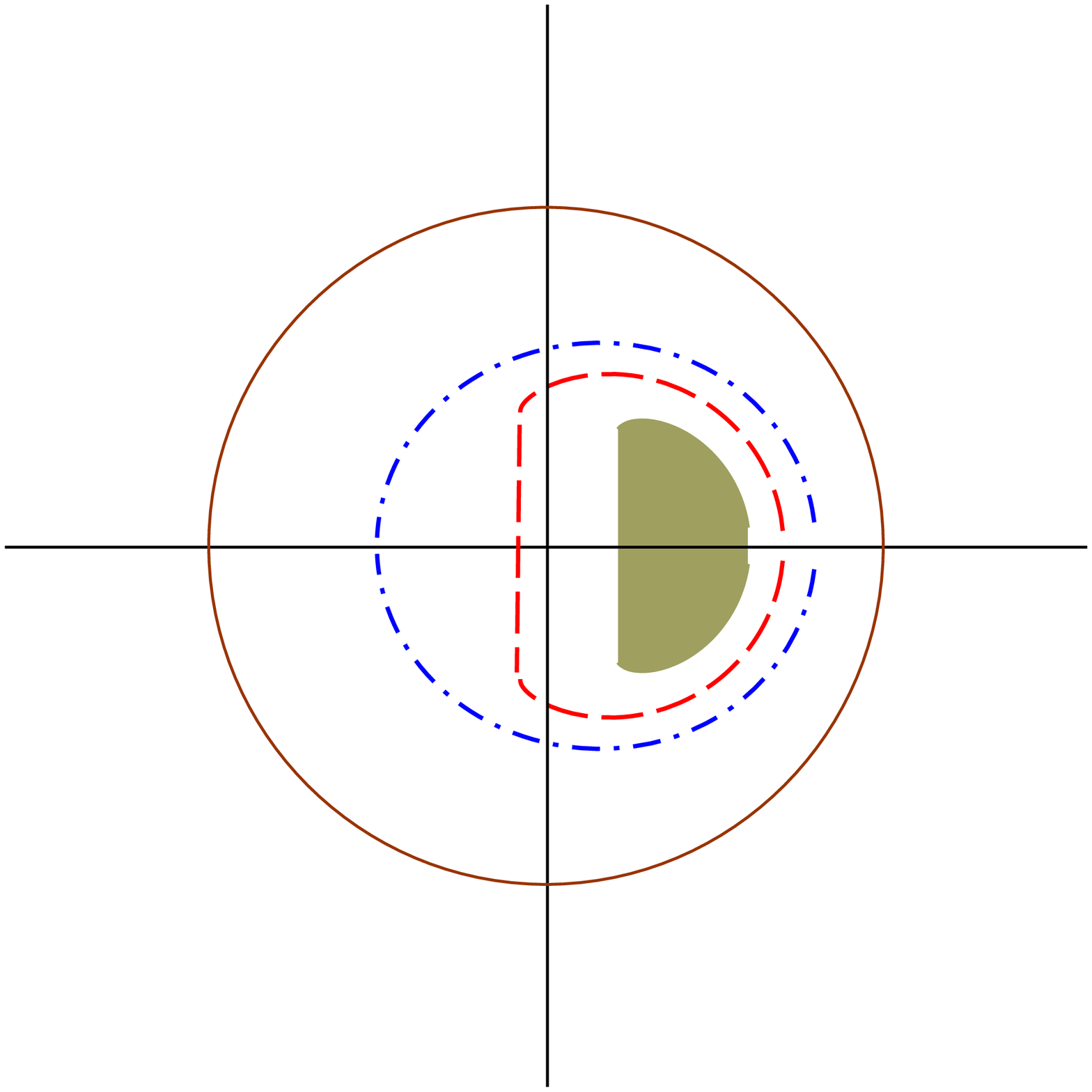}
	}
	\subfigure[]{
		\includegraphics[width=0.4\textwidth]{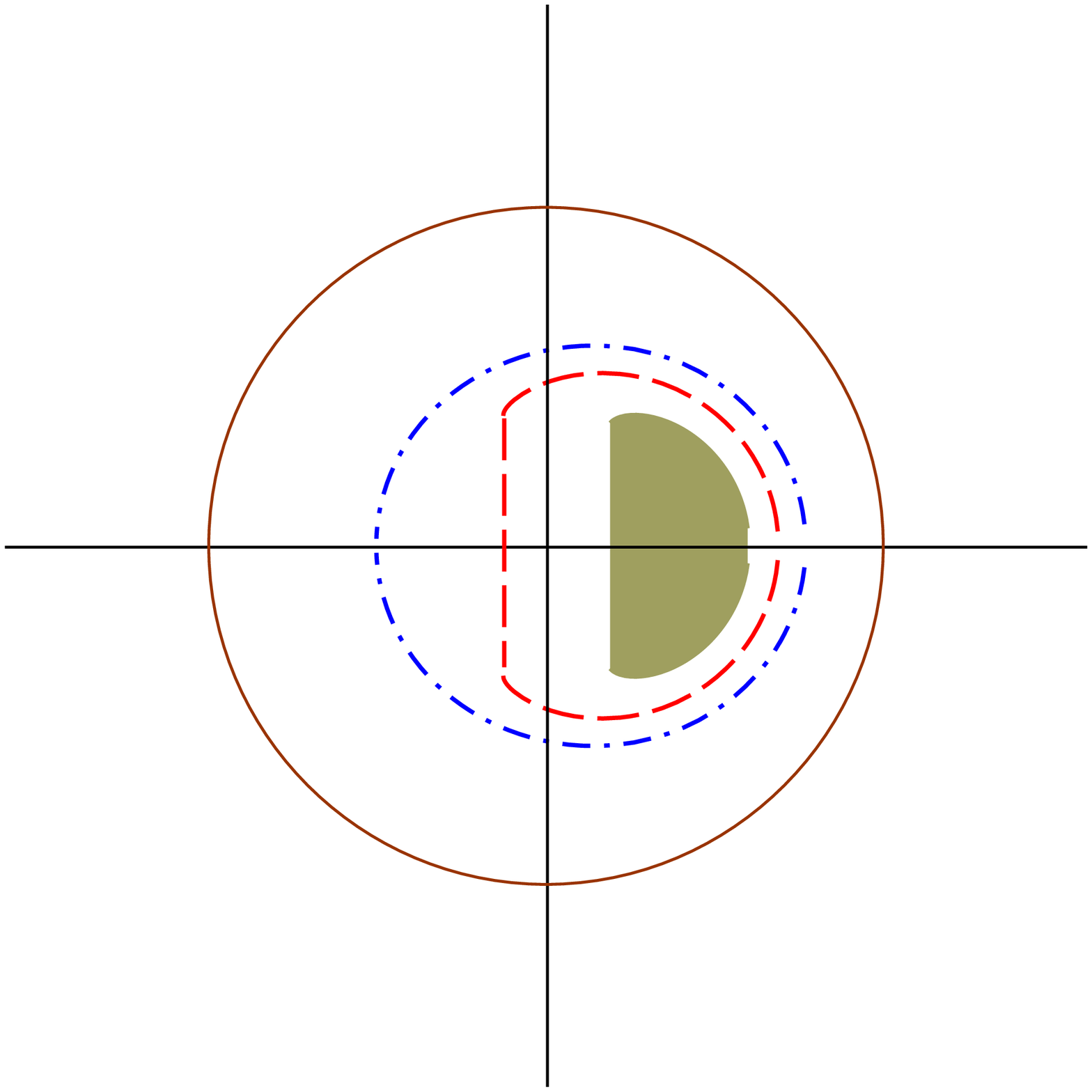}
	}
	\subfigure[]{
		\includegraphics[width=0.4\textwidth]{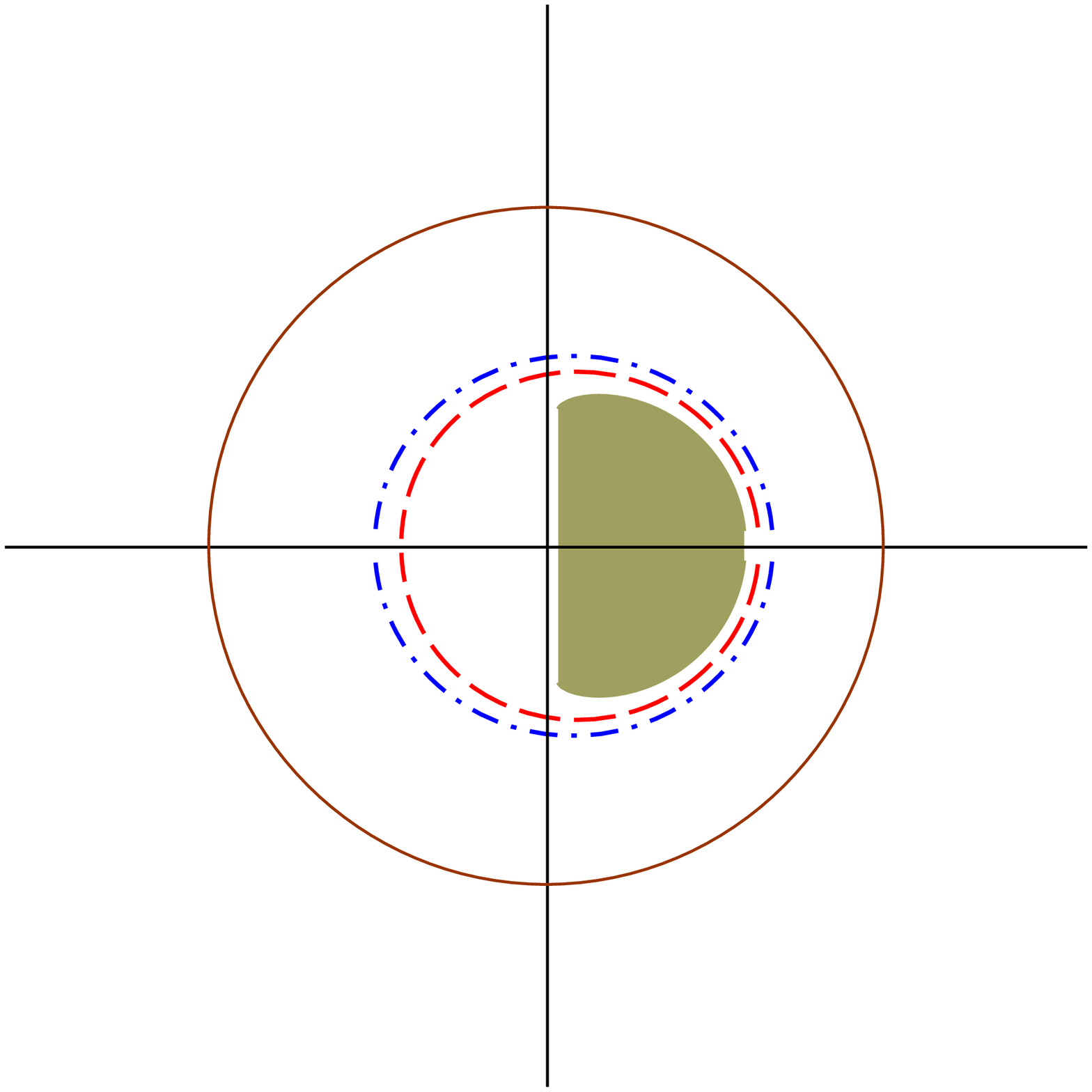}
	}
	\subfigure[]{
		\includegraphics[width=0.4\textwidth]{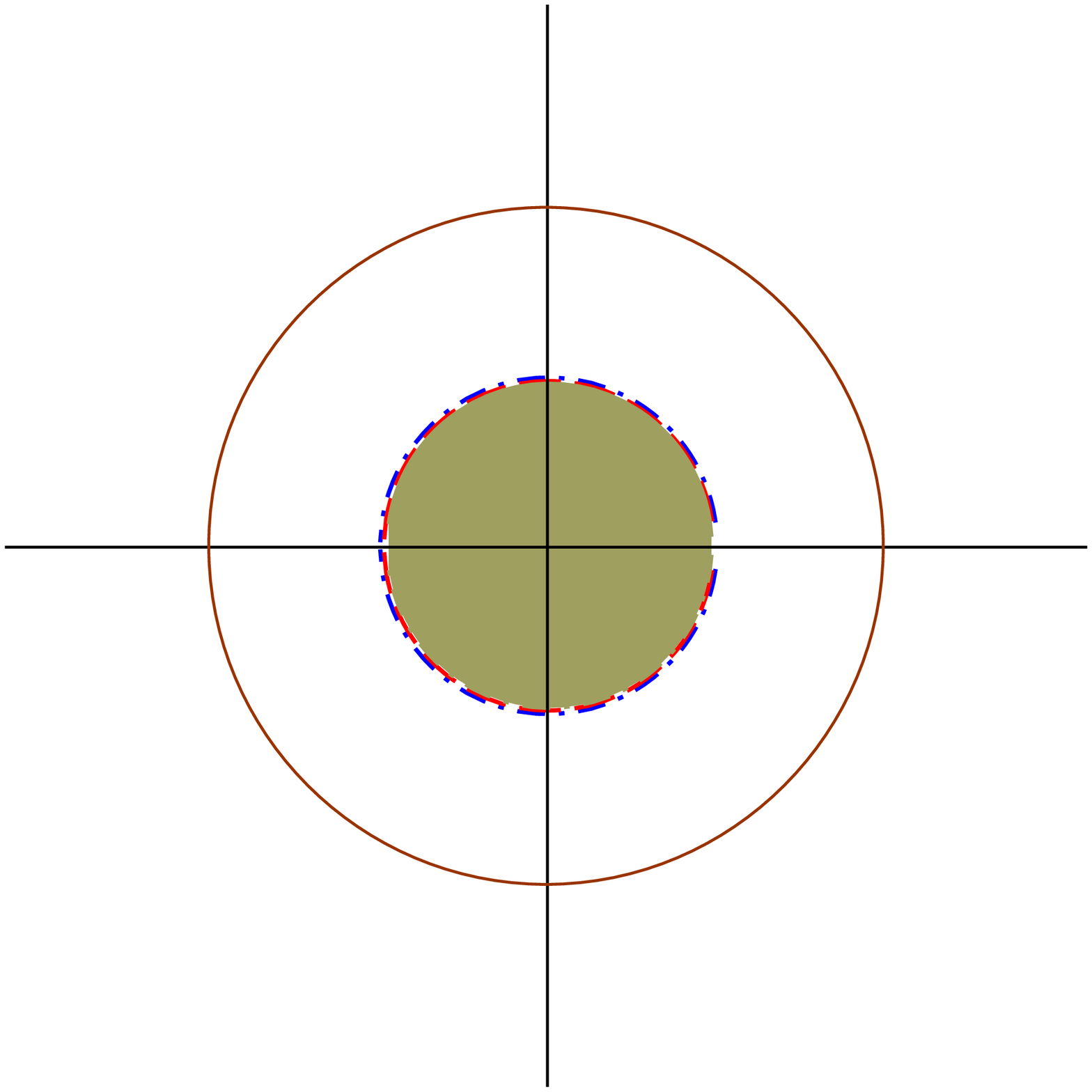}
	}
	
	\caption{\label{fig5}Shadow of black holes for $a=\frac{70}{100}a_{max}$. $Q=0$, $Q=0.75$, $Q=1.35$ and $Q=1.49$ in (a), (b), (c) and (d) respectively. Each figure is shown for different value of $\Lambda$. The blue dash-dot line is for $\Lambda=0$, in the dashed red line $\Lambda=10^{-2}$ and in the green solid filled line $\Lambda=6\times10^{-2}$, the detail of parameters is shown in table \ref{tab4}.}
 \label{pic:shadow}
\end{figure}

\clearpage
\section{shadow of black hole in the presence of plasma at $r \longrightarrow \infty$}\label{plasma}
In this section, we will discuss shadow of black hole in the presence of plasma at $r \longrightarrow \infty$.
In 1975, Bicak and Hadrava~\cite{Bick:1975}, had been investigated the travel of radiation for a dispersive and isotropic environment in General Relativity.
In addition, the shadow of black hole in the presence of plasma has been discussed in Refs.~\cite{Er:2013efa,Morozova:2012,Tsupko:2012}. 
In this section, we analyze the effect of the plasma around the black holes in $f(R)$ gravity. The plasma has the refraction index equal to $n=n(x^{i},\omega)$. This refraction index is attached to the photon four-momentum as~\cite{synge:11}
\begin{equation}\label{32}
n^{2} = 1 +\frac{p_{\alpha}p^{\alpha}}{(p_{\beta} u^{\beta})^{2}} ,
\end{equation}
where $u^{\beta}$, is the observer velocity. Note that, for the vacuum environment, $n=1$. Introducing the specific form of the plasma frequency for analytical results, we have~\cite{Atamurotov:2015nra}
\begin{equation}\label{33}
n^{2}= 1-\frac{\omega^{2}_{e}}{\omega_{v}^{2}} ,
\end{equation}
where, $\omega_{v}$, is the photon frequency and $\omega_{e}$, is the plasma frequency. Using the Hamilton-Jacobi equation for this geometry~\cite{BisnovatyiKogan:2010ar}
\begin{equation}\label{34}
\frac{\partial S}{\partial \tau}=-\frac{1}{2}[g^{ij}p_{i}p_{j}-(n^{2}-1)(p_{0}\sqrt{-g^{00}})^{2}] . 
\end{equation}
The equations of motion of photons in the presence of plasma can be obtained as
\begin{align}\label{35}
\rho^{4}(\dfrac{dr}{d\tau})^{2}=-\Delta_{r}(K+\varepsilon r^{2})+\big[(a^{2}+r^{2})E-aL\Xi \big]^{2}+(r^{2}+a^{2})^{2}(n^{2}-1)E^{2}=R(r),
\end{align}
\begin{align}\label{36}
\rho^{4}(\dfrac{d\theta}{d\tau})^{2}=\Delta_{\theta}(K-\varepsilon a^{2}cos^{2}\theta)-\dfrac{1}{sin^{2}\theta}\big(aE sin^{2}\theta -L\Xi \big)^{2}-(n^{2}-1)a^{2}E^{2}\sin ^{2}\theta =\Theta(\theta),
\end{align}
\begin{align}\label{37}
\rho^{2}(\dfrac{d\varphi}{d\tau})=\dfrac{aE\Xi (a^{2}+r^{2})-a^{2}\Xi^{2}L}{\Delta_{r}}-\dfrac{1}{\Delta_{\theta}sin^{2}\theta}(a\Xi E sin^{2}\theta -\Xi^{2}L),
\end{align}
\begin{align}\label{38}
\rho^{2}(\dfrac{dt}{d\tau})=\dfrac{n^{2}E(r^{2}+a^{2})^{2}-aL\Xi(r^{2}+a^{2})}{\Delta_{r}}-\dfrac{sin^{2}\theta}{\Delta_{\theta}}(n^{2}E a^{2}-\dfrac{L\Xi a}{sin^{2}\theta}).
\end{align}
Now, we consider the plasma frequency as~\cite{Abdujabbarov:2015pqp}
\begin{equation}\label{39}
\omega^{2}_{e}=\frac{4\pi e^{2}N(r)}{m_{e}},
\end{equation}
where, $m$ and $e$ are the mass and electron charge respectively. Also, in Eq.~(\ref{39}), $N(r)$, is the plasma number density, which is considered as below
\begin{equation}\label{40}
N(r) = \frac{N_{0}}{r^{h}},
\end{equation}
So, we have 
\begin{equation}
\omega_{e}^{2}=\frac{4\pi e^{2}N_{0}}{m_{e}r^{h}}=\frac{k}{r^{h}},
\end{equation}
in which, $h\geq0$. In following, for this case, we consider $h=1$~\cite{Rogers:2015dla}, and $n$ is equal to $\sqrt{1-\frac{k}{r}}$.
Therefore, we obtain the constants of motion (i.e. $\eta$ and $\xi$), using $R(r)=0$ and $\dot{R}(r)=0$ conditions in Eq.~(\ref{35}). Then, for an observer in $\theta_{o}$=$\pi/2$, the celestial coordinates ~(\ref{alpha})--~(\ref{beta}) will take the forms
\begin{eqnarray}\label{42}
\alpha &=&-\frac{\xi}{n},\nonumber \\                     
\beta &=&\frac{\sqrt{\eta+a^{2}-n^{2}a^{2}}}{n}.
\end{eqnarray}
Now, using $\alpha$ and $\beta$, we plot some examples of black holes shadow in the presence of plasma, which are shown in Fig.\ref{fig9}. It can be seen from Fig.\ref{fig9} that the shape and size of the shadow are dependent to values of $a$, $Q$ and plasma parameters. 
In addition, for this situation, again we put $\Lambda=0$, because our observer is placed at infinity.
\begin{figure}[h]\label{fig9}
	\centering
	\subfigure[$a=0$]{
		\includegraphics[width=0.4\textwidth]{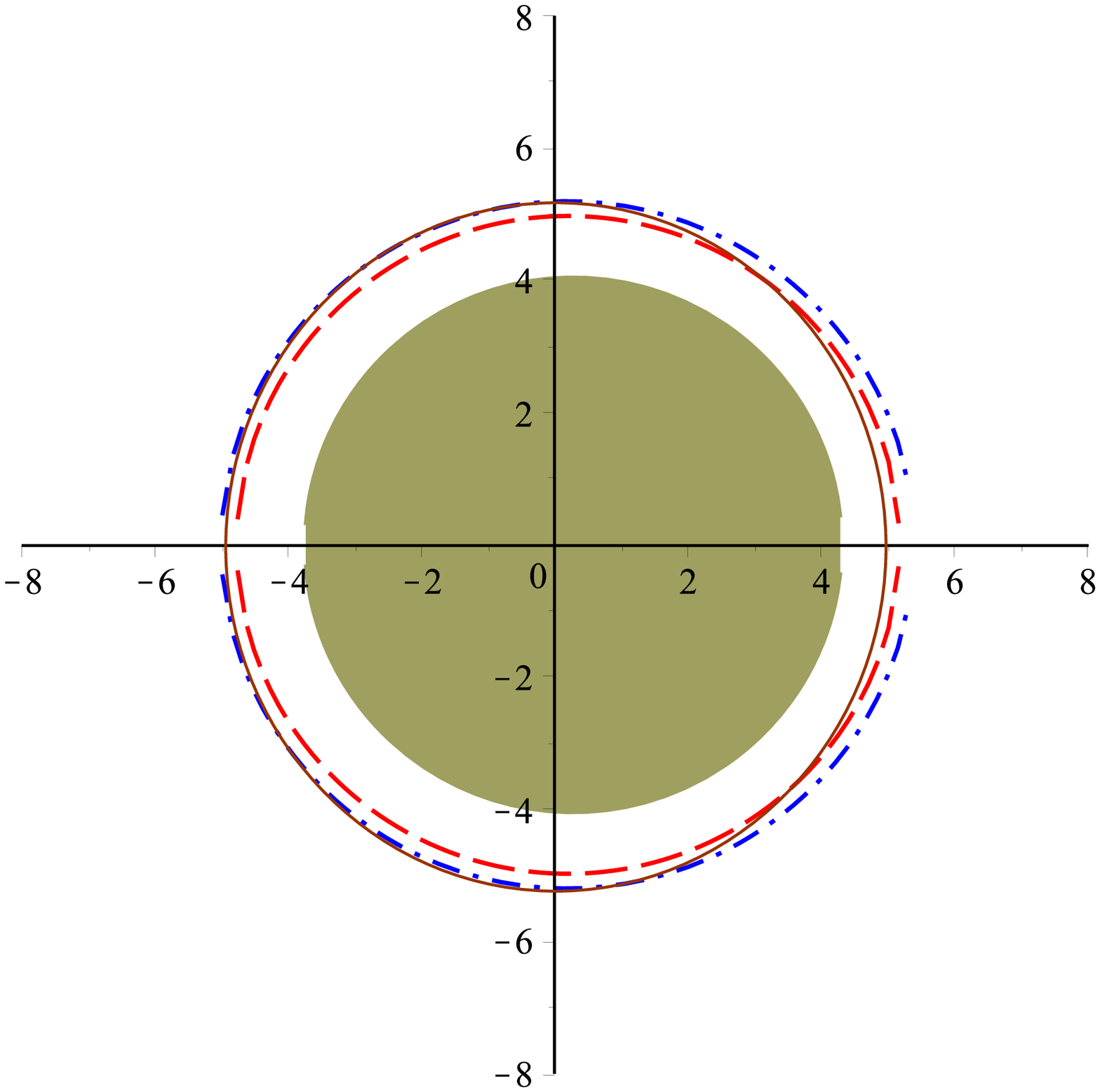}
	}
	\subfigure[$a=0.5$]{
		\includegraphics[width=0.4\textwidth]{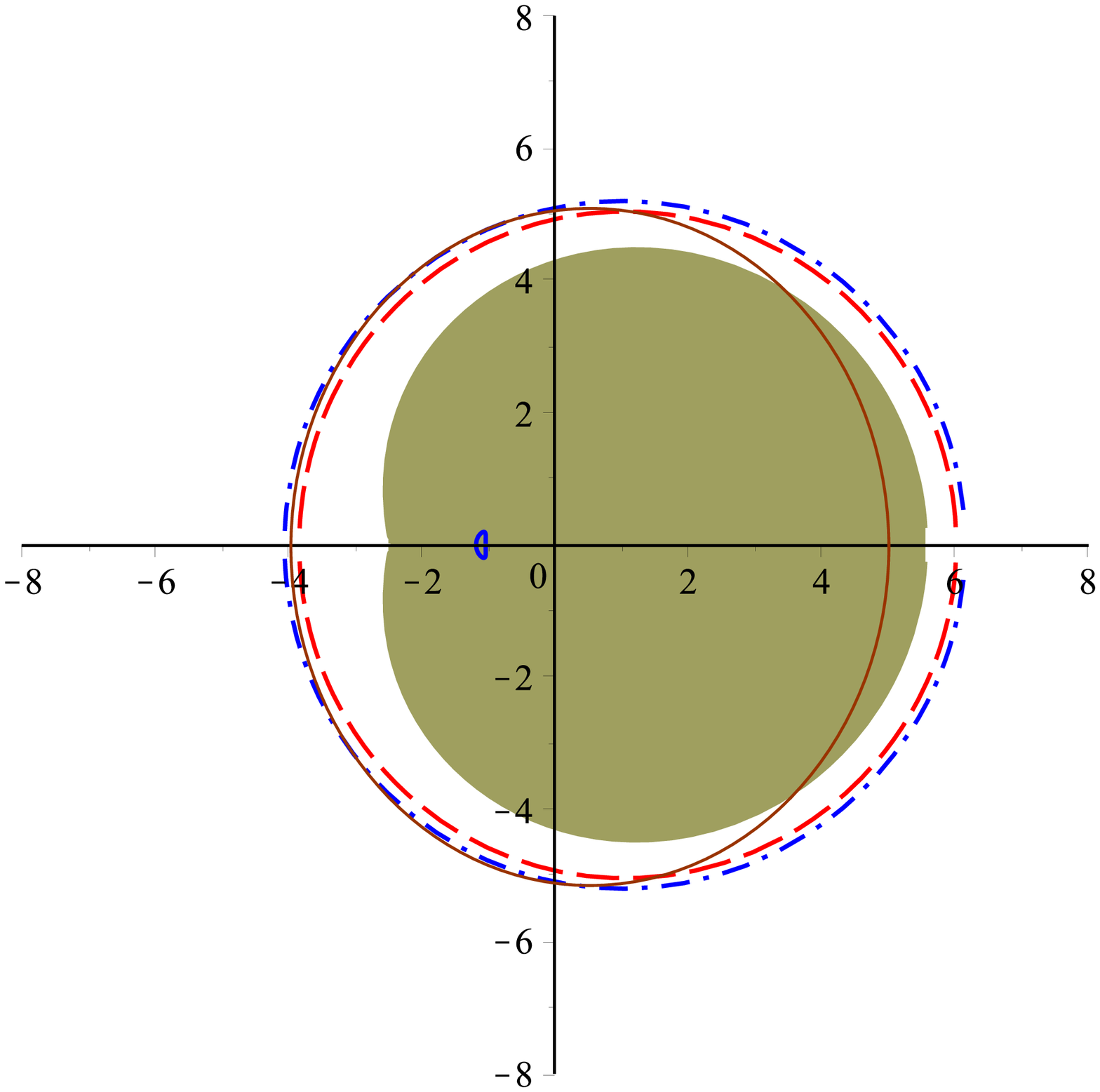}
	}
	\subfigure[$a=0.7$]{
		\includegraphics[width=0.4\textwidth]{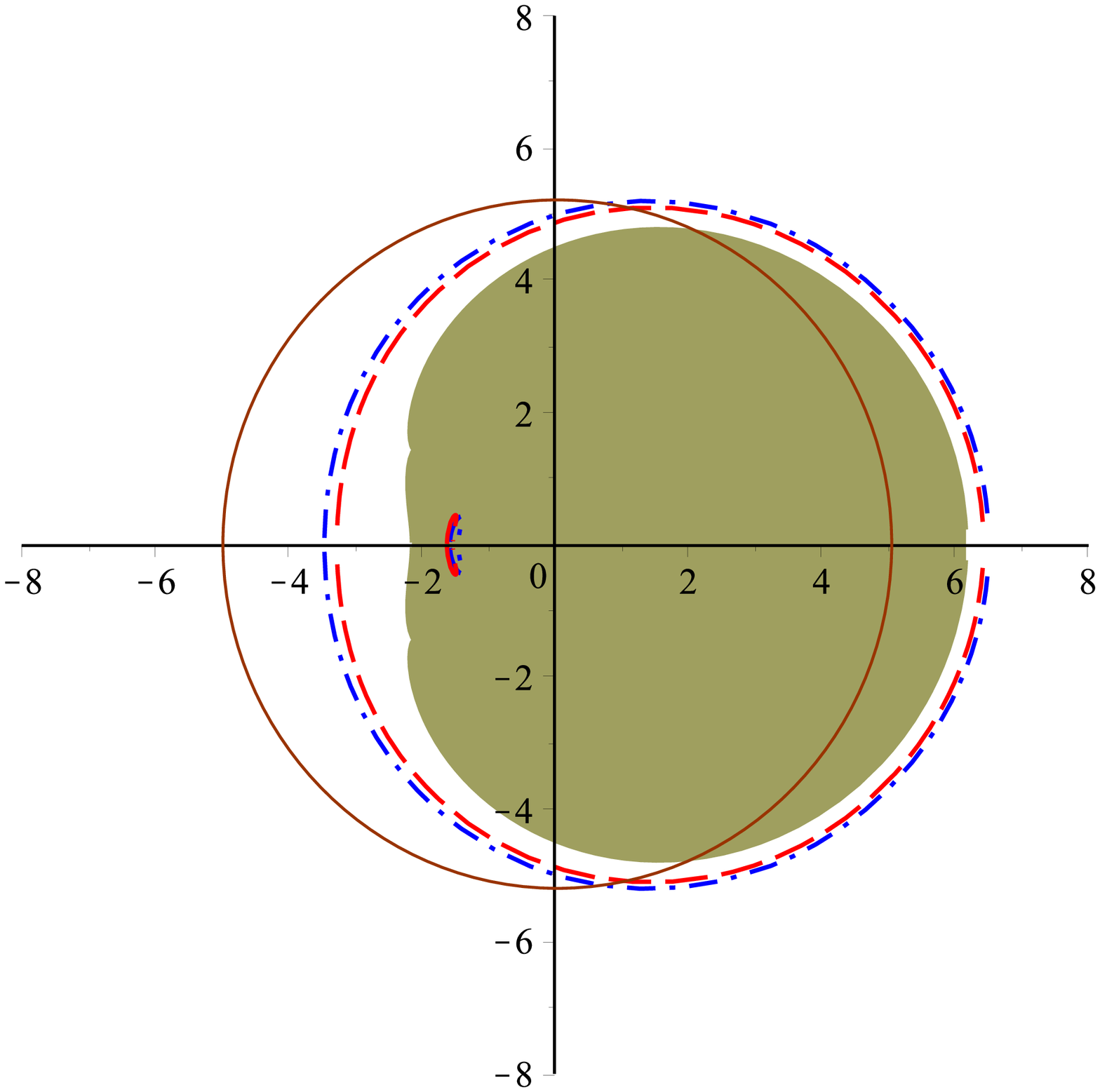}
		}
		\subfigure[$a=1$]{
		\includegraphics[width=0.4\textwidth]{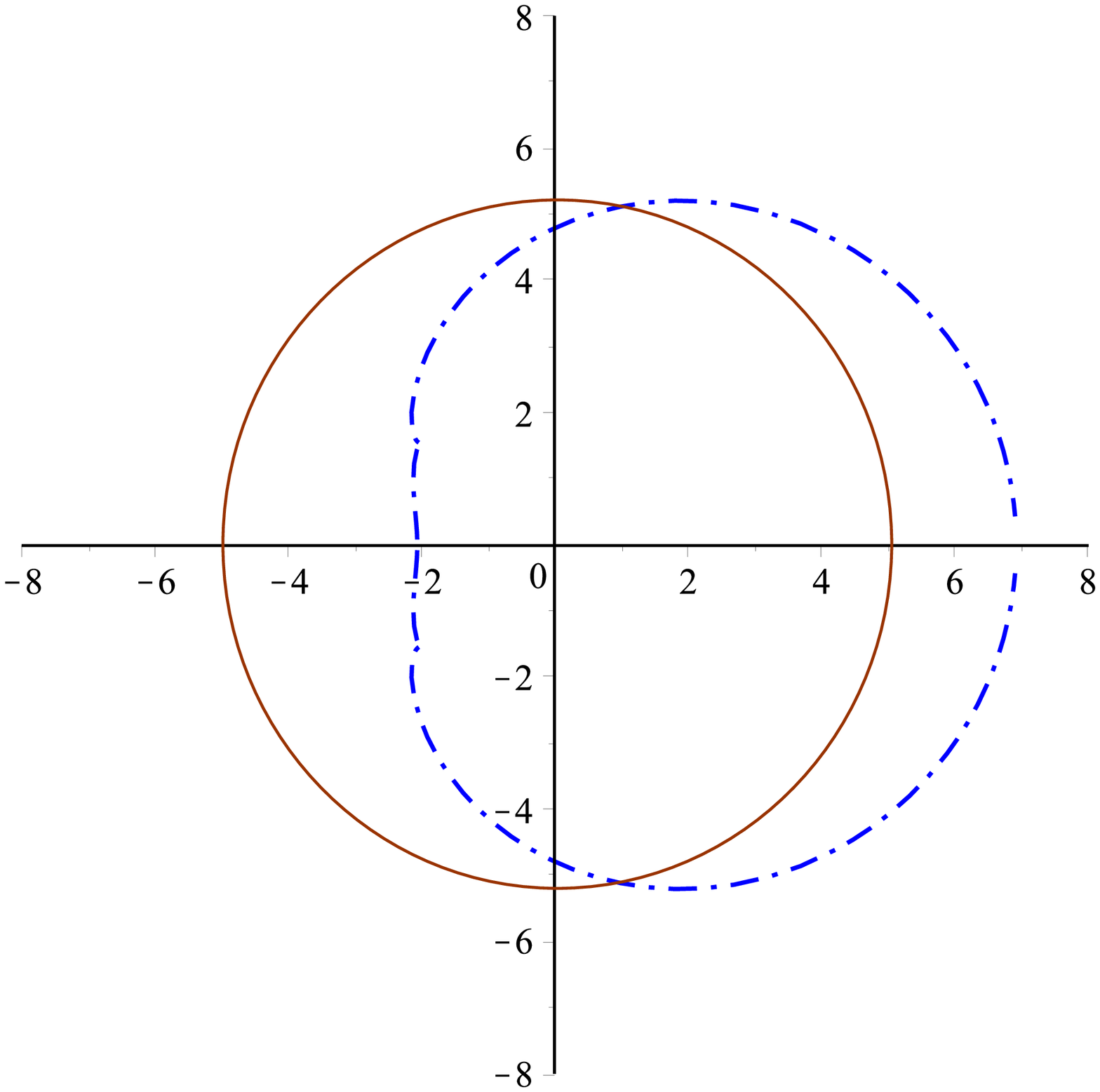}
		}
	\caption{\label{fig9}Shadow of the black hole in the presence of plasma for different values of the rotation parameter $a=0$, $a=0.5$, $a=0.7$ and $a=1$ in (a), (b), (c) and (d) respectively. In each figure, $Q=0$, $Q=\frac{Q_{crit}}{2}$ and $Q=Q_{crit}$ is shown by blue dash dot line, red dash line and green solid filled line respectively. The solid brown circle is reference circle. The detail of parameters is shown in table~\ref{tab5}(see Appendix).}
 \label{pic:shadow}
\end{figure}
 
 In Fig. \ref{fig11}, the effects of different plasma parameter $k$ on the shape of shadow are investigated. One can see that, the size of black hole's shadow decreases in the presence of plasma, in other words, the size of shadow decreases when $k$ increases.
\begin{figure}[h]\label{fig11}
	\centering
	{
		\includegraphics[width=0.4\textwidth]{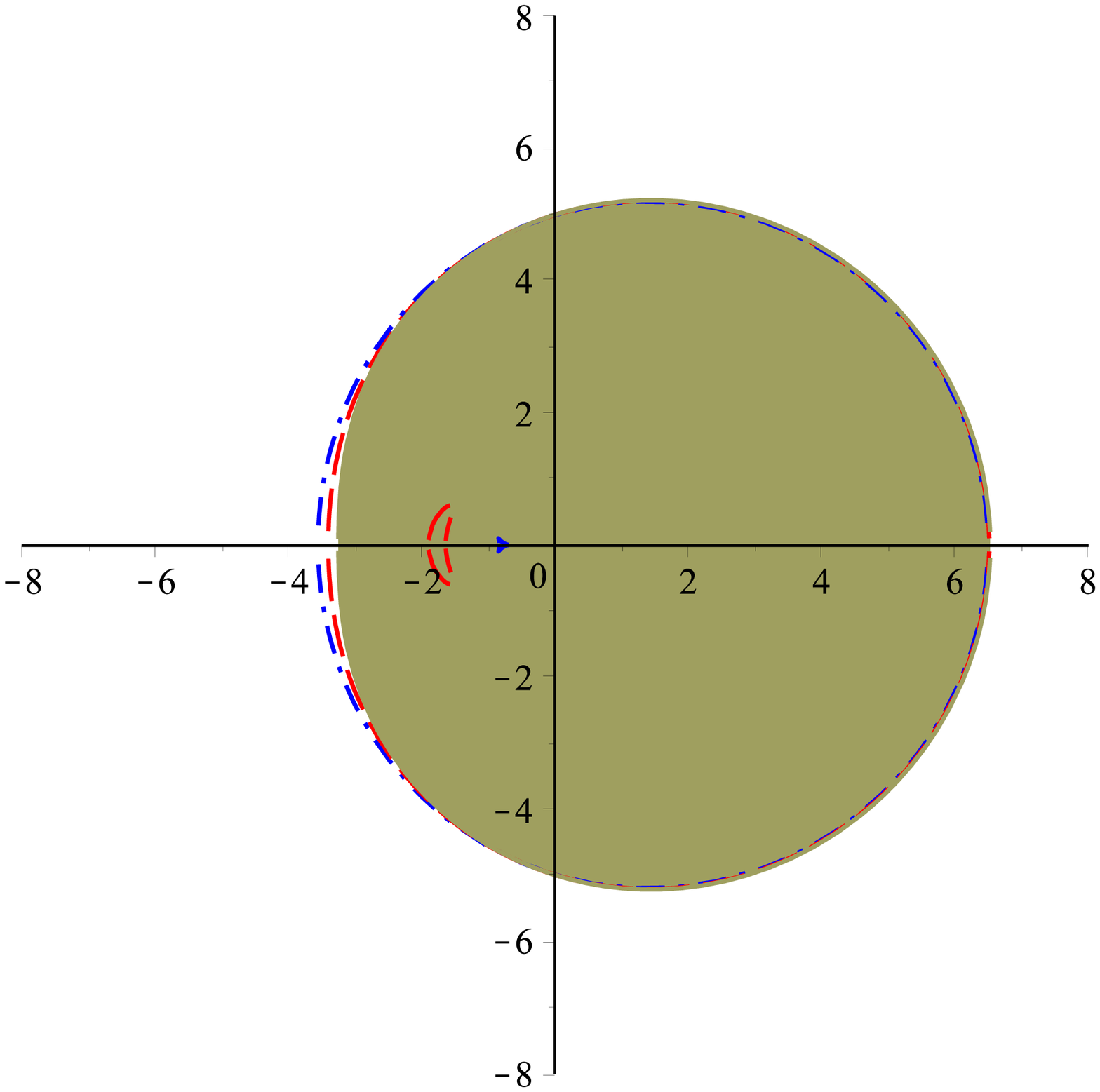}
	}
	\caption{\label{fig11}plot showing the influence of $k$ for $a=0.7$ and $Q=0$. The blue dash-dot line indicate $k=0$, the dash red line show $k=0.15$ and the green filled line show $k=0.25$.}
 \label{pic:shadow}
\end{figure}
Moreover, the influence of charge and rotation parameters in the presence of plasma, are similar to vacumm state (i.e. by increasing electric charge, the size of shadow become smaller, as well as, by increasing spin parameter, the symmetry of black hole's shadow decreases). On the other hand, the radius of shadow in the presence of plasma, is always less than or equal to vacuum state.

\section{the shadow for an observer in ($r_{O}$ ,$\theta_{O}$) in the presence of plasma}\label{plasma limited}
In this section, we plot the shadow of black hole in $f(R)$ gravity for an observer in $r$=$r_{O}$ in the presence of plasma. For this purpose, by the conditions $R(r)=0$ and $\frac{dR(r)}{dr}=0$ for Eq.~(\ref{35}), we can obtain the constant of motion, i.e. ($\xi$ and $\eta$) as
\begin{equation}\label{18.5.3}
\eta=\frac{4r(a^{2}n^{2}\Delta_{r}^{'}+r^{2}n^{2}\Delta_{r}^{'}-a^{2}\Delta_{r}^{'}-r^{2}\Delta_{r}^{'}+2r\Delta_{r}+\sqrt{\mathcal{N}})}{\Delta_{r}^{'2}},
\end{equation}
\begin{align}
\mathcal{N}=(4a^{2}n^{2}r\Delta_{r}\Delta_{r}^{'}+4r^{3}n^{2}\Delta_{r}\Delta_{r}^{'}-a^{2}n^{2}\Delta_{r}^{'2}\nonumber\\
-4a^{2}r\Delta_{r}\Delta_{r}^{'}-4r^{3}\Delta_{r}\Delta_{r}^{'}+a^{2}\Delta_{r}^{'2}+4r^{2}\Delta_{r}^{2}+r^{2}\Delta_{r}^{'2}-n^{2}r^{2}\Delta_{r}^{'2}).
\end{align}
and 
\begin{equation}\label{18.5.4}
\xi=\frac{(a^{2}+r^{2})}{a\Xi}-\frac{2r\Delta_{r}}{a\Xi\Delta_{r}^{'}}-\frac{\sqrt{\mathcal{N^{'}}}}{a\Xi\Delta_{r}^{'}}.
\end{equation}
\begin{align}
\mathcal{N^{'}}=(4\Delta_{r}^{'}\Delta_{r}(a^{2}n^{2}r+n^{2}r^{3}-a^{2}r-r^{3})+\Delta_{r}^{'2}(r^{2}-a^{2}n^{2}-n^{2}r^{2}+a^{2})+4r^{2}\Delta_{r}^{2}) .
\end{align}
The equations, ~(\ref{18.5.3}) and ~(\ref{18.5.4}) convert to ~(\ref{18.5.1}) and ~(\ref{18.5.2}) when $n$ is equal to 1(vacuum case).
Using equations~(\ref{24}),~(\ref{28}),~(\ref{29}), the geodesic equations in the presence of plasma~(\ref{35})--~(\ref{38}) and constant of motion,~(\ref{18.5.3}) and ~(\ref{18.5.4}), we obtain the cartesian coordinates as
\begin{eqnarray}\label{42}
x(r_{p})&=&−2\tan (\frac{\omega(r_{p})}{2}) \sin (\psi (r_{p})), \nonumber \\
y(r_{p})&=&−2\tan (\frac{\omega(r_{p})}{2}) \cos  (\psi (r_{p})).
\end{eqnarray}
Now, we use of $x(r_{p})$ and $y(r_{p})$ parameters to plot some examples of black hole's shadow.
In this situation, the observer is in ($r_{O}$ ,$\theta_{O}$) in the domain of outer communication, so we can put $\Lambda>0$.
The shadows for different value of spin $a$, and electric charge $Q$, are shown in Figs.\ref{fig120} and \ref{fig131}.

\begin{figure}[h]\label{fig120}
	\centering
	\subfigure[]{
		\includegraphics[width=0.3\textwidth]{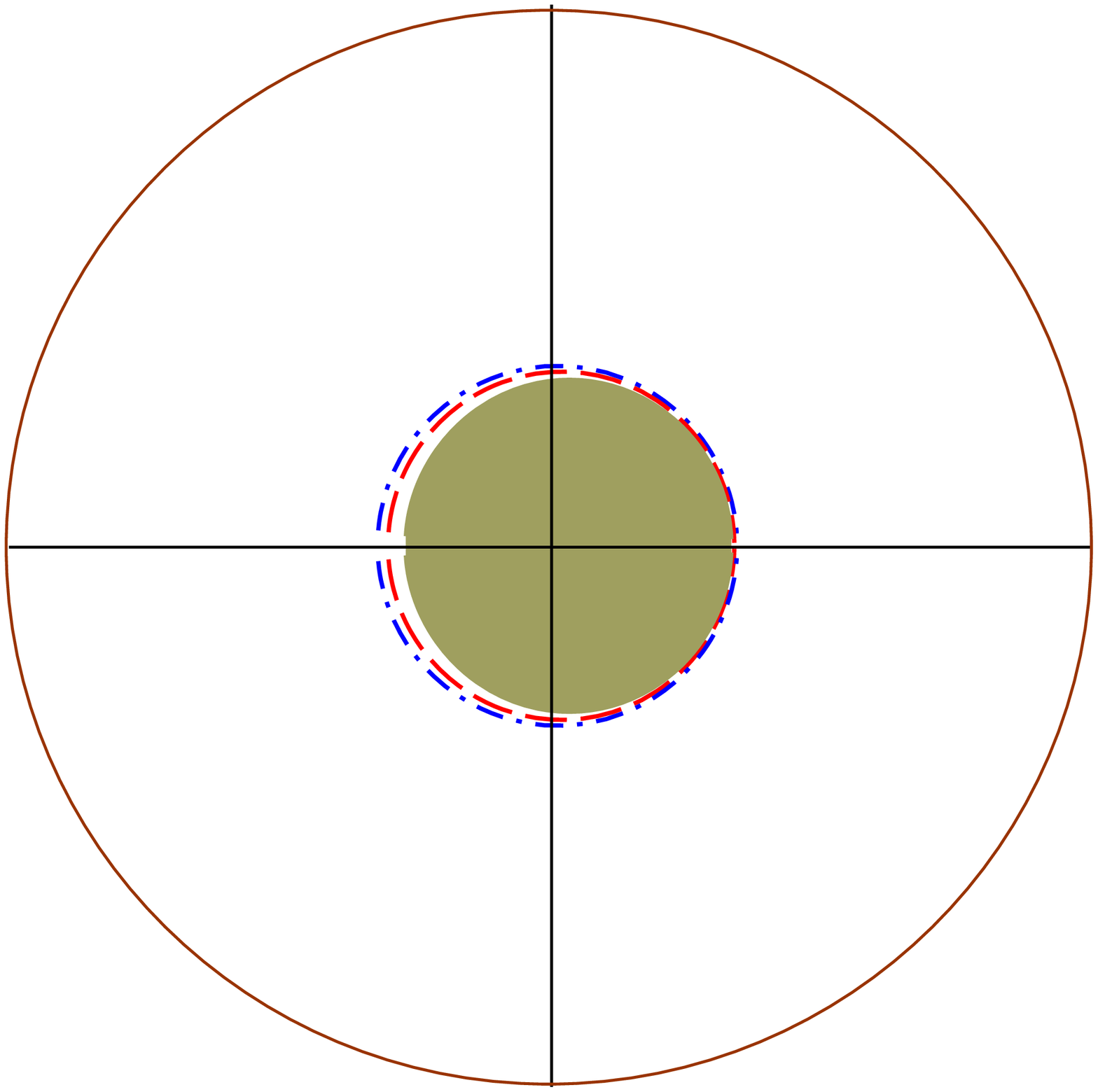}
	}
	\subfigure[]{
		\includegraphics[width=0.3\textwidth]{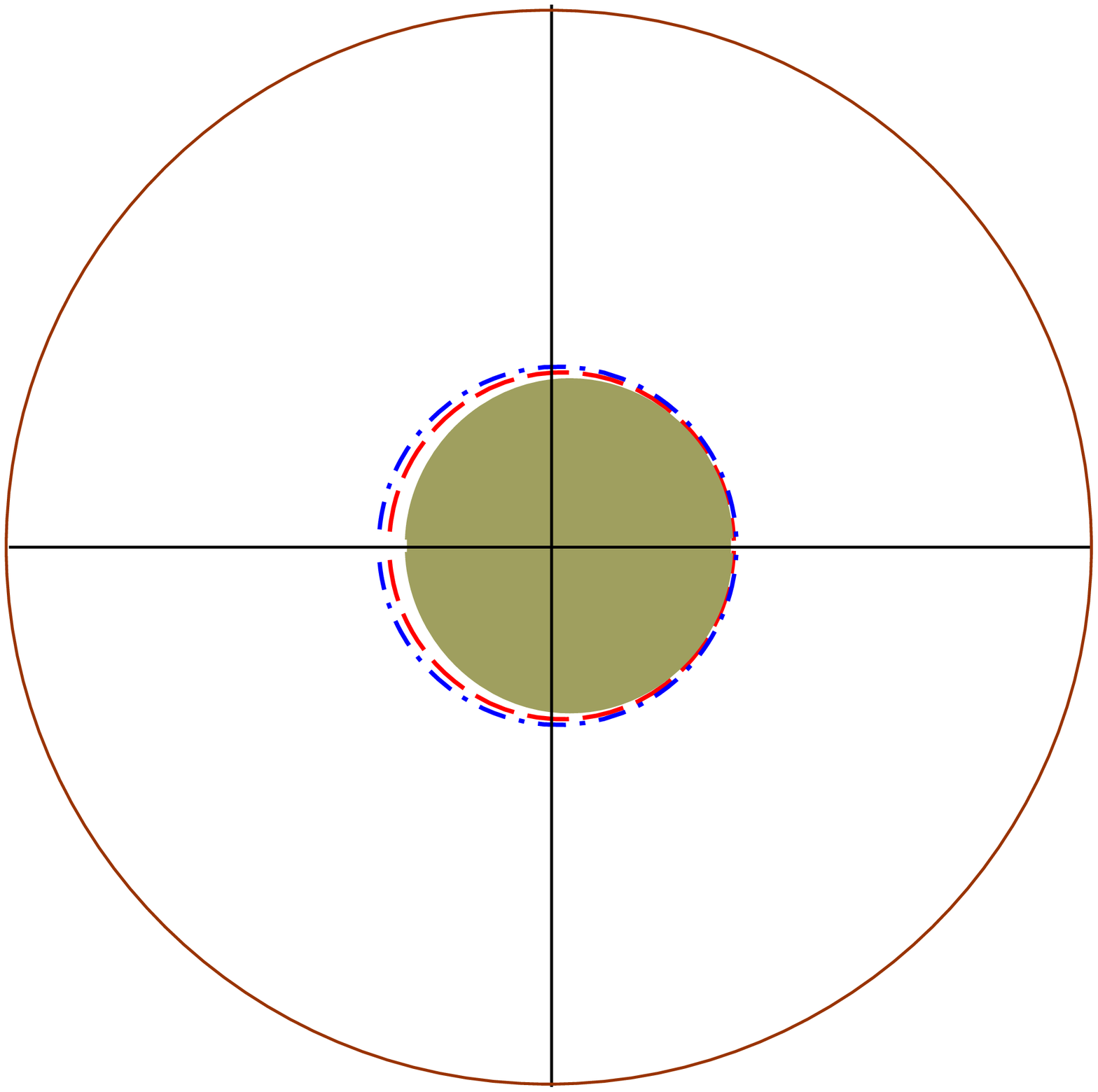}
	}	
	\subfigure[]{
		\includegraphics[width=0.3\textwidth]{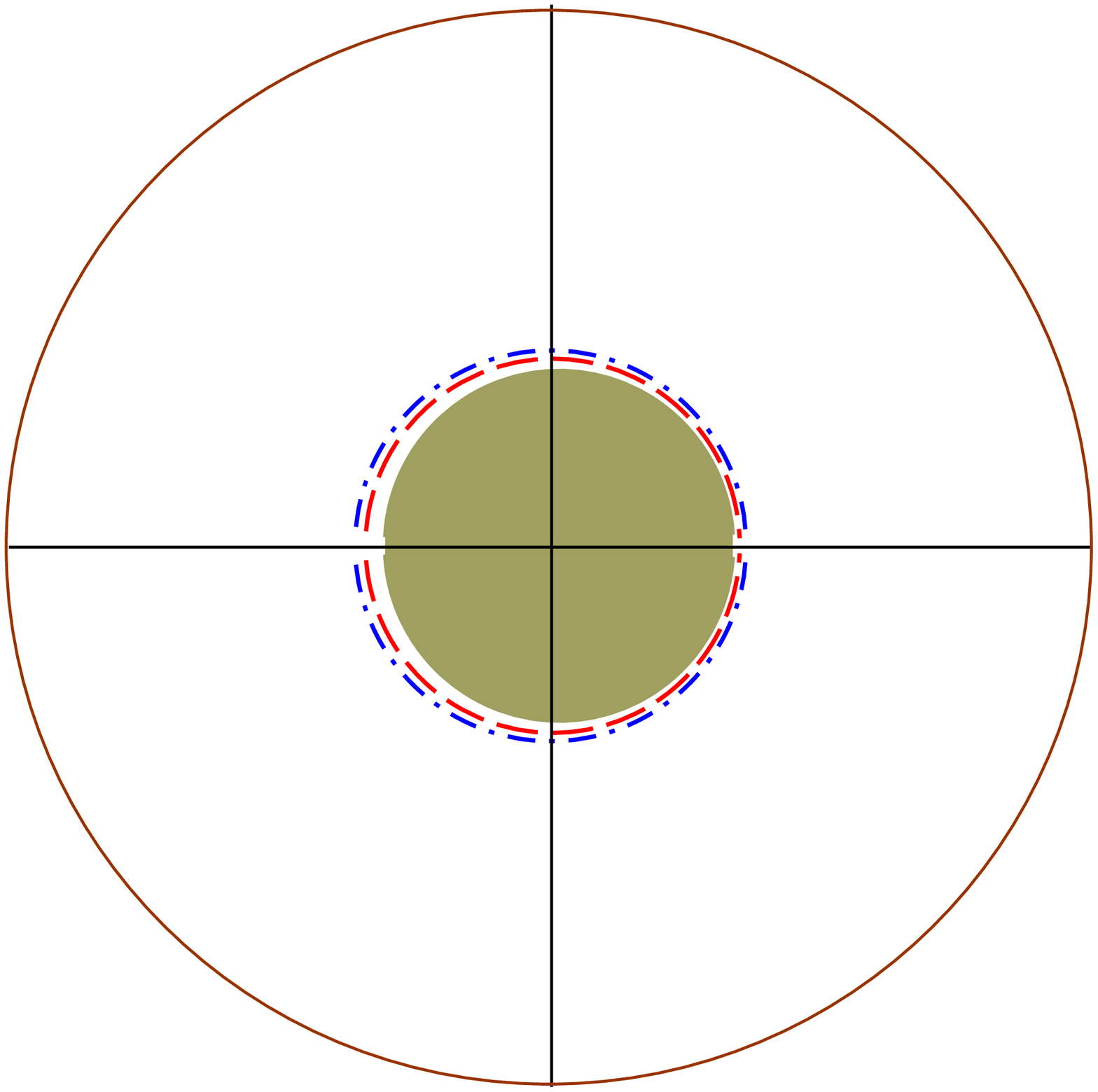}
	}
	\caption{\label{fig120}Shadow of black hole in the presence of plasma for specific observer. In this figure $Q=0.75$. $\Lambda=0$, $\Lambda=0.01$ and $\Lambda=0.06$ in (a), (b) and (c) respectively. In (a) and (b), shows $\frac{40}{100}a_{max}$, $\frac{50}{100}a_{max}$ and $\frac{65}{100}a_{max}$ by the blue dash-dot line, the red dash line and the green filled shape respectively. In (c), the blue dash-dot line, shows $\frac{10}{100}a_{max}$, the red dash line shows $\frac{20}{100}a_{max}$ and the green filled shape shows $\frac{40}{100}a_{max}$. The solid brown circle is reference circle. The detail of parameter are shown in table~\ref{tab7} (see Appendix).}
 \label{pic:shadow}
\end{figure}

\begin{figure}[h]\label{fig131}
	\centering
	\subfigure[]{
		\includegraphics[width=0.3\textwidth]{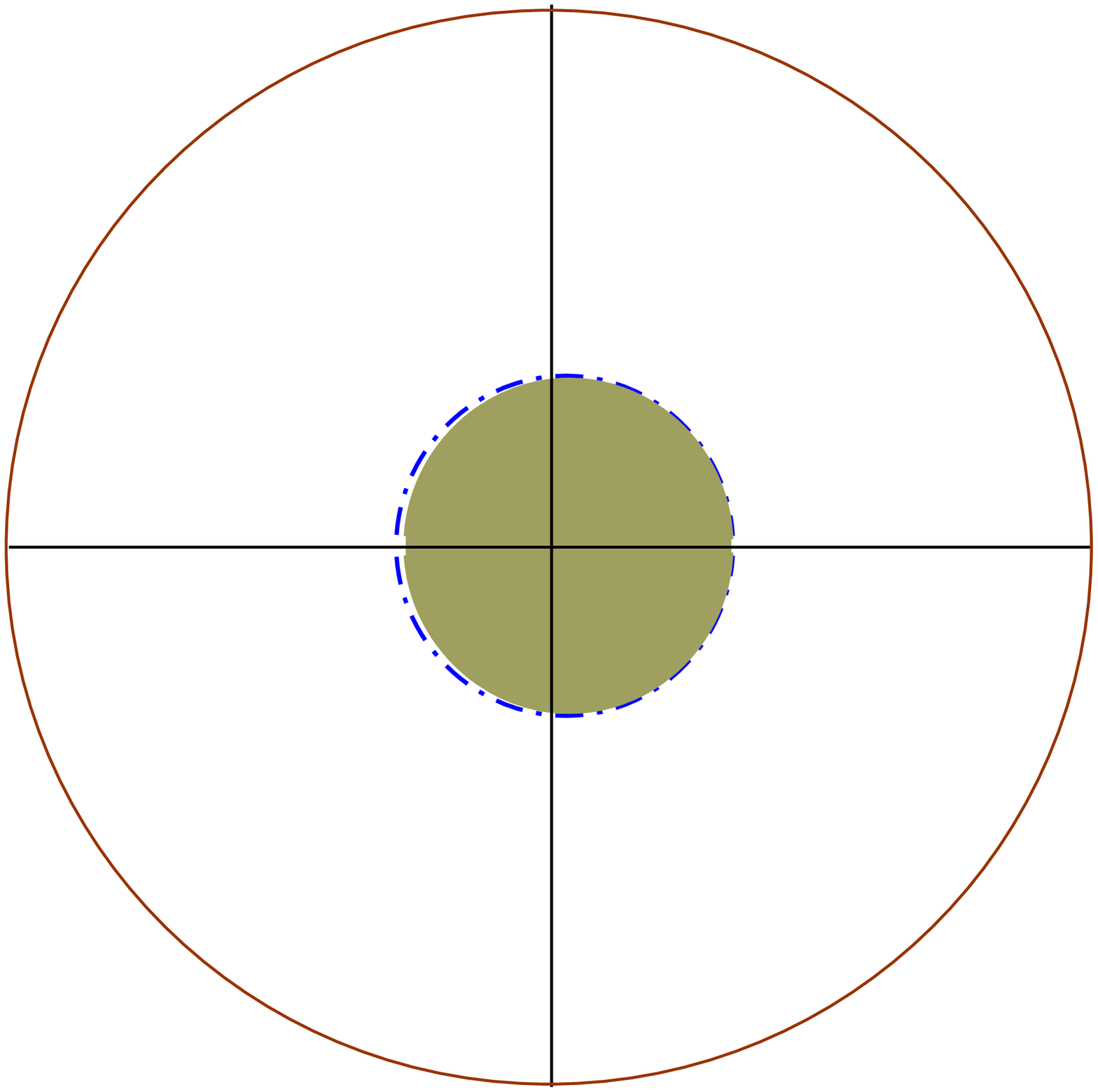}
	}
	\subfigure[]{
		\includegraphics[width=0.3\textwidth]{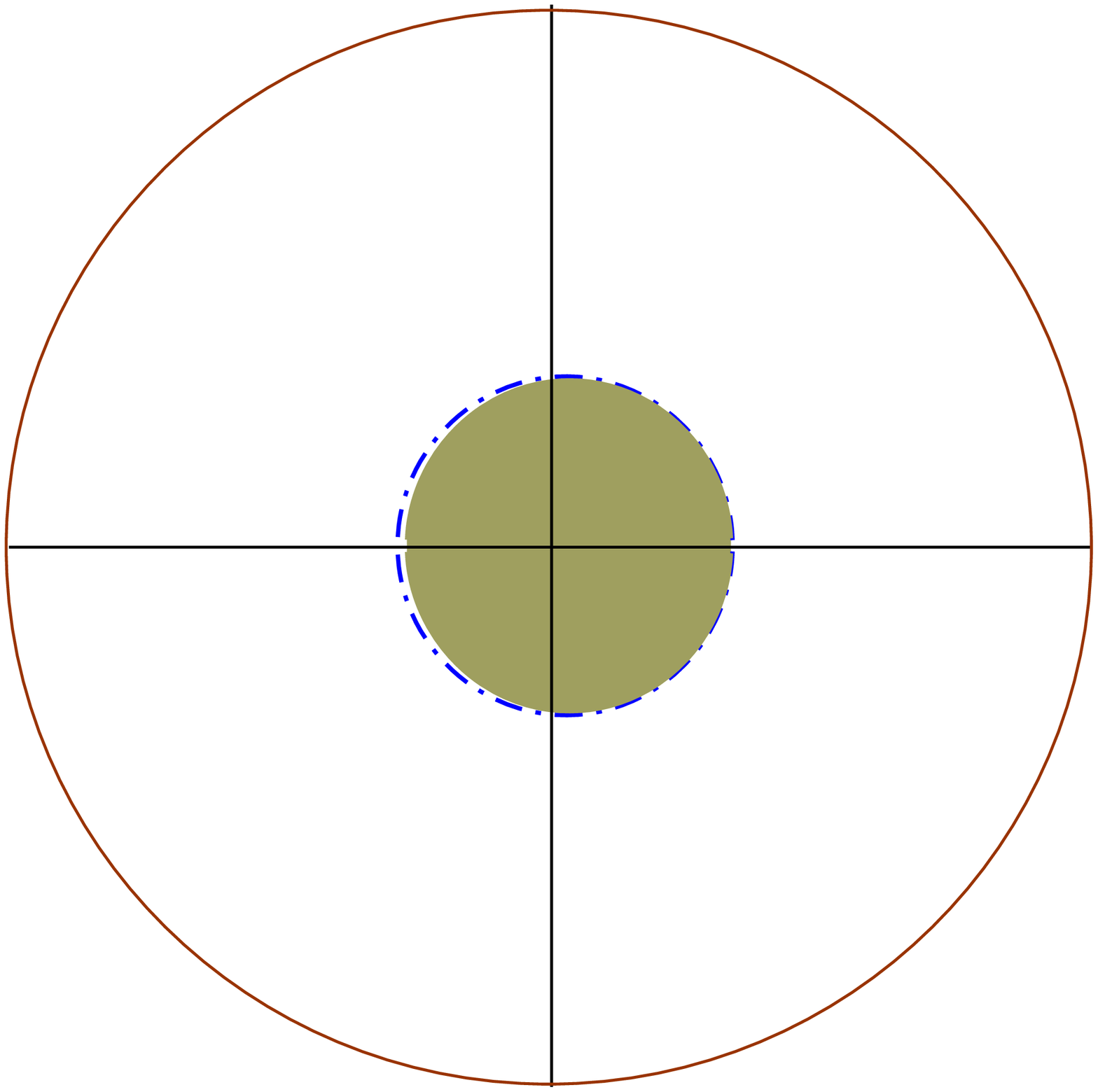}
		}
		\subfigure[]{
		\includegraphics[width=0.3\textwidth]{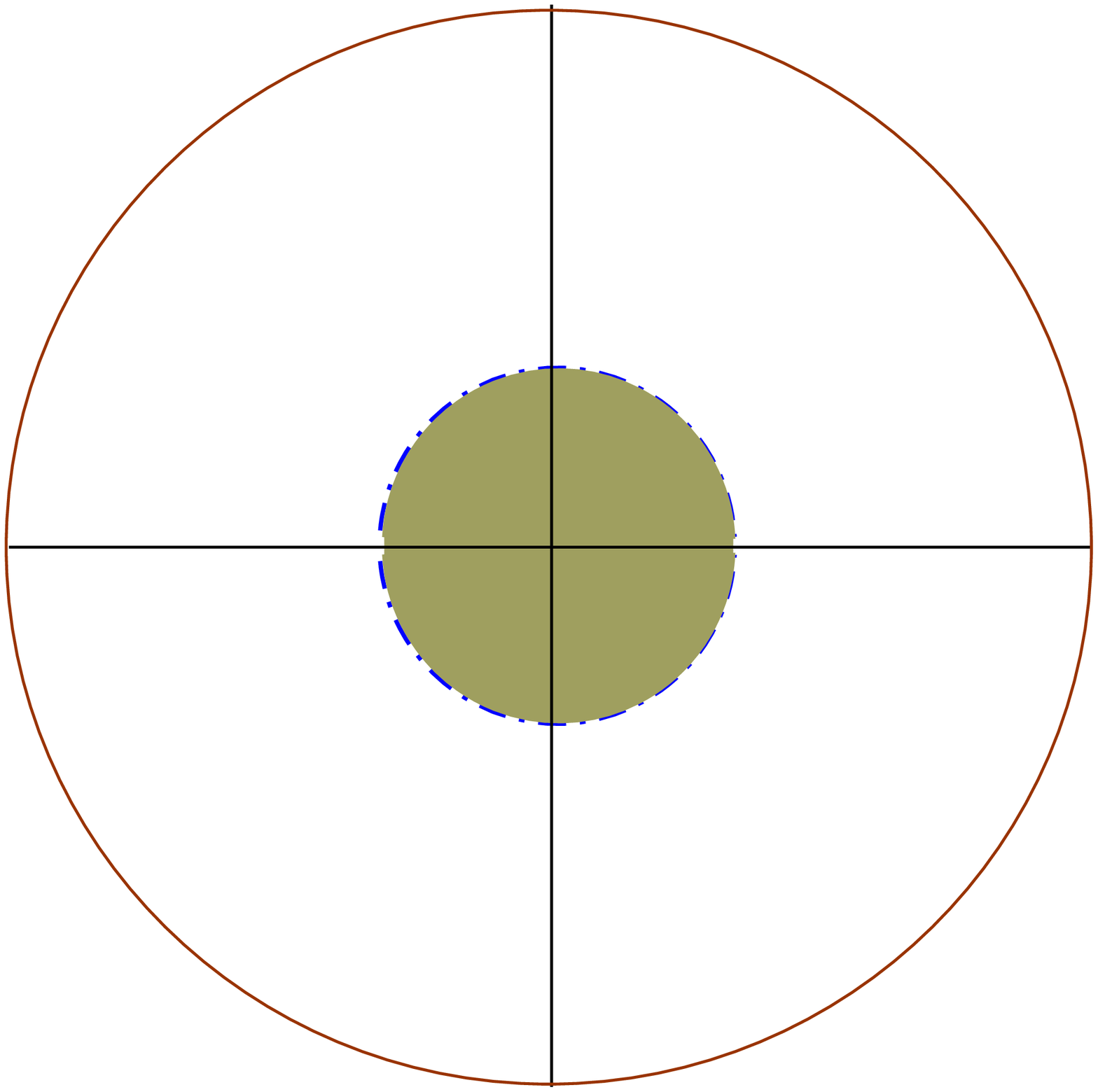}
	}
		
	\caption{\label{fig131}Shadow of black hole in the presence of plasma for specific observer. In this figure $Q_{crit}=0.75$ and $a=\frac{65}{100}a_{max}$. $\Lambda=0$, $\Lambda=0.01$ and $\Lambda=0.06$ in (a), (b) and (c) respectively. In each figure, the blue dash-dot line, shows $Q=0$ and the green filled shape, shows $Q=0.75$. The solid brown circle is reference circle. The detail of parameter are shown in table~\ref{tab8} (see Appendix).}
 \label{pic:shadow}
\end{figure}

One can see that by increasing $Q$ the size of shadow become smaller, and by increasing $a$ the shape of shadow deviate into the horizontal axis, which effect of spin parameter in this situation is similar to vacuum case for an observer in specific coordinate ($r_{O}$ ,$\theta_{O}$), but it is important to note that in similar circumstances, the radius of shadow are smaller than or equal to radius of black hole's shadow in vacuum.

In addition, we analyze the effect of inclination $\theta_{O}$ on the shape of black hole's shadow in Fig. \ref{fig14}. We see that the symmetry into the vertical axis becomes better, if the observer approaches the axis, in the limit of $\theta_{O} \longrightarrow 0$.
\begin{figure}[h]\label{fig14}
	\centering
	\subfigure[$\theta_{O}\simeq 0$]{
		\includegraphics[width=0.3\textwidth]{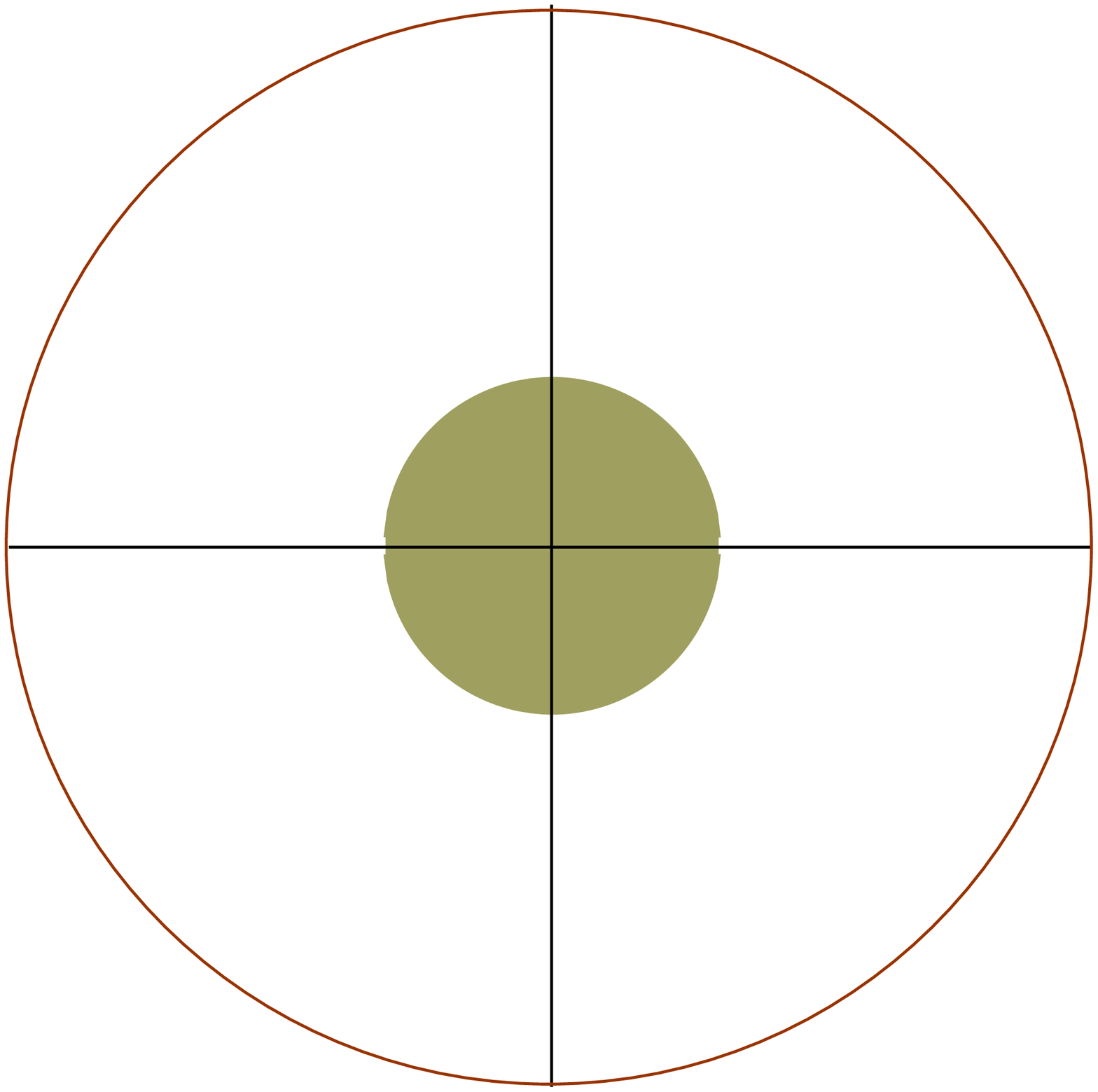}
	}
	\subfigure[$\theta_{O}$=$\frac{\pi}{6}$ ]{
		\includegraphics[width=0.3\textwidth]{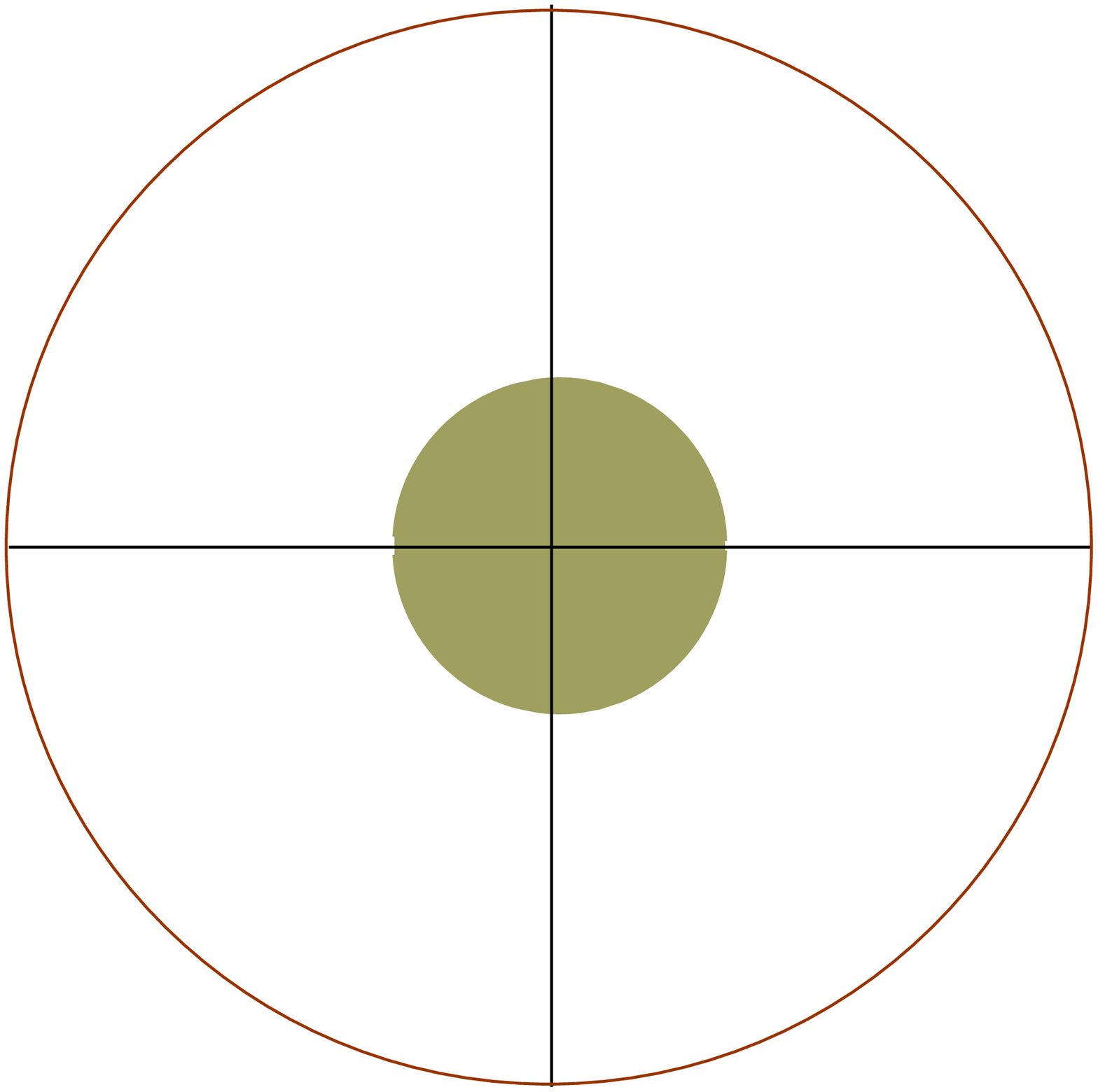}
	}
	\subfigure[$\theta_{O}$=$\frac{\pi}{4}$]{
		\includegraphics[width=0.3\textwidth]{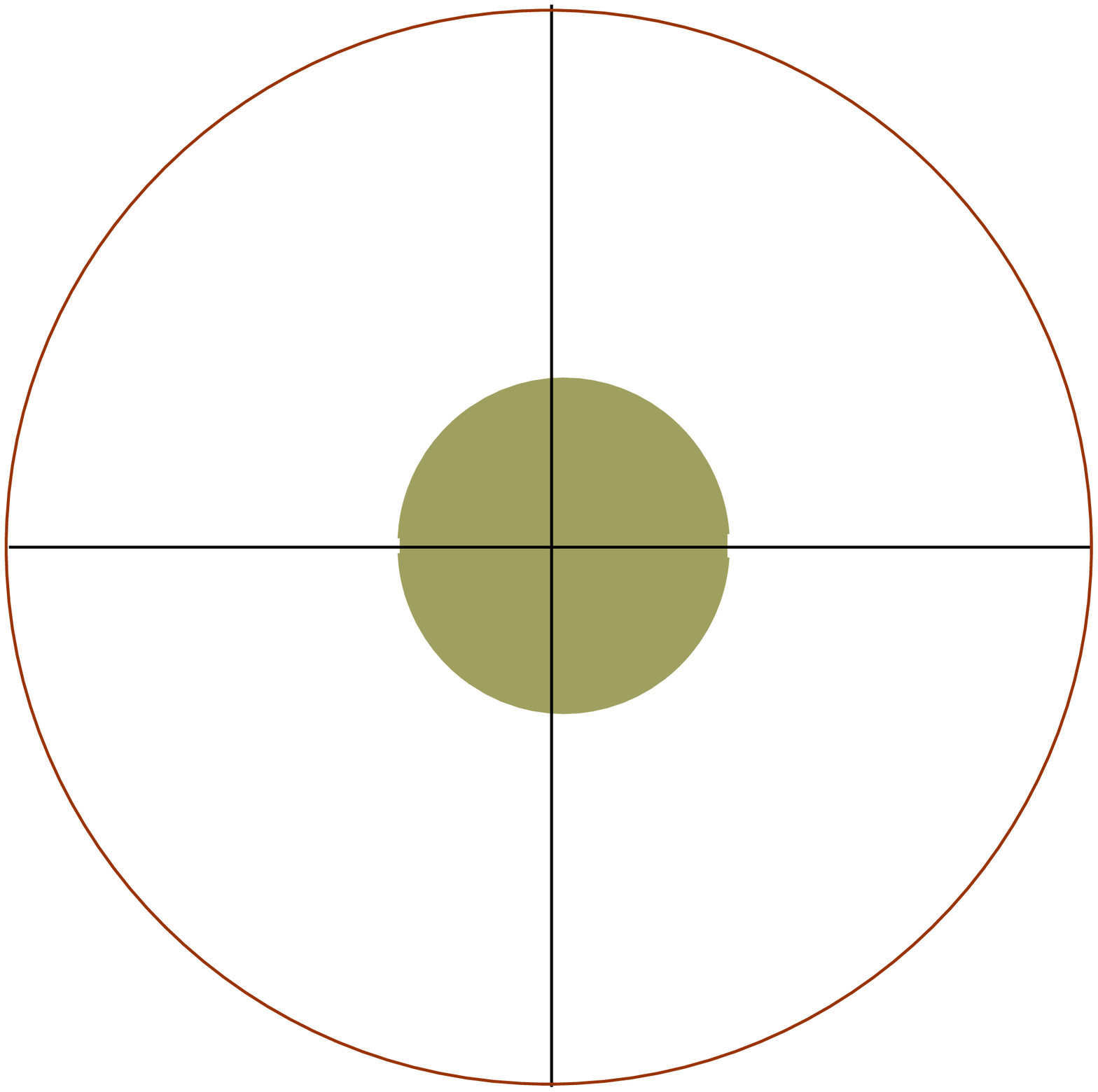}
	}
	\subfigure[$\theta_{O}$=$\frac{pi}{3}$]{
		\includegraphics[width=0.3\textwidth]{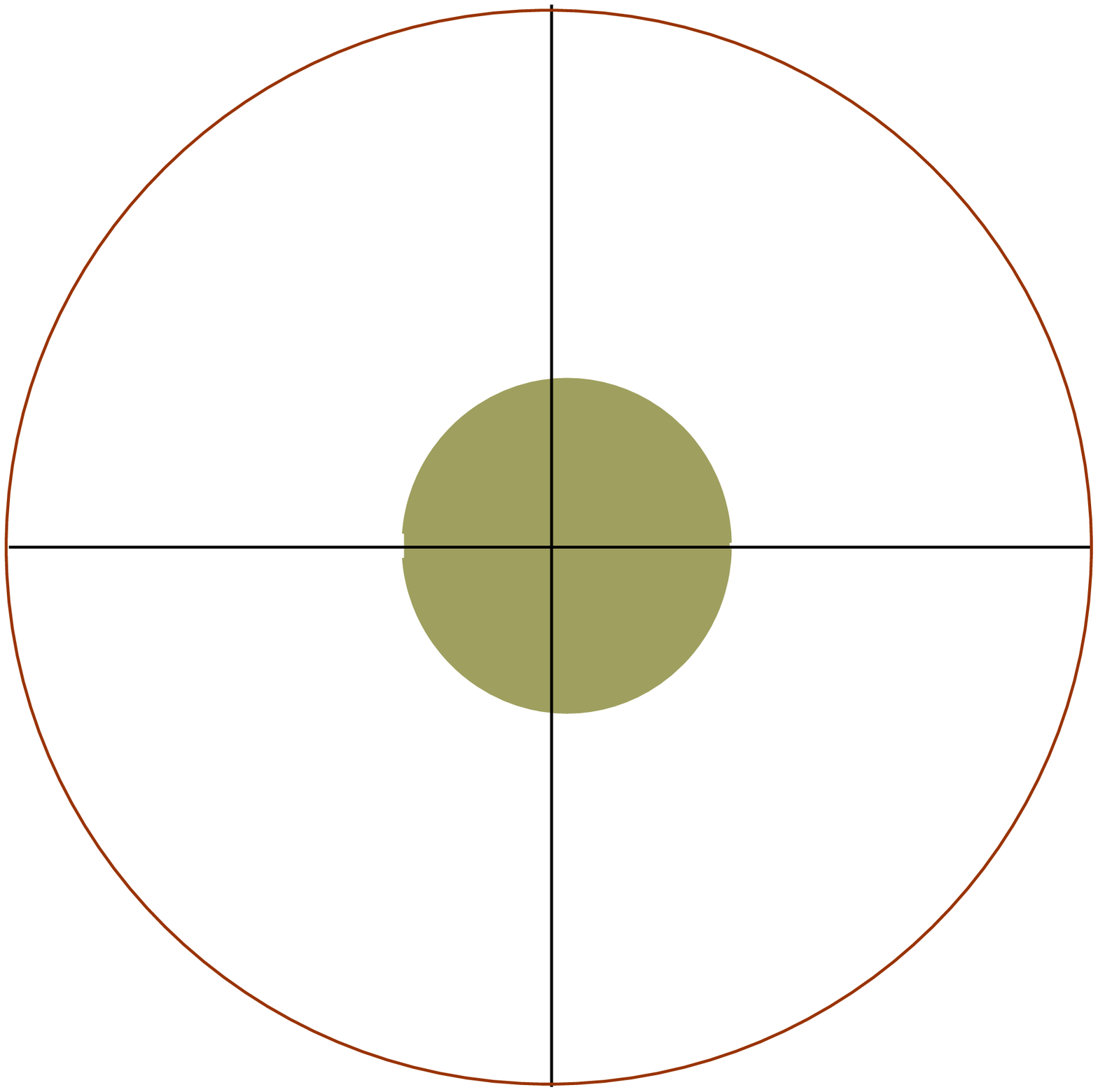}
	}
	\subfigure[$\theta_{O}$=$\frac{\pi}{2}$]{
		\includegraphics[width=0.3\textwidth]{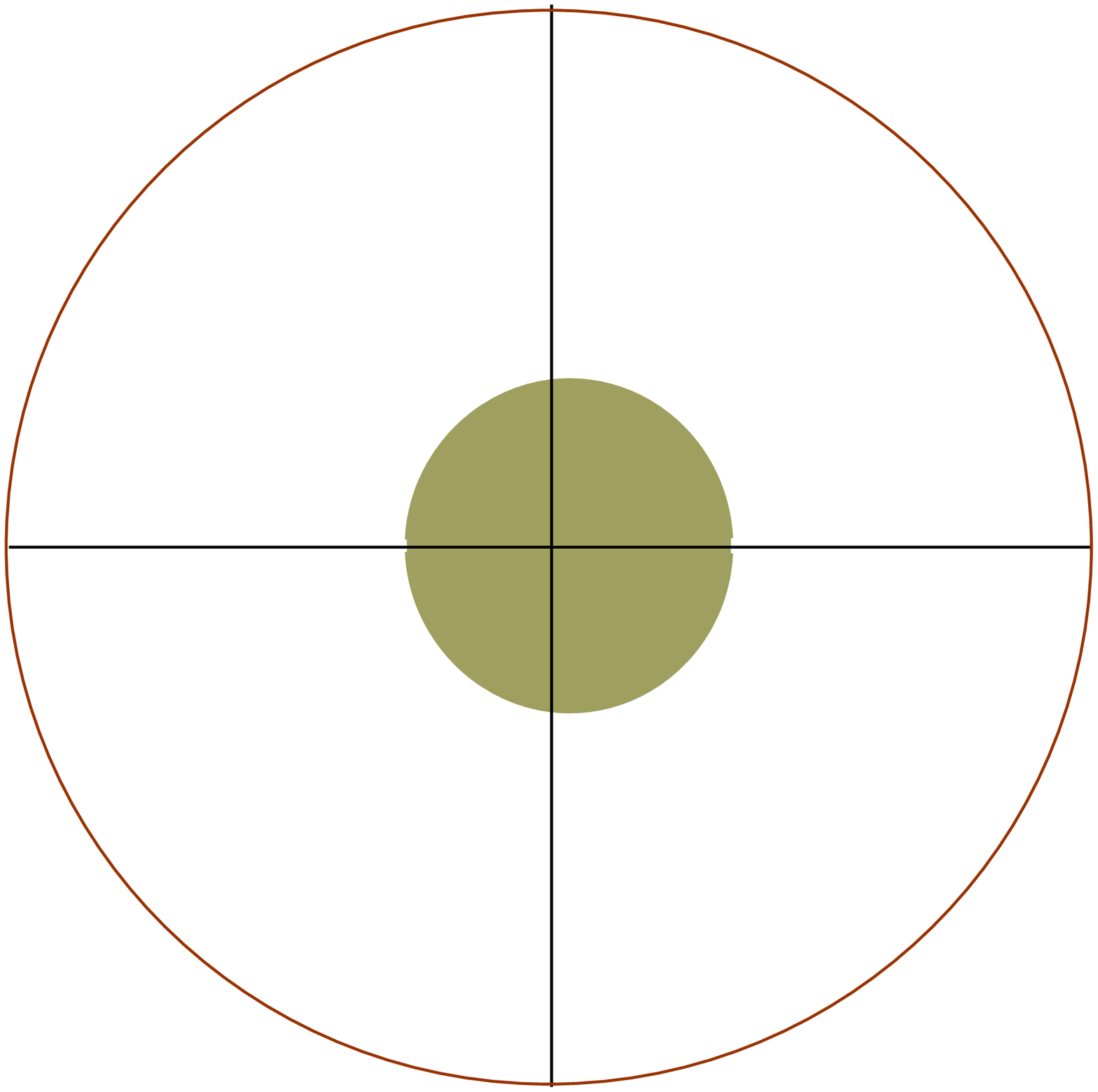}
	}
	\caption{\label{fig14}Shadow of a black hole for an observer at $r_{O}=5M$ and different inclination angles $\theta_{O}$, with fixed $Q=0.75$,$\Lambda=0.01$ and $a=\frac{65}{100}a_{max}$.}
 \label{pic:shadow}
\end{figure}

\clearpage
\section{CONCLUSIONS}\label{conclusion}
In this paper, we investigated a charged rotating black holes in $f(R)$ gravity. First, we obtained geodesic equation for this space time. Then, using geodesic equations, we studied image of black hole's shadow in the absence and the presence of plasma for an observer at infinity and limited distance.
The results show that space time parameters, effect on the size and symmetry of black hole's shadow. It can be seen that for all cases, increasing the rotation parameter, reduces circular symmetry of shadow image of the black hole. Also, it is shown that by increasing the electric charge $Q$, the size of shadow, decreases. Furthermore, the effect of modified gravity parameter $f'(R_{0})$, was investigated for an observer who was placed at infinity in vacuum. We observed that, by increasing $f'(R_{0})$, the size of shadow image, increases and the symmetry of black hole's shadow will improve. 
In addition, for an observer at infinity, size of shadow in the presence of plasma is less than or equal to the size of shadow in vacuum.
Also, changes of inclination angle $\theta_{O}$ and $\Lambda$ parameter, were investigated for an observer at limited distance. We show that, by changing $\theta_{O}$ from $\frac{\pi}{2}$ to limit of zero, deviation of shadow image decreases in vacuum and in the presence of plasma. In addition, we show that for an obserever at limited distance in vacuum, increasing $\Lambda$, will effect on the size of shadow and it becomes smaller. All detail data of figures, have been shown in table~(\ref{tab1}--\ref{tab8}) (see Appendix). 
For future research, it would be interesting to study black hole's shadow in Kerr-Sen Dilaton-Axion space time.

\clearpage
\section*{Appendix}
\begin{table}[h]
\begin{center}
\begin{tabular}{|c|c|c|c|c|}

\hline
$\qquad$Constant parameters$\qquad$&Fig.&$\qquad$ $a$ $\qquad$&$\qquad$ $Q_{crit}$ $\qquad$& $\qquad$ $\Delta_{r}(roots)$ $\qquad$ \\
\hline\hline

 $\Lambda=0 $ & 1(a) &0 & 1.49 & 1.11,0.88\\                    
$\theta_{O}=\frac{\pi}{2}$ & 1(b) & 0.5 &  1.29 & 1.10,0.89 \\
 $M=1$& 1(c) & 0.7 & 1.07 & 1.03,0.96\\
 $f'(R_{0})=1.25$ & 1(d) &1 &   0 & 1

\\ \hline
\end{tabular}
\caption{\label{tab1}The details of parameter in Fig.\ref{infinity}, for an observer at infinity, the $Q$ and $a$ parameter was investigated in this figure in vacuum.}
\label{tab1}
\end{center}
\end{table}

\begin{table}[h]
\begin{center}
\begin{tabular}{|c|c|c|c|c|}
\hline
Constant parameters&Fig.&$\qquad$ $Q$ $\qquad$&$\qquad$ $a_{max}$ $\qquad$& $\Delta_{r}(roots)$\\
\hline\hline

 $\theta_{O}=\frac{\pi}{2}$&6(a)&0&1&1\\
 $\Lambda=0$&6(b)&0.75&0.86&1.10,0.89\\
 $M=1$&6(c)&1.35&0.43&1.07,0.92                  
                       
\\ \hline

$\theta_{O}=\frac{\pi}{2}$&7(a)&0&1&0.92,1.09,16.22,-18.24\\
 $\Lambda=0.01$&7(b)&0.75&0.86&0.88,1.13,16.23,-18.24\\
 $M=1$&7(c)&1.35&0.44&0.98,1.02,16.24,-18.26
\\ \hline
 $\theta_{O}=\frac{\pi}{2}$&8(a)&0&1.02&0.99,1.14,5.75,-7.90\\
 $\Lambda=0.06$&8(b)&0.75&0.88&0.93,1.19,5.78,-7.91\\
 $M=1$&8(c)&1.35&0.46&0.97,1.12,5.84,-7.94
\\ \hline

\end{tabular}
\caption{\label{tab2}The details of parameter in Fig.\ref{fig2}--\ref{fig4}, for an observer at limited distance, effect of spin parameter, $a$ was investigated in this figure in vacuum.}
\label{tab2}
\end{center}
\end{table}


\begin{table}[h]
\begin{center}
\begin{tabular}{|c|c|c|c|c|c|c|c|c|c|}
\hline
Constant parameters&Fig.&$\Lambda$ &$\qquad$ $Q$ $\qquad$&$\frac{a}{a_{max}}$&$\qquad$ $a_{max}$ $\qquad$& $\Delta_{r}(roots)$\\
\hline\hline

$\theta_{O}=\frac{\pi}{2}$&&0&&&&1.71,0.29\\
 $f'(R_{0})=1.25$&10(a)&0.01&0&$\frac{70}{100}$&1&0.28,1.73,16.21,-18.24\\
 $M=1$&&0.06&&&&0.28,1.89,5.72,-7.90              
                       
\\ \hline
$\theta_{O}=\frac{\pi}{2}$&&0&&&&1.62,0.37\\
 $f'(R_{0})=1.25$&10(b)&0.01&0.75&$\frac{70}{100}$&0.86&0.37,1.64,16.22,-18.25\\
 $M=1$&&0.06&&&&0.37,1.78,5.76,-7.92             
                       
\\ \hline
$\theta_{O}=\frac{\pi}{2}$&&0&&&&1.29,0.70\\
 $f'(R_{0})=1.25$&10(c)&0.01&1.35&$\frac{70}{100}$&0.46&0.70,1.31,16.24,-18.26\\
 $M=1$&&0.06&&&&0.69,1.41,5.84,-7.94 
                       
\\ \hline
$\theta_{O}=\frac{\pi}{2}$&&0&&&&1.10,0.89\\
 $f'(R_{0})=1.25$&10(d)&0.01&1.49&$\frac{70}{100}$&0&0.88,1.13,16.25,-18.26\\
 $M=1$&&0.06&&&&0.85,1.24,5.86,-7.95               
                       
\\ \hline
\end{tabular}
\caption{The details of parameter in Fig.\ref{fig5}, for an observer at limited distance. The effect of $\Lambda$, was investigated in this figure in vacuum.}
\label{tab4}
\end{center}
\end{table}

\begin{table}[h]
\begin{center}
\begin{tabular}{|c|c|c|c|c|c|c|c|c|c|}
\hline
Constant parameter&Fig.&$\qquad$ $Q_{crit}$ $\qquad$&$\qquad$ $a$ $\qquad$& $\Delta_{r}(roots)$\\
\hline\hline

 $\Lambda=0$&11(a)&1.48&0&1.08,0.89\\

  $\theta_{O}=\frac{\pi}{2}$&11(b)&1.26&0.5&1.21,0.78\\

$M=1$&11(c)&1&0.7&1.24,0.75\\

$f^{'}(R_{0})=1.25$&11(d)&0&1&1.34,0.65               
                       
\\ \hline
\end{tabular}
\caption{The details of parameter in Fig.\ref{fig9} for an observer at $\infty$, the effect of $a$ and $Q$, was investigated in this figure in the presence of plasma. Plasma parameter is $k=0.1$.}
\label{tab5}
\end{center}
\end{table}

\begin{table}[h]
\begin{center}
\begin{tabular}{|c|c|c|c|c|c|c|c|c|c|}
\hline
Constant parameters&Fig.&$\Lambda$&$\qquad$ $Q$ $\qquad$&$\qquad$ $a_{max}$ $\qquad$& $\Delta_{r}(roots)$\\
\hline\hline
$\theta_{O}=\frac{\pi}{2}$&13(a)&0&& 0.86&1.62,0.37\\                       

 $f^{'}(R_{0})=1.25$&13(b)&0.01&0.75& 0.86&1.64,0.37,16.22,-18.25 \\                      

$M=1$&13(c)&0.06&& 0.88&1.78,0.37,5.76,-7.92                       
\\ \hline
\end{tabular}
\caption{The details of parameter in Fig.\ref{fig120}, the effect of spin parameter, $a$ for an observer at limited distance in the presence of plasma. Plasma parameter is $k=0.1$.}
\label{tab7}
\end{center}
\end{table}

\begin{table}[h]
\begin{center}
\begin{tabular}{|c|c|c|c|c|c|c|c|c|c|}
\hline
Constant parameter&Fig.&$\Lambda$ &$\qquad$ $Q_{crit}$ $\qquad$&$\qquad$ $a$ $\qquad$& $\Delta_{r}(roots)$\\
\hline\hline
$\theta_{O}=\frac{\pi}{2}$&14(a)&0&& 0.559&1.66,0.33\\                       

$f^{'}(R_{0})=1.25$&14(b)&0.01&0.75& 0.559&1.85,0.17,16.21,-18.24\\                       

$M=1$&14(c)&0.06&& 0.344&1.97,0.20,5.74,-7.92                       
\\ \hline
\end{tabular}
\caption{The details of parameter in Fig.\ref{fig131}, the effect of electric charge parameter, $Q$ for an observer at limited distance in the presence of plasma. The plasma parameter is $k=0.1$.}
\label{tab8}
\end{center}
\end{table}

\clearpage

\bibliographystyle{amsplain}

\begin{thebibliography}{5}

\bibitem{Antonucci:2011zza} 
  F.~Antonucci {\it et al.},
  Class.\ Quant.\ Grav.\  {\bf 28}, 094001 (2011).
  doi:10.1088/0264-9381/28/9/094001


\bibitem{Doeleman:2008qh}
  S.~Doeleman {\it et al.},
  Nature {\bf 455} (2008) 78
  [arXiv:0809.2442 [astro-ph]].
  
\bibitem{EHT:2008qh}
project website:www.Eventhorizontelescope.org.


\bibitem{BHC:2008qh}
project website:BlackHoleCam.org.

\bibitem{synge}
J. L. Synge. 1966. Mon.Not.Roy.Astron.Soc.,131,463.

\bibitem{Chandrasekhar:1985kt} 
  S.~Chandrasekhar,
  OXFORD, UK: CLARENDON (1985) 646 P.
  
\bibitem{Virbhadra:1999nm} 
  K.~S.~Virbhadra and G.~F.~R.~Ellis,
  Phys.\ Rev.\ D {\bf 62}, 084003 (2000)
  [astro-ph/9904193].
 
\bibitem{Claudel:2000yi} 
  C.~M.~Claudel, K.~S.~Virbhadra and G.~F.~R.~Ellis,
  J.\ Math.\ Phys.\  {\bf 42}, 818 (2001)
  [gr-qc/0005050].

\bibitem{Bardeen}
J. M. Bardeen, in Black Holes (Les Astres Occlus), edited by
C. DeWitt and B. S. DeWitt (Gordon and Breach, New York,
1973) p. 215

\bibitem{kerr-newman}
A. de Vries, Class. Quantum Grav. 17, 123 (2000)

\bibitem{Abdujabbarov:2012bn}
  A.~Abdujabbarov, F.~Atamurotov, Y.~Kucukakca, B.~Ahmedov and U.~Camci,
  Astrophys.\ Space Sci.\  {\bf 344} (2013) 429
  [arXiv:1212.4949 [physics.gen-ph]].
  
\bibitem{Li:2013jra}
  Z.~Li and C.~Bambi,
  JCAP {\bf 1401} (2014) 041
  [arXiv:1309.1606 [gr-qc]].

\bibitem{Yumoto:2012kz}
  A.~Yumoto, D.~Nitta, T.~Chiba and N.~Sugiyama,
  Phys.\ Rev.\ D {\bf 86} (2012) 103001
  [arXiv:1208.0635 [gr-qc]].

\bibitem{Amarilla:2010zq}
  L.~Amarilla, E.~F.~Eiroa and G.~Giribet,
  Phys.\ Rev.\ D {\bf 81} (2010) 124045
  [arXiv:1005.0607 [gr-qc]].

\bibitem{Amarilla:2011fx}
  L.~Amarilla and E.~F.~Eiroa,
  Phys.\ Rev.\ D {\bf 85} (2012) 064019
  [arXiv:1112.6349 [gr-qc]].
  
\bibitem{Cunha:2015yba}
  P.~V.~P.~Cunha, C.~A.~R.~Herdeiro, E.~Radu and H.~F.~Runarsson,
  Phys.\ Rev.\ Lett.\  {\bf 115} (2015) no.21,  211102
  doi:10.1103/PhysRevLett.115.211102
  [arXiv:1509.00021 [gr-qc]].
  
\bibitem{Vincent:2016sjq}
  F.~H.~Vincent, E.~Gourgoulhon, C.~Herdeiro and E.~Radu,
  arXiv:1606.04246 [gr-qc].

\bibitem{Hioki:2009na}
  K.~Hioki and K.~i.~Maeda,
  Phys.\ Rev.\ D {\bf 80} (2009) 024042
  [arXiv:0904.3575 [astro-ph.HE]].
  
\bibitem{Grenzebach:2014fha}
  A.~Grenzebach, V.~Perlick and C.~Lämmerzahl,
  Phys.\ Rev.\ D {\bf 89} (2014) no.12,  124004
  [arXiv:1403.5234 [gr-qc]].
\bibitem{Brax:2003fv} 
  P.~Brax and C.~van de Bruck,
  Class.\ Quant.\ Grav.\  {\bf 20}, R201 (2003)
  [hep-th/0303095].
  L.~A.~Gergely,
  Phys.\ Rev.\ D {\bf 74}, 024002 (2006)
  [hep-th/0603244].
  M.~Demetrian,
  Gen.\ Rel.\ Grav.\  {\bf 38}, 953 (2006)
  [gr-qc/0506028].
  
\bibitem{Lovelock:1971yv} 
  D.~Lovelock,
  J.\ Math.\ Phys.\  {\bf 12}, 498 (1971).
  D.~Lovelock,
  J.\ Math.\ Phys.\  {\bf 13}, 874 (1972).
  S.~H.~Hendi and M.~H.~Dehghani,
  Phys.\ Lett.\ B {\bf 666}, 116 (2008)
  [arXiv:0802.1813 [hep-th]].
  M.~H.~Dehghani and R.~Pourhasan,
  Phys.\ Rev.\ D {\bf 79}, 064015 (2009)
  [arXiv:0903.4260 [gr-qc]].
  S.~H.~Hendi, S.~Panahiyan and H.~Mohammadpour,
  Eur.\ Phys.\ J.\ C {\bf 72}, 2184 (2012)
  [arXiv:1501.05841 [gr-qc]].
  A.~Sheykhi, H.~Moradpour and N.~Riazi,
  Gen.\ Rel.\ Grav.\  {\bf 45}, 1033 (2013)
  [arXiv:1109.3631 [physics.gen-ph]].
\bibitem{Brans:1961sx} 
  C.~Brans and R.~H.~Dicke,
  Phys.\ Rev.\  {\bf 124}, 925 (1961).
  Y.~Fujii and K.~Maeda.~\textit{The Scalar-Tensor Theory of Gravitation}~(Cambridge University Press, Cambridge, 2003).
  T.~P.~Sotiriou,
  Class.\ Quant.\ Grav.\  {\bf 23}, 5117 (2006)
  [gr-qc/0604028].
   \bibitem{Riess:1998cb}
  A.~G.~Riess {\it et al.}  [Supernova Search Team Collaboration],
  Astron.\ J.\  {\bf 116}, 1009 (1998)
  [astro-ph/9805201].
  S.~Perlmutter {\it et al.}  [Supernova Cosmology Project Collaboration],
  Astrophys.\ J.\  {\bf 517}, 565 (1999)
  [astro-ph/9812133].
  J.~L.~Tonry {\it et al.}  [Supernova Search Team Collaboration],
  Astrophys.\ J.\  {\bf 594}, 1 (2003)
  [astro-ph/0305008].
  C.~L.~Bennett {\it et al.}  [WMAP Collaboration],
  Astrophys.\ J.\ Suppl.\  {\bf 148}, 1 (2003)
  [astro-ph/0302207].
  G.~Hinshaw {\it et al.}  [WMAP Collaboration],
  Astrophys.\ J.\ Suppl.\  {\bf 170}, 288 (2007)
  [astro-ph/0603451].

\bibitem{Buchdahl:1983zz}
  H.~A.~Buchdahl,
  Mon.\ Not.\ Roy.\ Astron.\ Soc.\  {\bf 150}, 1 (1970).

\bibitem{Starobinsky:1980te}
  A.~A.~Starobinsky,
  Phys.\ Lett.\ B {\bf 91}, 99 (1980).

  \bibitem{Bamba:2008ja}
  K.~Bamba and S.~D.~Odintsov,
  JCAP {\bf 0804}, 024 (2008)
  [arXiv:0801.0954 [astro-ph]].

\bibitem{Akbar:2006mq}
  M.~Akbar and R.~G.~Cai,
  Phys.\ Lett.\ B {\bf 648}, 243 (2007)
  [gr-qc/0612089].
  K.~Bamba and S.~D.~Odintsov,
  JCAP {\bf 0804}, 024 (2008)
  [arXiv:0801.0954 [astro-ph]].
  G.~Cognola, E.~Elizalde, S.~Nojiri, S.~D.~Odintsov, L.~Sebastiani and S.~Zerbini,
  Phys.\ Rev.\ D {\bf 77}, 046009 (2008)
  [arXiv:0712.4017 [hep-th]].
  C.~Corda,
  Int.\ J.\ Mod.\ Phys.\ D {\bf 18}, 2275 (2009)
  [arXiv:0905.2502 [gr-qc]].
  S.~Capozziello, F.~Darabi and D.~Vernieri,
  Mod.\ Phys.\ Lett.\ A {\bf 25}, 3279 (2010)
  [arXiv:1009.2580 [gr-qc]].
  S.~H.~Hendi and D.~Momeni,
  Eur.\ Phys.\ J.\ C {\bf 71}, 1823 (2011)
  [arXiv:1201.0061 [gr-qc]].
  S.~Asgari and R.~Saffari,
  Gen.\ Rel.\ Grav.\  {\bf 44}, 737 (2012)
  [arXiv:1104.5108 [gr-qc]].
  S.~H.~Mazharimousavi, M.~Halilsoy and T.~Tahamtan,
  Eur.\ Phys.\ J.\ C {\bf 72}, 1958 (2012)
  [arXiv:1109.3655 [gr-qc]].
  S.~G.~Ghosh, S.~D.~Maharaj and U.~Papnoi,
  Eur.\ Phys.\ J.\ C {\bf 73}, no. 6, 2473 (2013)
  [arXiv:1208.3028 [gr-qc]].
  S.~H.~Hendi, B.~Eslam Panah and R.~Saffari,
  Int.\ J.\ Mod.\ Phys.\ D {\bf 23} (2014)
  [arXiv:1408.5570 [hep-th]].


\bibitem{Saffari:2007zt}
  R.~Saffari and S.~Rahvar,
  Phys.\ Rev.\ D {\bf 77}, 104028 (2008)
  [arXiv:0708.1482 [astro-ph]].
  
\bibitem{Chakraborty:2014xla} 
  S.~Chakraborty and S.~SenGupta,
  Eur.\ Phys.\ J.\ C {\bf 75}, no. 1, 11 (2015)
  [arXiv:1409.4115 [gr-qc]].
  
\bibitem{Chakraborty:2015bja} 
  S.~Chakraborty and S.~SenGupta,
  Eur.\ Phys.\ J.\ C {\bf 75}, no. 11, 538 (2015)
  [arXiv:1504.07519 [gr-qc]].
  
\bibitem{Soroushfar:2015wqa}
  S.~Soroushfar, R.~Saffari, J.~Kunz and C.~L\"ammerzahl,
  Phys.\ Rev.\ D {\bf 92}, no. 4, 044010 (2015)
  [arXiv:1504.07854 [gr-qc]].
  
\bibitem{Soroushfar:2016nbu} 
  S.~Soroushfar, R.~Saffari and N.~Kamvar,
  arXiv:1605.00767 [gr-qc].
  
\bibitem{Larranaga:2011fv}
  A.~Larranaga,
  Pramana {\bf 78} (2012) 697
  [arXiv:1108.6325 [gr-qc]].
  
  \bibitem{B.Carter}
  B. Carter. 
   Phys.Rev., 174, 5:1559, 1968.
   
\bibitem{Wei:2013kza}
  S.~W.~Wei and Y.~X.~Liu,
  JCAP {\bf 1311} (2013) 063
  doi:10.1088/1475-7516/2013/11/063
  [arXiv:1311.4251 [gr-qc]].

\bibitem{Bardeen:1972fi}
  J.~M.~Bardeen, W.~H.~Press and S.~A.~Teukolsky,
  Astrophys.\ J.\  {\bf 178} (1972) 347.
  
\bibitem{Vazquez:2003zm}
  S.~E.~Vazquez and E.~P.~Esteban,
  Nuovo Cim.\ B {\bf 119} (2004) 489
  [gr-qc/0308023].
  
\bibitem{Abdujabbarov:2015xqa} 
  A.~A.~Abdujabbarov, L.~Rezzolla and B.~J.~Ahmedov,
  Mon.\ Not.\ Roy.\ Astron.\ Soc.\  {\bf 454}, no. 3, 2423 (2015)
  doi:10.1093/mnras/stv2079
  [arXiv:1503.09054 [gr-qc]].  
  
	\bibitem{Griffiths}
 J. B. Griffiths and J. Podolský. 2009. \textit{Exact Space-Times in Einstein’s General Relativity.} (Cambridge University Press, Cambridge).
 
\bibitem{Bick:1975}
  J. Bicak, P. Hadrava, \textit{Astronomy and Astrophysics} 44, 389 (1975)
  
  \bibitem{Tsupko:2012}
  O.Y. Tsupko, G.S. Bisnovatyi-Kogan, \textit{Gravitation
and Cosmology} 18, 117 (2012).


  \bibitem{Morozova:2012}
V.S. Morozova, B.J. Ahmedov, A.A. Tursunov, \textit{Astrophys Space Sci} 346, 513 (2013).

\bibitem{Er:2013efa}
  X.~Er and S.~Mao,
  Mon.\ Not.\ Roy.\ Astron.\ Soc.\  {\bf 437} (2014) no.3,  2180
  [arXiv:1310.5825 [astro-ph.CO]].
  
\bibitem{synge:11}
  J. L. Synge, Relativity: \textit{The General Theory}. (NorthHolland, Amsterdam, 1960)



\bibitem{Atamurotov:2015nra}
  F.~Atamurotov, B.~Ahmedov and A.~Abdujabbarov
  Phys.\ Rev.\ D {\bf 92} (2015) 084005
  [arXiv:1507.08131 [gr-qc]].
  
\bibitem{BisnovatyiKogan:2010ar}
  G.~S.~Bisnovatyi-Kogan and O.~Y.~Tsupko,
  Mon.\ Not.\ Roy.\ Astron.\ Soc.\  {\bf 404} (2010) 1790
  doi:10.1111/j.1365-2966.2010.16290.x
  [arXiv:1006.2321 [astro-ph.CO]].
  
 
\bibitem{Abdujabbarov:2015pqp} 
  A.~Abdujabbarov, B.~Toshmatov, Z.~Stuchlík and B.~Ahmedov,
  arXiv:1512.05206 [gr-qc].
  
\bibitem{Rogers:2015dla}
  A.~Rogers,
  Mon.\ Not.\ Roy.\ Astron.\ Soc.\  {\bf 451} (2015) no.1,  17
  [arXiv:1505.06790 [gr-qc]].
   





  
 
  
   
  
 \end{thebibliography}

 \end{document}